\documentclass[apj]{emulateapj}

\pdfoutput=1
\usepackage{url}
\usepackage{amsmath}
\usepackage{multirow}

\def\kms{km~s$^{-1}$}
\def\degpnt{^{\circ}\kern-1.7mm.\kern+.35mm}
\def\arcpnt{"\kern-1.7mm.\kern+.35mm}
\def\minpnt{'\kern-1.0mm.\kern+.30mm}
\def\deg{^{\circ}}

\slugcomment{Accepted to ApJ}

\shorttitle{{\it WISE} Tully-Fisher Relation Calibration}
\shortauthors{Neill, et al.}

\begin{document}


\title{The Calibration of the {\it WISE} W1 and W2 Tully-Fisher Relation}


\author{J.~D.~Neill\altaffilmark{1}, Mark~Seibert\altaffilmark{2}, 
	R. Brent Tully\altaffilmark{3}, H\'{e}l\`{e}ne Courtois\altaffilmark{4},
	Jenny G. Sorce\altaffilmark{4,5}, T. H. Jarrett\altaffilmark{6}, 
	Victoria Scowcroft\altaffilmark{2}, and Frank J. Masci\altaffilmark{7}
}

\altaffiltext{1}{California Institute of Technology, 1200 E. California
Blvd. MC 278-17, Pasadena, CA 91125, USA}
\altaffiltext{2}{The Observatories of the Carnegie Institute of Washington, 813 Santa Barbara Street, Pasadena, CA  91101, USA}
\altaffiltext{3}{Institute for Astronomy, University of Hawaii, 2680 Woodlawn Drive, HI 96822, USA}
\altaffiltext{4}{Institut de Physique Nucleaire, Universit\'e Claude Bernard Lyon I, Lyon, France}
\altaffiltext{5}{Leibniz-Institut f\"{u}r Astrophysik, Potsdam, Germany}
\altaffiltext{6}{University of Cape Town, Private Bag X3, Rondebosch 7701, Republic of South Africa}
\altaffiltext{7}{Image Processing and Analysis Center (IPAC), California
Institute of Technology, 1200 E. California Blvd. MC 100-22, Pasadena, CA  91125, USA}


\begin{abstract}

In order to explore local large-scale structures and velocity fields,
accurate galaxy distance measures are needed.  We now extend the well-tested
recipe for calibrating the correlation between galaxy rotation
rates and luminosities -- capable of providing such distance measures -- to
the all-sky, space-based imaging data from the Wide-field Infrared Survey
Explorer ({\it WISE}\,) W1 ($3.4\mu$m) and W2 ($4.6\mu$m) filters.  We find
a linewidth to absolute magnitude correlation (known as the Tully-Fisher
Relation, TFR) of $\mathcal{M}^{b,i,k,a}_{W1} = -20.35 - 9.56 (\log
W^i_{mx} - 2.5)$ (0.54 magnitudes rms) and $\mathcal{M}^{b,i,k,a}_{W2}
= -19.76 - 9.74 (\log W^i_{mx} - 2.5)$ (0.56 magnitudes rms)
from 310 galaxies in 13 clusters. We update the I-band TFR using a sample
9\% larger than in \citet{Tully:12:78}.  We derive $\mathcal{M}^{b,i,k}_I =
-21.34 - 8.95 (\log W^i_{mx} - 2.5)$ (0.46 magnitudes rms).  The {\it WISE}
TFRs show evidence of curvature.  Quadratic fits give
$\mathcal{M}^{b,i,k,a}_{W1} = -20.48 - 8.36 (\log W^i_{mx} - 2.5) + 3.60
(\log W^i_{mx} - 2.5)^2$ (0.52 magnitudes rms) and
$\mathcal{M}^{b,i,k,a}_{W2} = -19.91 - 8.40 (\log W^i_{mx} - 2.5) + 4.32
(\log W^i_{mx} - 2.5)^2$ (0.55 magnitudes rms).  We apply an I-band --
{\it WISE} color correction to lower the scatter and derive
$\mathcal{M}_{C_{W1}} = -20.22 - 9.12 (\log W^i_{mx} - 2.5)$
and $\mathcal{M}_{C_{W2}} = -19.63 - 9.11 (\log W^i_{mx} - 2.5)$ (both 
0.46 magnitudes rms).  Using our three independent TFRs (W1 curved, W2
curved and I-band), we calibrate the UNION2 supernova Type Ia sample
distance scale and derive $H_0 = 74.4 \pm 1.4$(stat) $\pm\
2.4$(sys) \kms\,Mpc$^{-1}$ with 4\% total error.

\end{abstract}





\section{Introduction\label{SEC:INTRO}}

The utility of calibrating the power-law correlation between galaxy
rotation rates and their luminosities \citep{Tully:77:661} in the
mid-infrared (MIR) has been clearly demonstrated by \citet{Sorce:13:94}.
Their use of the 3.6 $\mu$m {\it SPITZER} Infrared Array Camera (IRAC)
photometry provided a calibration of the Tully-Fisher relation (TFR) with a
scatter comparable to that seen in the I-band \citep{Tully:12:78}.  Having
space-based photometry in the MIR mitigates the effects of dust and removes
possible systematics when attempting to constrain the motions of galaxies
across the entire sky.

While the IRAC 3.6 $\mu$m calibration is useful, there are a limited number
of galaxies that have been observed through the camera's 4-arcminute
field-of-view.  The W1 band of the {\it Wide-field Infrared Survey
Explorer} \citep[{\it WISE},][]{Wright:10:1868} is similar in wavelength
coverage ($\lambda_{eff} = 3.4\ \mu$m), is also space-based and thus enjoys
all the benefits of the IRAC calibration.  In addition, the {\it WISE}
mission has covered the entire sky to a depth similar to the IRAC coverage
of selected nearby galaxies.  The number of calibrator galaxies with {\it
WISE} imaging is 310, an increase of 46\% over the sample available to
\citet{Sorce:13:94}, although the number of calibrator galaxies observed in
the IRAC [3.6] band continues to increase.  This additional utility of
all-sky coverage motivated this work.

The {\it WISE} imaging represents the opportunity for providing
high-quality 3.4 $\mu$m (W1) and 4.6 $\mu$m (W2) photometry over the entire
sky, however, the automated catalog photometry available from the mission
has not been optimized for extended galaxies.  Corrections can be made to
the catalog photometry, however when applied to the TFR the resulting
scatter is significantly larger \citep[0.69 mag,][]{Lagattuta:13:88} than
for the I-band calibration \citep[0.41 mag,][]{Tully:12:78}.  We have instigated a separate project to provide
high-quality surface photometry of all {\it WISE} galaxies larger than 0.8
arcminutes on the sky.  The {\it WISE} Nearby Galaxy Atlas
\citep[WNGA,][]{Seibert:14b} will provide photometry that is quality
controlled for over 20,000 galaxies.  This photometry, optimized for
extended sources, significantly reduces the resulting scatter in the TFR
calibration, and thus improves the resulting distances.  Having an accurate
calibration of the TFR for these two WISE passbands will allow the use of
this large sample to explore the structure and dynamics of local galaxy
bulk flows.  This calibration has been completed and is presented herein.

The focus of this paper is the calibration of the TFR using photometry in
the {\it WISE} W1 and W2 bands, however, we take the occasion to update the
I-band calibration and present it in \S\ref{SEC:ICALIBRATION}.  We
introduce a significant number of new calibration candidates by considering
all galaxies associated with the calibrating clusters (see
\S\ref{SEC:CALIBRATORS}) contained in the 2MASS redshift survey complete to
K=11.75 \citep{Huchra:12:26}.

\citet{Sorce:13:94} found a reduction in the scatter of the TFR when
applying a correction to the IRAC photometry based on the optical-MIR color
with the optical measure being provided by I-band photometry.
Unfortunately, there is no space-based all-sky survey in the I-band.  Thus,
the uncorrected calibration for W1 and W2 may be useful for those wishing
to extend their catalogs to as many galaxies as possible (those without
I-band photometry), even though the scatter will be slightly larger.  While
the TFR in the optical has proven to be a straight power-law, there is
evidence that in the MIR there is curvature in the relation.  We
investigate that possibility and present our results in \S\ref{SEC:CURVE}.

Once an accurate calibration is derived, we can use the distances derived
thereby for calculating the Hubble constant, $H_0$ \citep{Courtois:12:174,
Sorce:12:L12}.  We do this in \S\ref{SEC:H0_CLUSTERS} using the subset of
clusters from the calibration set that have recession velocities that place
them in the Hubble flow \citep[$> 4000$\,\kms,][]{Tully:12:78,
Sorce:13:94}.  For a more robust measure of $H_0$ that extends well into
the Hubble-flow (to $z>1$) we use TFR distances to re-normalize the
distance scale for the UNION2 SN~Ia sample \citep{Amanullah:10:712} and
calculate $H_0$ directly from the re-normalization in
\S\ref{SEC:H0_SUPERNOVAE}.

\section{Data\label{SEC:DATA}}

\subsection{Calibrators\label{SEC:CALIBRATORS}}

We adopt the galaxy cluster technique for deriving the calibration
described by \citet{Tully:12:78} (see also \S\ref{SEC:RELATIVEDISTANCES}).
We take advantage of the fact that the galaxies within a given cluster are
at the same distance and that the galaxy masses, and hence HI linewidths,
span a range large enough to determine the slope of the correlation for
each cluster.  We then shift each cluster along the luminosity axis such
that their data appear to be from a single cluster.  We iteratively combine
the galaxy data derived from a set of thirteen nearby clusters to derive a
universal slope, and then set the zero-point of the relation using the
universal slope applied to nearby galaxies with accurate distance
measurements derived from independent techniques.  To minimize the effect
of the Malmquist bias, the slopes are derived from fitting the inverse
Tully-Fisher relation \citep[ITFR,][]{Willick:94:1}.  Details on the method
and the calibrator and cluster sample can be found in \citet{Tully:12:78}
and \citet{Sorce:13:94}.  The appendix in \citet{Tully:12:78} discusses
issues specific to each of the 13 clusters.

In order to avoid excessive noise in the calibrations we apply several cuts
to the input sample \citep{Tully:12:78}.  Because we must de-project the HI
linewidths based on the observed inclination, we exclude galaxies more
face-on than $45^{\circ}$, the limit where typical errors in the
de-projections begin to exceed 8\%.  Morphological types earlier than Sa
greatly increase the scatter in the relation, most likely due to the mass
of the bulge not contributing to the HI linewidth, and are excluded.
Systems with insufficient or confused HI, and galaxies that appear
disrupted are also excluded. Following \citet{Sorce:13:94}, the
\citet{Tully:12:78} sample for Abell 2634 has been extended to include the
adjacent Abell 2666 which is, within measurement uncertainties, at the same
distance.

\tabletypesize{\tiny}
\setlength{\tabcolsep}{0.01cm}
\begin{deluxetable*}{cccccccccccccccl}
\tablecolumns{16}
\tablecaption{Calibrator Data\\
(full table in online version)\label{tab_data}}
\tablehead{
\colhead{PGC\tablenotemark{1}} & 
\colhead{Name\tablenotemark{2}} & 
\colhead{$I_T$\tablenotemark{3}} & 
\colhead{$I_T^{b,i,k}$\tablenotemark{4}} & 
\colhead{$W1_T$\tablenotemark{5}} & 
\colhead{$W1_T^{b,i,k,a}$\tablenotemark{6}} &
\colhead{$W2_T$\tablenotemark{7}} & 
\colhead{$W2_T^{b,i,k,a}$\tablenotemark{8}} & 
\colhead{$C_{I-W1}$\tablenotemark{9}} & 
\colhead{$C_{I-W2}$\tablenotemark{10}} & 
\colhead{b/a\tablenotemark{11}} & 
\colhead{Inc\tablenotemark{12}} &
\colhead{$W_{mx}$\tablenotemark{13}} & 
\colhead{$W^i_{mx}$\tablenotemark{14}} & 
\colhead{$\log$($W^i_{mx}$)\tablenotemark{15}} & 
\colhead{Sam\tablenotemark{16}}
}
\startdata
40095 & NGC4312    & 10.650 $\pm$ 0.152 & 10.230 & 11.252 $\pm$ 0.001 & 11.233 & 11.893 $\pm$ 0.001 & 11.899 &  -0.661 &  -1.327 &  0.27 & 79 &  217 &  221 & 2.344 $\pm$ 0.036 & Virgo \\
40105 & NGC4313    & 10.537 $\pm$ 0.184 &  9.970 & 11.056 $\pm$ 0.001 & 11.028 & 11.653 $\pm$ 0.001 & 11.657 &  -0.716 &  -1.345 &  0.22 & 85 &  257 &  258 & 2.412 $\pm$ 0.028 & Virgo \\
40201 & NGC4330    & 11.429 $\pm$ 0.189 & 10.810 & 11.936 $\pm$ 0.001 & 11.902 & 12.503 $\pm$ 0.001 & 12.503 &  -0.750 &  -1.351 &  0.17 & 90 &  251 &  251 & 2.400 $\pm$ 0.026 & Virgo \\
40507 & NGC4380    & 10.109 $\pm$ 0.104 &  9.820 & 11.038 $\pm$ 0.001 & 11.043 & 11.651 $\pm$ 0.001 & 11.675 &  -0.881 &  -1.513 &  0.52 & 61 &  265 &  304 & 2.483 $\pm$ 0.042 & Virgo \\
\nodata \\
\enddata
\tablenotetext{1}{Principal Galaxies Catalog Number, $^2$Common name,}
\tablenotetext{3}{I-band mag (Vega), $^4$I-band mag with $A^I_{b,i,k}$ applied,}
\tablenotetext{5}{{\it WISE} W1 mag (AB), $^6$W1 mag with $A^{W1}_{b,i,k,a}$ applied,}
\tablenotetext{7}{{\it WISE} W2 mag (AB), $^8$W2 mag with $A^{W2}_{b,i,k,a}$ applied,}
\tablenotetext{9}{$I_T^{b,i,k} - W1_T^{b,i,k,a}$ color (AB), $^{10}$$I_T^{b,i,k} - W2_T^{b,i,k,a}$ color (AB),}
\tablenotetext{11}{Axial ratio b/a, $^{12}$Inclination in degrees,}
\tablenotetext{13}{Uncorrected linewidth, $^{14}$Inclination-corrected linewidth,}
\tablenotetext{15}{Logarithm of the inclination-corrected HI linewidth,}
\tablenotetext{16}{Sample name}
\end{deluxetable*}

{\it WISE} photometry is available for all targets (see
Table~\ref{tab_data}).  The Spitzer photometry \citep{Sorce:13:94} was
acquired from pointed observations with the consequence that a significant
fraction of calibration candidates remained unobserved (although the number
observed continues to increase). Likewise, the I-band photometry, acquired
by pointed observations, remains incomplete.  With the current tally, there
are 310 cluster calibrators with {\it WISE} W1 and W2 photometry, compared
with 213 available to \citet{Sorce:13:94} for the Spitzer calibration, and
291 of the 310 {\it WISE} calibrators have I-band photometry, compared with
the 267 available to \citet{Tully:12:78} for the previous I-band
calibration.

A minor update with the current work is the conversion of Galactic
obscuration values to \citet{Schlafly:11:103} from \citet{Schlegel:98:525}.
This change has negligible impact on the {\it WISE} W1 magnitudes (and even
less on W2) and only a $1-2\%$ impact at I-band.

A small number of ambiguous cases are being rejected from the cluster
calibration sample.  One case is now not ambiguous: PGC~42081 was flagged
in the earlier calibrations as possibly foreground to the Virgo Cluster.
Recent Hubble Space Telescope observations provide a distance of 9.5 Mpc
from a tip of the red giant branch measurement \citep{Karachentsev:14:4},
confirming that this galaxy is in the foreground.  Further in the case of
the Virgo Cluster, the galaxies PGC~41531 and 43601 are considered probable
background galaxies.  Their velocities (1626 and 1783 \kms\ respectively)
and distances are consistent with membership in the structure including the
Virgo W Cluster and M Cloud at roughly twice the Virgo distance.
Similarly, PGC~30498, which was considered as a candidate for the Antlia
Cluster because of proximity on the sky, is now considered an outlying
associate of the more distant Hydra Cluster.  The velocity range of the two
clusters overlap.  PGC~30498, located between the two clusters, $3^{\circ}$
from Antlia, has a velocity and distance compatible with Hydra.  It is
$5^{\circ}$ from Hydra; too removed to be taken into the Hydra sample.  See
the appendix of \citet{Tully:12:78} for discussions of the environments of
these clusters.

The same inclination, morphology, and HI quality criteria described above
are applied to our zero-point calibrator sample along with the additional
constraint that each zero-point galaxy have a well-known distance derived
either from Cepheid or TRGB measurements.  To set the Cepheid distance
scale, we use the recently updated LMC distance modulus of $18.48\pm0.04$
\citep{Scowcroft:11:76, Scowcroft:12:84, Monson:12:146, Freedman:12:24}.
We use a TRGB calibration that has been demonstrated to be consistent with
the Cepheid scale by \citet{Rizzi:07:815} and \citet{Tully:08:184}.

\subsection{HI Line Widths\label{SEC:HILINEWIDTHS}}

We use HI linewidth measurements from the Cosmic Flows project
\citep{Tully:13:86} that contains over 14,000 galaxies with measurements of
$W_{m50}$, the width at 50\% of the mean flux within the velocity range in
the HI line that encompasses 90\% of the total line flux.  These data are
available at the Extragalactic Distance Database (EDD)
website\footnote{\url{http://edd.ifa.hawaii.edu}; catalog ``All Digital
HI''}.  This observed parameter is de-projected and corrected to a
measurement of the intrinsic maximum rotation velocity width, $W^i_{mx}$.
This is accomplished using a method that accounts for galaxy inclination,
relativistic broadening and finite spectral resolution as described in
\citet{Courtois:09:1938, Courtois:11:2005} and reviewed in
\citet{Tully:12:78}.  The error in $W^i_{mx}$ is derived from the signal at
the 50\% level divided by the noise measured outside the line in regions of
no signal.  An error threshold of 20 \kms\ is applied to remove noisy
measurements.  Retained profiles meet a minimum per-channel signal-to-noise
requirement of $S/N \geq 2$ and are also visually inspected to remove
pathological cases.

In subsequent plots, it becomes obvious that the errors in the HI
line-widths dominate the observational errors.  Slow rotators exhibit a
higher fractional error because of their small line widths.  Lower
inclination systems are also prone to higher errors motivating our
inclination threshold of 45$\deg$, below which a 5$\deg$ error in
inclination results in a $> 8$\% error in linewidth.

\subsection{W1 and W2 Data and Photometry\label{SEC:W1PHOTOMETRY}}

Thanks to the {\it WISE} public data release, available from the NASA/IPAC
infrared science archive
(IRSA)\footnote{\url{http://irsa.ipac.caltech.edu/Missions/wise.html}}, all
of the galaxies in our sample have imaging in the {\it WISE} W1 and W2
bands.  Image cutouts combining the level 1b (single) image products were
drizzled using version 3.8.3 of the Image Co-addition with Optional
Resolution Enhancement (ICORE) software \citep{Masci:09:67,Masci:13:02010}.
To minimize background problems, we selected the 1b images with moon angles
greater than 25$\deg$, and with epochs at least 2000 seconds from an
annealing event.  We combined the resulting image set on an output scale of
1.0 arcseconds per pixel.

Photometry of the calibrator galaxies was performed using the photometry
routines developed for the WNGA \citep{Seibert:14b}.  This method uses
elliptical apertures with fixed shapes, orientations, and centers but
varying major axes in steps approximately equal to a resolution element in
the W1 band (6 arcseconds) to measure the flux of the galaxy within each
annulus from the center to the edge of the galaxy.  Foreground stars and
contaminating neighbor galaxies are masked prior to measurement and this
masking is accounted for in computing the flux within each annulus.  The
influence of partially resolved and unresolved background galaxies is
mitigated by allowing our sky value to contain flux from these objects.
This is achieved by setting a masking limit in the sky annulus fainter
than which objects are not masked.  This produces an accurate sky that
accounts for these fainter galaxies that will be present in the measurement
annulus, but are very difficult to detect and mask.  Without this observed sky
value, these faint, barely resolved galaxies prevent the surface photometry
growth curve from converging.

The default axial ratios for the measurement ellipses for the WNGA are
those given by HyperLEDA\footnote{\url{http://leda.univ-lyon1.fr}}
\citep{Paturel:03:45}.  However, since the dominant source of error in
calibrating the TFR arises from errors in the HI linewidth inclination
correction, much effort has gone into determining accurate axial ratios and
from them inclinations.  For this paper, we chose to use the axial ratios
that were used to determine the correction to the HI linewidths.  These are
derived from optical imaging, mostly I-band \citep[see
\S\ref{SEC:IPHOTOMETRY} and \S4.4 in][]{Courtois:11:1935}.  We found that,
in the mean, the difference in the W1 photometry between using the default
axial ratios and using the HI linewidth correction axial ratios was on the
order of 4 milli-magnitudes, well below our photometric error threshold.

In order to derive the total magnitudes of the galaxy in the {\it WISE}
bands, $W1_T$ and $W2_T$, the radial photometric profile is analyzed and
two versions of ``total magnitude'' are derived: (1) an asymptotic total
magnitude that is the integration of the galaxy radial profile up to the
point where the profile curve of growth has mathematically converged within
the errors and (2) a procedure that starts with the isophotal magnitude
within 25.5 mag arcsec$^{-2}$ then adds a small extrapolation derived by
extending an exponential disk fit to infinity \citep{Tully:96:2471}.  The
extrapolation is given by the formula
\begin{equation}
\Delta m_{ext} = 2.5\ {\rm log}[1-(1+\Delta n) e^{-\Delta n}]
\label{delm}
\end{equation}
where $\Delta n = (\mu_{25.5}-\mu_0)/1.086$ is the number of disk
exponential scalelengths between the central surface brightness $\mu_0$ and
the limiting isophotal surface brightness $\mu_{25.5}$.  The exponential
disk central surface brightness $\mu_0$ excludes the bulge by defining the
exponential disk fit over the range from the effective radius (enclosing
half the light of the galaxy) to the $\mu_{25.5}$ isophotal radius.  If the
disk central surface brightness is brighter than $\mu_0= 20$ then the
correction $\Delta m_{ext}$ is less than $0.03$ mag.

The ensemble difference between these two types of magnitudes is
characterized by a mean offset of 0.0003 magnitudes and a standard
deviation of 0.0234 magnitudes.  To check for a systematic trend with
magnitude we fit the differences as a function of asymptotic magnitude and
derived a line with a slope of 0.0013 $\pm$ 0.0006 and a zero-point of
0.0109 $\pm$ 0.0073.  As a further check, we used both of these magnitudes
to carry out the calibration and the resulting set of coefficients were
statistically identical.  We have chosen to use the asymptotic magnitudes
for the calibration presented herein because they are a standard output
product of the WNGA and thus require no extra processing beyond our
photometry pipeline.  In addition, the extrapolated disk magnitudes are
only appropriate for disk galaxies, while the asymptotic magnitudes are
consistent regardless of galaxy type.

We convert our W1 and W2 magnitudes from the Vega to the AB system using
the Vega-AB offsets of 2.699 mag for W1 and 3.339 mag for W2 from Table~3
of Section~IV.4.h of the Explanatory Supplement to the {\it WISE} All-Sky
Data Release
Products\footnote{\url{http://wise2.ipac.caltech.edu/docs/release/allsky/expsup/ sec4\_4h.html}}.
Uncertainties in the observed W1 magnitudes are similar to or smaller than
those measured for the IRAC [3.6] magnitudes
\citep[$\pm0.05$,][]{Sorce:12:133}.  The smaller uncertainties arise for
galaxies that have a large number of individual images from the {\it WISE}
survey and thus when coadded are deeper than the IRAC [3.6] images.  This
variable depth coverage in the {\it WISE} survey is due to the fact that
the scans were conducted as great circles intersecting at the ecliptic
poles \citep{Wright:10:1868}, thus the frame coverage density increases
from a minimum at the ecliptic plane to a maximum at the ecliptic poles.

We apply the following corrections to our measured total magnitudes: 
\begin{enumerate}
	\item $A^{[W1,2]}_b$, a Milky Way extinction correction\citep{Schlafly:11:103,Fitzpatrick:99:63}
	\item $A^{[W1,2]}_i$, an internal extinction correction\citep{Giovanelli:95:1059,Giovanelli:97:22,Tully:98:2264},
	\item $A^{[W1,2]}_k$, a Doppler shift or k-correction \citep{Oke:68:21,Huang:07:840}.
	\item $A^{[W1,2]}_a$, a total flux aperture correction from Table 5 of Section IV.4.c of the {\it WISE} Explanatory Supplement\footnote{\url{http://wise2.ipac.caltech.edu/docs/release/allsky/expsup/ sec4\_4c.html}}.
\end{enumerate}

In these and subsequent equations the notation $W1,2$ means the values for
the {\it WISE} W1 and W2 bands.  All these corrections are discussed in
detail in \citet{Sorce:12:133}.  The internal extinction correction is
described by the formula $A_i^{[W1,2]} = \gamma_{W1,2}\log(a/b)$
\citep{Tully:98:2264}, where $a/b$ is the major to minor axial ratio and
$\gamma_{W1}$ has the form
\begin{equation}
	\gamma_{W1} = 0.12 + 0.21 (\log W^i_{mx} - 2.5).
\end{equation}
The factor $\gamma_{W2}$ can be obtained by multiplying $\gamma_{W1}$ by
the ratio of the reddening coefficients $R_{W2}/R_{W1} = 0.661$
\citep{Fitzpatrick:99:63}.  The k-corrections for $W1$ and $W2$ are very
small and roughly the same over the redshift range of interest.  The
correction is based on Figure~6 in \citet{Huang:07:840} and has the form
$A^{[W1,2]}_k = -2.27z$.  The {\it WISE} aperture correction,
$A^{[W1,2]}_a$, arises because the photometric calibration of {\it WISE} is
conducted with point sources within a fixed aperture that misses some of
the scattered light that is picked up in the extended apertures required to
measure galaxies.  The fixed apertures used for the {\it WISE} W1 and W2
photometric calibrations are 8.25 arcseconds in radius and therefore much
smaller than any of the galaxies used in this paper, thus each galaxy has
fixed corrections of $A^{W1}_a = -0.034$ mag and $A^{W2}_a = -0.041$ mag
applied.  The fully corrected {\it WISE} magnitude is then
\begin{equation}
\begin{split}
	W1,2^{b,i,k,a}_T = W1,2_T &-A^{[W1,2]}_b - A^{[W1,2]}_i \\
	&-A^{[W1,2]}_k - A^{[W1,2]}_a.
\end{split}
\end{equation}

\subsection{I-band Photometry\label{SEC:IPHOTOMETRY}}

The sources of the I-band photometry were discussed in \citet{Tully:12:78}.
There are contributions from \citet{Courtois:11:1935} and from the
literature.  The present calibration is augmented with 24 new galaxies, an
increase of 9\%.  Photometric corrections and analysis procedures are the
same as in the previous publication save for the small shift in reddening
due to our Galaxy in going from \citet{Schlegel:98:525} to
\citet{Schlafly:11:103} and the small shift in distance scale zero point
implicit in the shift of the LMC modulus from 18.50 to 18.48
\citep{Scowcroft:11:76, Scowcroft:12:84, Monson:12:146, Freedman:12:24}.

The main interest of the current paper is the calibration of the {\it WISE}
W1 and W2 band TFR, but an I-band re-calibration is worth presenting.  We
collect I-band magnitudes because, as will be discussed in
\S\ref{SEC:COLOR_TERM}, we can couple the I-band and {\it WISE} magnitudes
and recover the I-band scatter through an optical - {\it WISE} color
correction.  For determining the {\it WISE} color terms, we convert the
I-band Vega magnitudes to the AB system using the offset from
\citet{Frei:94:1476} of 0.342 magnitudes.  This publication provides an
opportunity to update the I-band calibration to assure consistency between
optical and MIR distance measurements.  For the I-band TFR re-calibration,
the native Vega system is used.  

The resulting input data for calibrating the Tully-Fisher relation are
presented in Table~\ref{tab_data}.  This table gives the input total W1, W2
AB photometry, $W1_T$ and $W2_T$, and the input total I-band Vega
photometry, $I_T$, and the corrected magnitudes, $W1_T^{b,i,k,a}$,
$W2_T^{b,i,k,a}$, and $I_T^{b,i,k}$ for each calibrator galaxy.  Also
presented are the optical to MIR AB colors, along with the axial ratios and
inclinations and input and corrected HI linewidths and the sample (ZeroPt
or cluster) each calibrator resides in.

\section{The W1 and W2 Calibration\label{SEC:W1CALIBRATION}}

The similarity of {\it WISE} W1 and W2 bands allows us to use identical
procedures for both bands.  Thus we will describe both calibrations and
present both sets of results together.

It has been shown that the Malmquist bias incurred by fitting the direct
TFR can be mitigated by fitting the inverse relation \citep{Willick:94:1}.
The major effect of the bias in fitting the direct relation is to flatten
the slope since fainter galaxies with the same linewidth are excluded by a
photometric or signal-to-noise cut.  Even with fitting the inverse relation
a residual bias due to scatter in the sample remains.  This is addressed in
section~\ref{SEC:BIAS}.

We use a linear regression fitting technique that uses the linewidth errors
as the input measurement error.  For the {\it WISE} data, this is sensible
since the formal measurement errors on the magnitudes are very small
compared to the linewidth errors.  The I-band magnitude errors are larger
and so we will make an adjustment to the linewidth errors that will account
for these larger photometric errors (see \S\ref{SEC:ICALIBRATION}).

\subsection{Relative Distances and TFR Slope\label{SEC:RELATIVEDISTANCES}}

\begin{figure}
	\includegraphics[width=\linewidth]{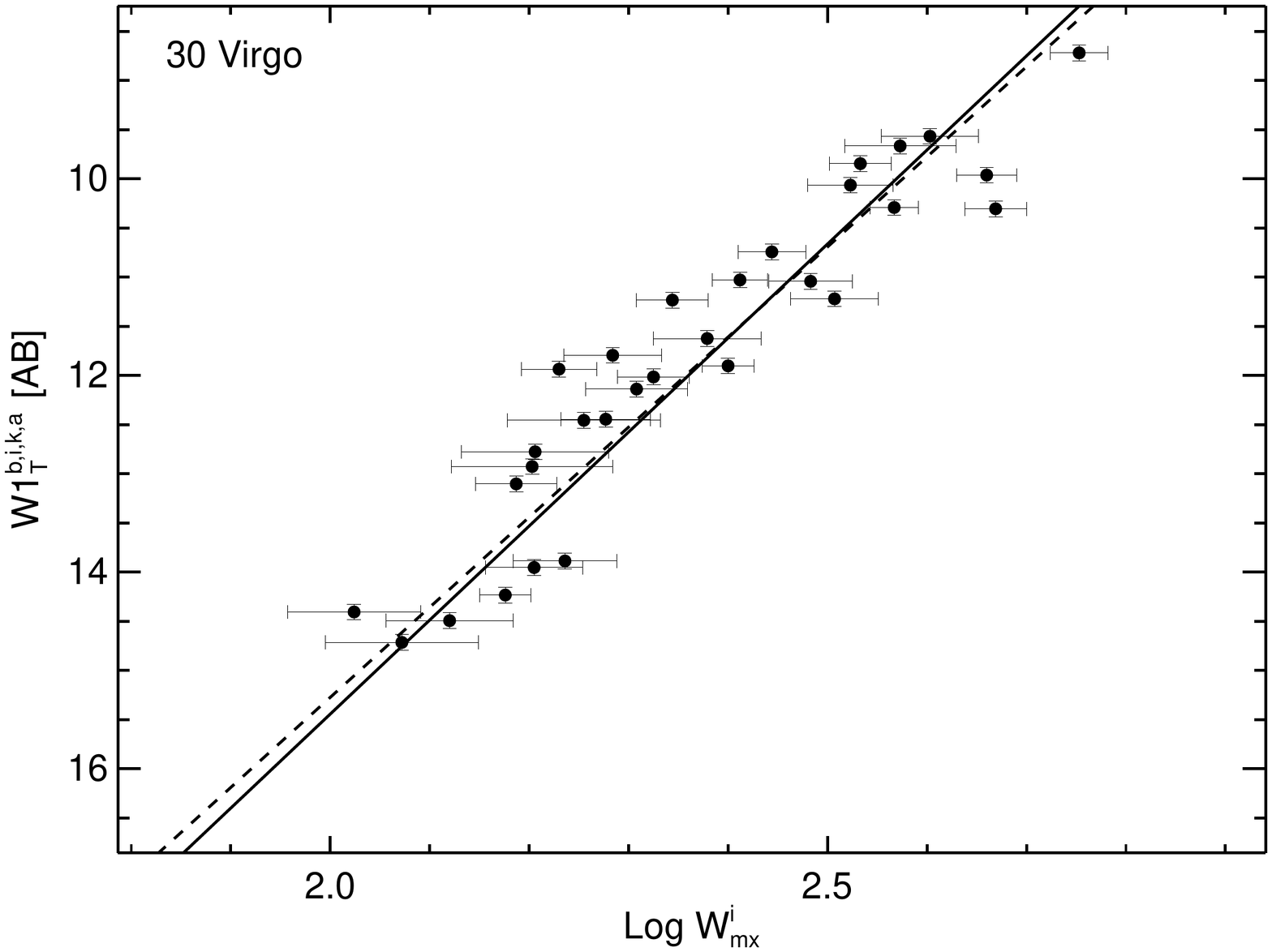}
	\includegraphics[width=\linewidth]{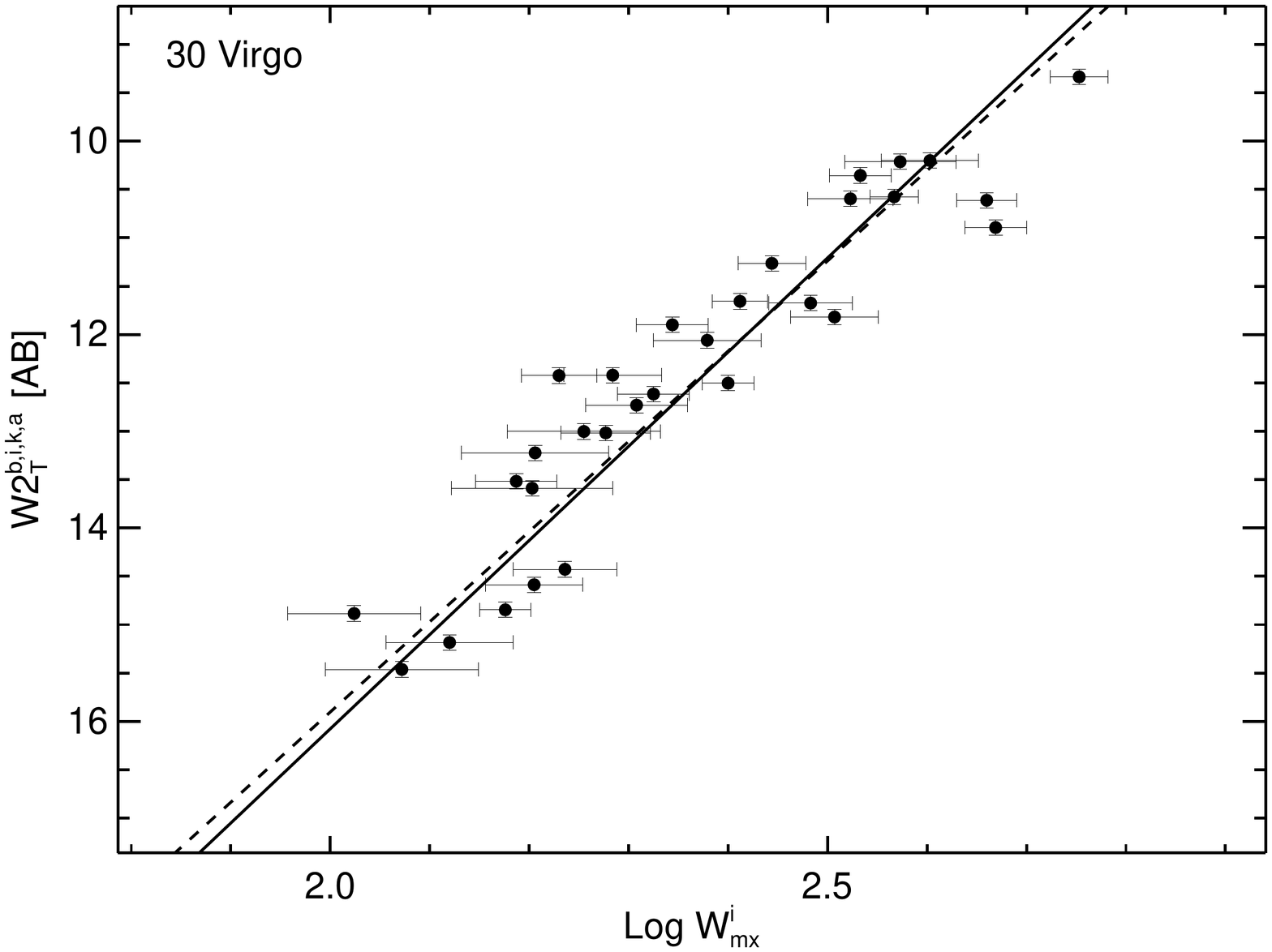}
	\caption{Linear Tully-Fisher relation (TFR) in the {\it WISE} W1
	(top) and W2 (bottom) bands for the Virgo Cluster.  The solid line
	is the inverse fit of the universal template correlation.  The
	dashed line is the fit to Virgo alone.}
	\label{fig_virgo}
\end{figure}

\begin{figure}
	\includegraphics[width=\linewidth]{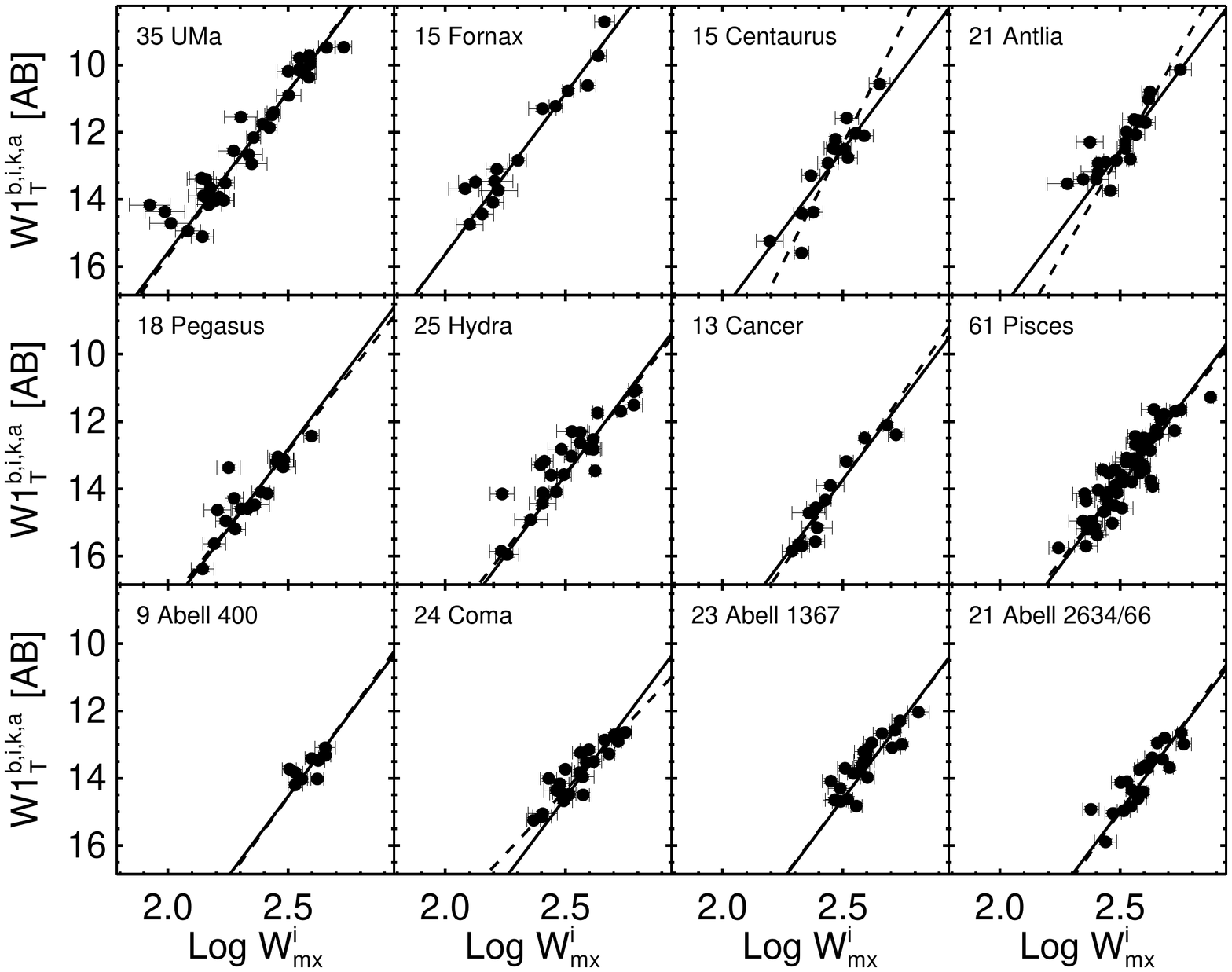}
	\includegraphics[width=\linewidth]{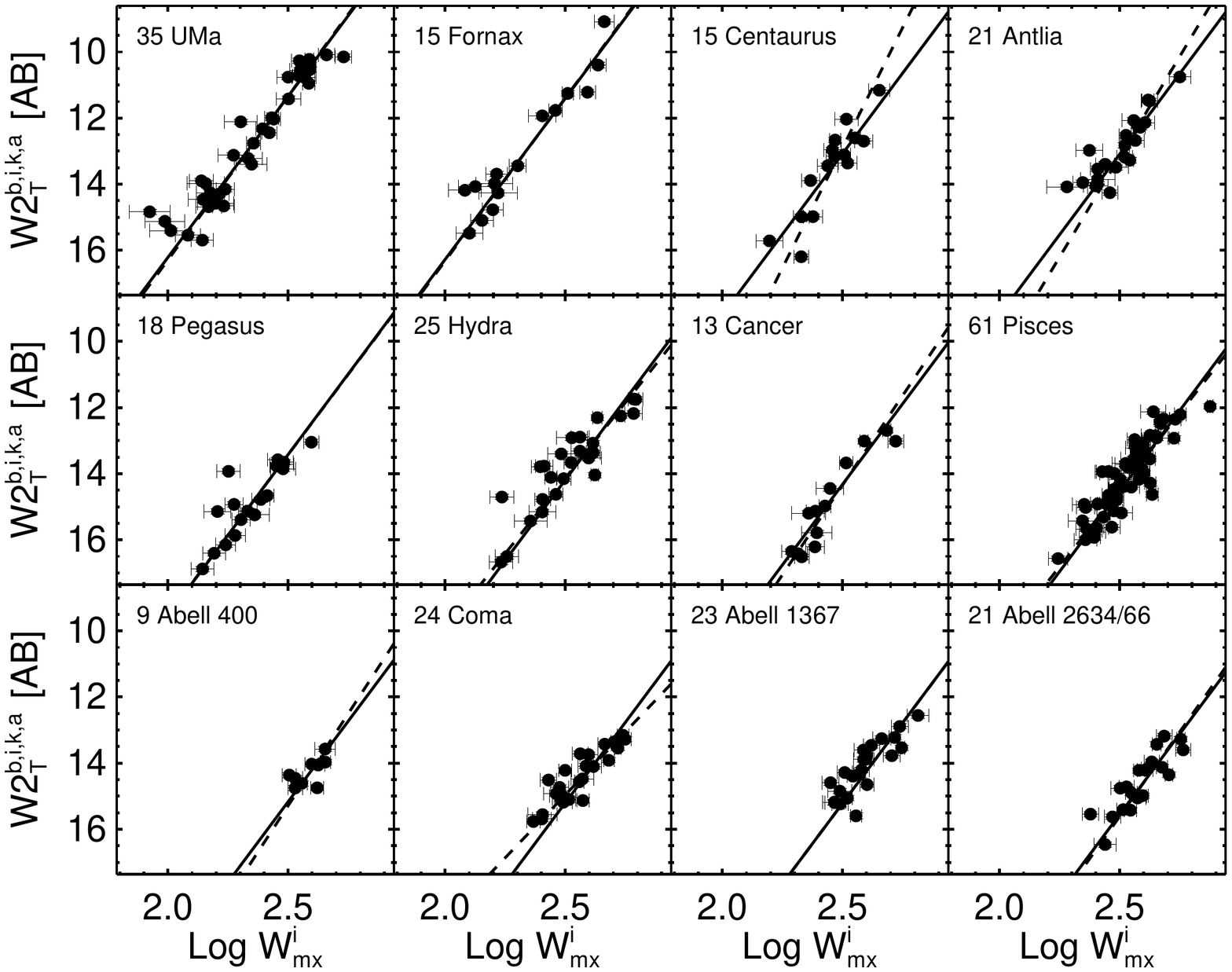}
	\caption{Linear TFR in the {\it WISE} W1 (top) and W2 (bottom)
	bands for Ursa Major, Fornax, Centaurus, Antlia, Pegasus, Hydra,
	Cancer, Pisces, Abell 400, Coma, Abell 1367, and Abell 2634/2666.
	Solid lines are the inverse fit of the universal template, while
	dashed lines are the fits for each cluster.}
	\label{fig_others}
\end{figure}

The TFR posits a universal slope in luminosity versus HI linewidth.  Our
first step in deriving this universal slope is to fit each cluster
individually.  The results of these fits are shown in
Figures~\ref{fig_virgo} and \ref{fig_others}.  Examining the dashed lines
in these figures shows how similar the individual slopes are.  In addition,
we see no significant trend in the slope with distance, a benefit of using
the ITFR which mitigates the Malmquist bias.  The slope values for the
individual clusters are given in column three of
Tables~\ref{tab_clusters_w1} and \ref{tab_clusters_w2}.

\tabletypesize{\scriptsize}
\tabcolsep=0.11cm
\begin{deluxetable*}{lrrrrrrrrr}
	\tablewidth{0in}
	\tabletypesize{\scriptsize}
	\setlength{\tabcolsep}{0.03in}
	\tablecaption{W1 Cluster Fit Properties\label{tab_clusters_w1}}
	\tablehead{
	\colhead{Cluster\tablenotemark{1}} &
	\colhead{N\tablenotemark{2}} &
	\colhead{Slope\tablenotemark{3}} &
	\colhead{ZP\tablenotemark{4}} &
	\colhead{rms\tablenotemark{5}} &
	\colhead{ZP$_{cur}$\tablenotemark{6}} &
	\colhead{rms$_{cur}$\tablenotemark{7}} &
	\colhead{N$_{cc}$\tablenotemark{8}} &
	\colhead{ZP$_{cc}$\tablenotemark{9}} &
	\colhead{rms$_{cc}$\tablenotemark{10}}
	}
\startdata
Virgo           &  30 &  -9.16 $\pm$  0.38 & 10.66 $\pm$ 0.11 & 0.60 & 10.67 $\pm$ 0.06 & 0.55 &  30 & 10.83 $\pm$ 0.09 & 0.52 \\
U Ma            &  35 &  -9.81 $\pm$  0.37 & 10.80 $\pm$ 0.11 & 0.63 & 10.73 $\pm$ 0.05 & 0.70 &  34 & 10.95 $\pm$ 0.10 & 0.58 \\
Fornax          &  15 &  -9.52 $\pm$  0.56 & 10.85 $\pm$ 0.13 & 0.50 & 10.81 $\pm$ 0.09 & 0.53 &  15 & 10.99 $\pm$ 0.11 & 0.44 \\
Antlia          &  21 & -12.40 $\pm$  1.30 & 12.54 $\pm$ 0.11 & 0.51 & 12.46 $\pm$ 0.05 & 0.50 &  16 & 12.70 $\pm$ 0.09 & 0.36 \\
Centaurus       &  15 & -14.16 $\pm$  1.43 & 12.53 $\pm$ 0.15 & 0.59 & 12.49 $\pm$ 0.07 & 0.62 &  13 & 12.71 $\pm$ 0.14 & 0.52 \\
Pegasus         &  18 &  -9.10 $\pm$  0.81 & 12.81 $\pm$ 0.14 & 0.58 & 12.90 $\pm$ 0.09 & 0.57 &  17 & 13.00 $\pm$ 0.10 & 0.39 \\
Hydra           &  25 &  -9.12 $\pm$  0.43 & 13.59 $\pm$ 0.13 & 0.67 & 13.40 $\pm$ 0.04 & 0.57 &  19 & 13.74 $\pm$ 0.12 & 0.54 \\
Pisces          &  61 &  -9.16 $\pm$  0.28 & 13.91 $\pm$ 0.07 & 0.53 & 13.73 $\pm$ 0.03 & 0.50 &  59 & 13.93 $\pm$ 0.06 & 0.47 \\
Cancer          &  13 & -10.35 $\pm$  0.63 & 13.73 $\pm$ 0.12 & 0.41 & 13.59 $\pm$ 0.06 & 0.41 &  13 & 13.80 $\pm$ 0.10 & 0.35 \\
A400            &   9 &  -9.84 $\pm$  2.67 & 14.54 $\pm$ 0.13 & 0.40 & 14.41 $\pm$ 0.07 & 0.34 &   8 & 14.67 $\pm$ 0.10 & 0.28 \\
A1367           &  23 &  -9.46 $\pm$  0.68 & 14.63 $\pm$ 0.09 & 0.46 & 14.43 $\pm$ 0.04 & 0.39 &  22 & 14.66 $\pm$ 0.09 & 0.42 \\
Coma            &  24 &  -7.62 $\pm$  0.40 & 14.58 $\pm$ 0.09 & 0.47 & 14.37 $\pm$ 0.04 & 0.36 &  24 & 14.60 $\pm$ 0.08 & 0.41 \\
A2634/66        &  21 &  -9.91 $\pm$  0.71 & 14.96 $\pm$ 0.10 & 0.48 & 14.73 $\pm$ 0.04 & 0.44 &  21 & 15.05 $\pm$ 0.09 & 0.43 \\
\enddata
\tablenotetext{1}{\ Cluster name}
\tablenotetext{2}{\ Number of galaxies measured in cluster}
\tablenotetext{3}{\ Slope of the fit to individual clusters}
\tablenotetext{4}{\ Zero-point with universal slope, no color correction (mag)}
\tablenotetext{5}{\ Scatter about universal slope, no color correction (mag)}
\tablenotetext{6}{\ Zero-point with universal curve, no color correction (mag)}
\tablenotetext{7}{\ Scatter about universal curve, no color correction (mag)}
\tablenotetext{8}{\ Number of color-corrected galaxies measured in cluster}
\tablenotetext{9}{\ Zero-point with universal slope after color correction (mag)}
\tablenotetext{10}{\ Scatter about universal slope after color correction (mag)}
\end{deluxetable*}

\begin{deluxetable*}{lrrrrrrrrr}
	\tablewidth{0in}
	\tabletypesize{\scriptsize}
	\setlength{\tabcolsep}{0.03in}
	\tablecaption{W2 Cluster Fit Properties\label{tab_clusters_w2}}
	\tablehead{
	\colhead{Cluster\tablenotemark{1}} &
	\colhead{N\tablenotemark{2}} &
	\colhead{Slope\tablenotemark{3}} &
	\colhead{ZP\tablenotemark{4}} &
	\colhead{rms\tablenotemark{5}} &
	\colhead{ZP$_{cur}$\tablenotemark{6}} &
	\colhead{rms$_{cur}$\tablenotemark{7}} &
	\colhead{N$_{cc}$\tablenotemark{8}} &
	\colhead{ZP$_{cc}$\tablenotemark{9}} &
	\colhead{rms$_{cc}$\tablenotemark{10}}
	}
\startdata
Virgo           &  30 &  -9.33 $\pm$  0.39 & 11.21 $\pm$ 0.12 & 0.64 & 11.21 $\pm$ 0.06 & 0.58 &  30 & 11.42 $\pm$ 0.10 & 0.52 \\
U Ma            &  35 &  -9.90 $\pm$  0.37 & 11.34 $\pm$ 0.11 & 0.63 & 11.27 $\pm$ 0.05 & 0.71 &  34 & 11.54 $\pm$ 0.10 & 0.57 \\
Fornax          &  15 &  -9.85 $\pm$  0.58 & 11.40 $\pm$ 0.14 & 0.56 & 11.34 $\pm$ 0.09 & 0.60 &  15 & 11.58 $\pm$ 0.12 & 0.45 \\
Antlia          &  21 & -12.13 $\pm$  1.25 & 13.09 $\pm$ 0.11 & 0.51 & 13.00 $\pm$ 0.05 & 0.49 &  16 & 13.28 $\pm$ 0.09 & 0.35 \\
Centaurus       &  15 & -14.53 $\pm$  1.49 & 13.08 $\pm$ 0.16 & 0.62 & 13.04 $\pm$ 0.07 & 0.65 &  13 & 13.30 $\pm$ 0.14 & 0.52 \\
Pegasus         &  18 &  -9.70 $\pm$  0.86 & 13.42 $\pm$ 0.14 & 0.61 & 13.50 $\pm$ 0.09 & 0.60 &  17 & 13.61 $\pm$ 0.09 & 0.38 \\
Hydra           &  25 &  -9.13 $\pm$  0.44 & 14.19 $\pm$ 0.14 & 0.70 & 13.97 $\pm$ 0.04 & 0.59 &  19 & 14.33 $\pm$ 0.12 & 0.55 \\
Pisces          &  61 &  -9.28 $\pm$  0.28 & 14.52 $\pm$ 0.07 & 0.54 & 14.32 $\pm$ 0.03 & 0.51 &  59 & 14.53 $\pm$ 0.06 & 0.47 \\
Cancer          &  13 & -10.84 $\pm$  0.67 & 14.32 $\pm$ 0.13 & 0.48 & 14.16 $\pm$ 0.06 & 0.47 &  13 & 14.39 $\pm$ 0.10 & 0.37 \\
A400            &   9 & -11.14 $\pm$  3.29 & 15.17 $\pm$ 0.14 & 0.43 & 15.03 $\pm$ 0.07 & 0.36 &   8 & 15.27 $\pm$ 0.10 & 0.29 \\
A1367           &  23 &  -9.73 $\pm$  0.72 & 15.22 $\pm$ 0.11 & 0.51 & 14.99 $\pm$ 0.04 & 0.43 &  22 & 15.24 $\pm$ 0.09 & 0.43 \\
Coma            &  24 &  -7.61 $\pm$  0.41 & 15.20 $\pm$ 0.11 & 0.52 & 14.97 $\pm$ 0.04 & 0.39 &  24 & 15.20 $\pm$ 0.09 & 0.42 \\
A2634/66        &  21 & -10.23 $\pm$  0.76 & 15.55 $\pm$ 0.11 & 0.50 & 15.30 $\pm$ 0.04 & 0.45 &  21 & 15.64 $\pm$ 0.09 & 0.43 \\
\enddata
\tablenotetext{1}{\ Cluster name}
\tablenotetext{2}{\ Number of galaxies measured in cluster}
\tablenotetext{3}{\ Slope of the fit to individual clusters}
\tablenotetext{4}{\ Zero-point with universal slope, no color correction (mag)}
\tablenotetext{5}{\ Scatter about universal slope, no color correction (mag)}
\tablenotetext{6}{\ Zero-point with universal curve, no color correction (mag)}
\tablenotetext{7}{\ Scatter about universal curve, no color correction (mag)}
\tablenotetext{8}{\ Number of color-corrected galaxies measured in cluster}
\tablenotetext{9}{\ Zero-point with universal slope after color correction (mag)}
\tablenotetext{10}{\ Scatter about universal slope after color correction (mag)}
\end{deluxetable*}

In order to find the universal TFR, we must combine all 13 clusters by
shifting the data along the magnitude axis, in effect moving each cluster
to the same distance.  Virgo is nearest and most complete and offers a
natural choice for the reference cluster.  The individual fits to each
cluster provide an estimate of the relative distances from Virgo through
comparing the TFR zero points (column four of Tables~\ref{tab_clusters_w1}
and \ref{tab_clusters_w2}).

These zero points recommend the following groups.  The first group is
comprised of Virgo, Ursa Major, and Fornax, a set that we consider the most
complete because they are all nearby.  This group is followed by the
Centaurus--Antlia--Pegasus group, then the Hydra--Cancer--Pisces group, and
finally the group comprised of Coma and the three Abell clusters, A0400,
A1367, and A2634/66.  As discussed in \citet{Tully:12:78} and
\citet{Sorce:13:94}, we adopt an iterative procedure for combining the
clusters.  Starting with the nearest group, we use the zero points for
Fornax and Ursa Major to shift the galaxy magnitudes within those clusters
to align with Virgo.  A least-squares fit to the ITFR is then made to this
aligned group.  The resulting ensemble slope is then assumed in fitting all
the individual clusters with only the zero points allowed to vary.  Using
these new zero points, we then shift the next group to align with Virgo and
add it to the ensemble fit.  This procedure is repeated, adding each of the
groups in turn until we have a final ensemble fit for all 13 clusters.
This procedure has been proven to work \citep{Sorce:13:94,Tully:12:78}
because the slope of the TFR is independent of the magnitude cutoff of each
cluster.

\begin{figure}
	\includegraphics[width=\linewidth]{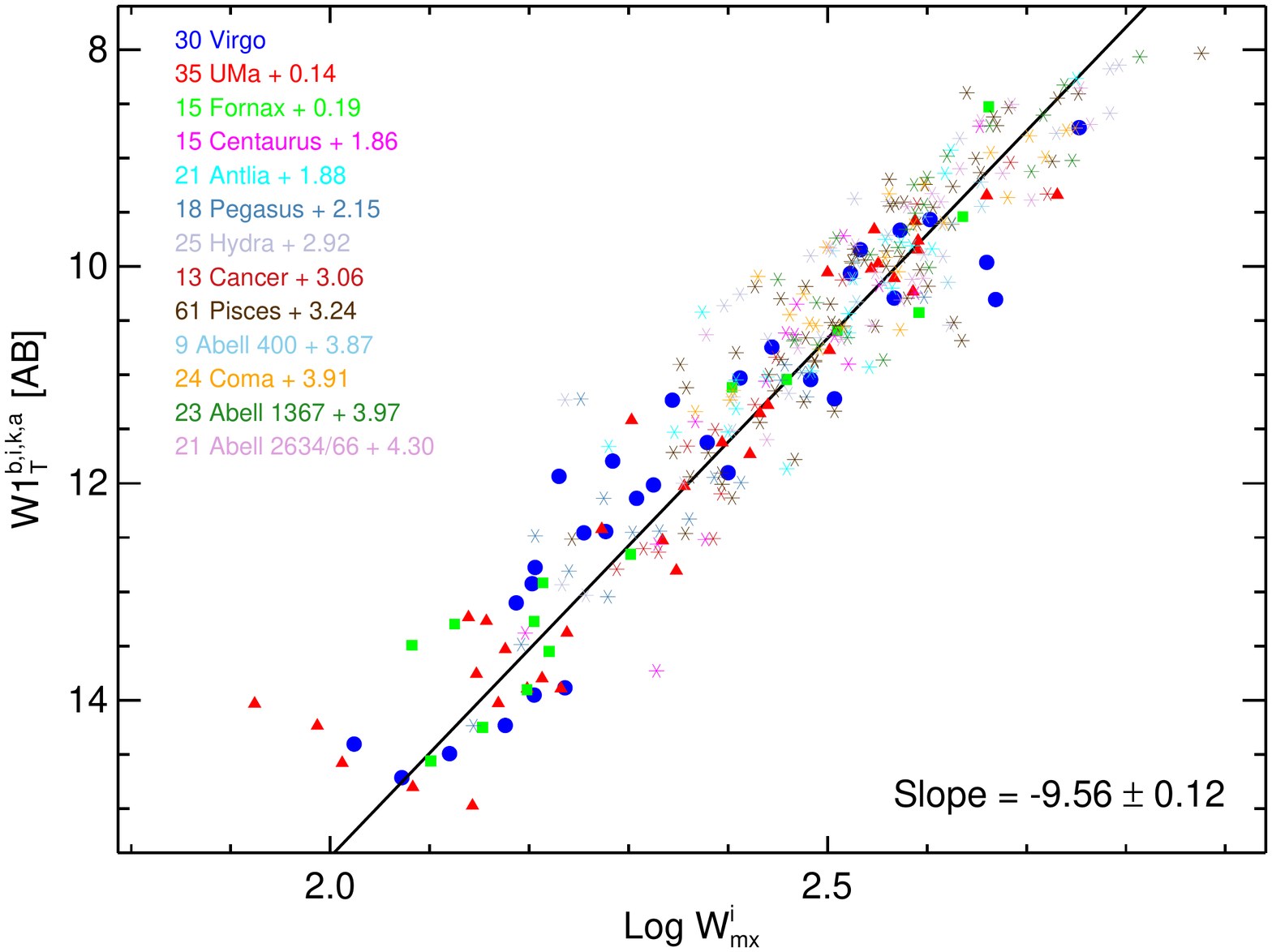}
	\includegraphics[width=\linewidth]{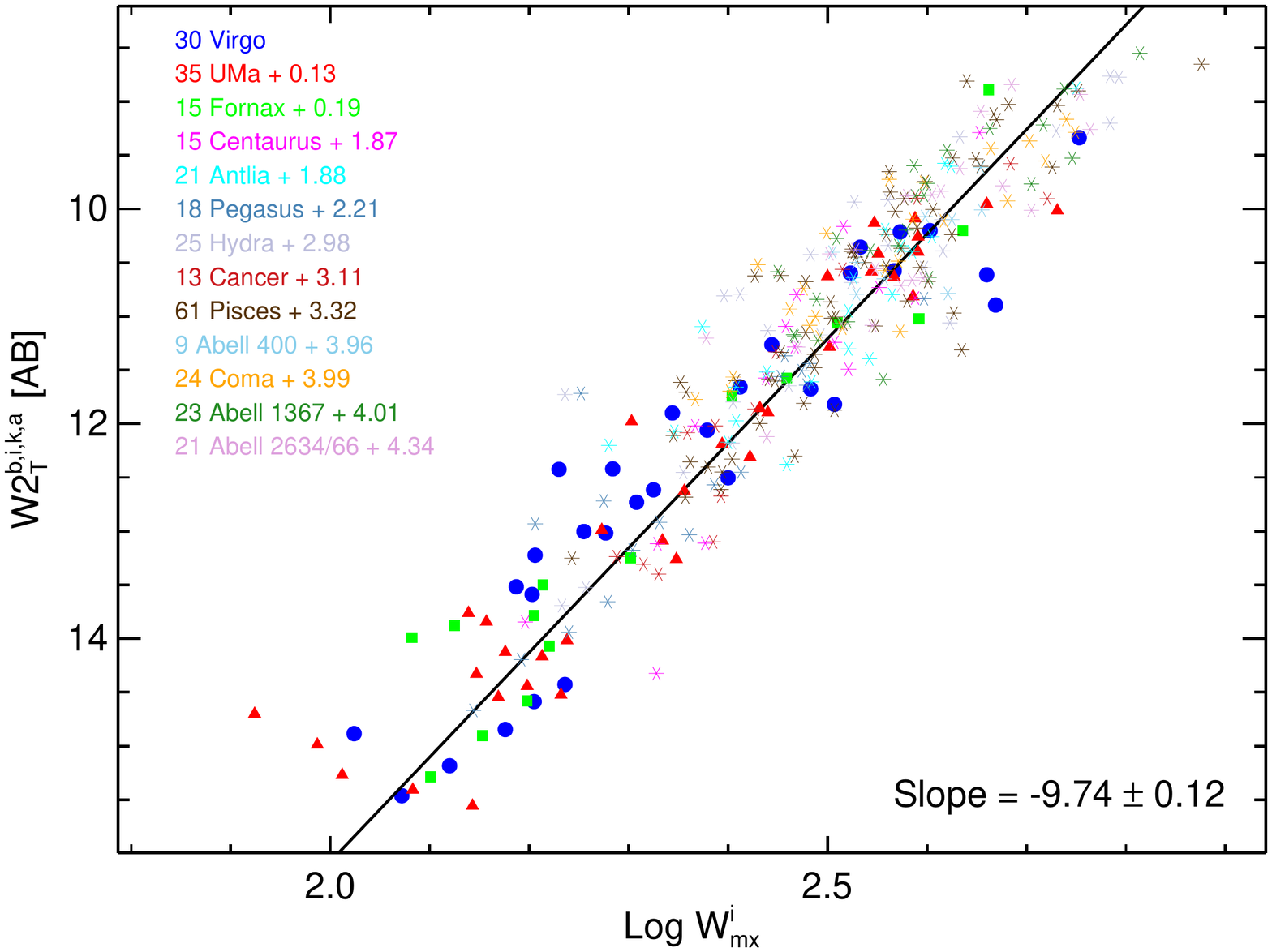}
	\caption{Linear TFR in the {\it WISE} W1 (top) and W2 (bottom)
	bands obtained from the galaxies in 13 clusters.  Offsets given
	with respect to the Virgo Cluster represent distance modulus
	differences between each cluster and Virgo. The solid line is the
	least-squares fit to all of the offset shifted galaxies with errors
	entirely in linewidths, the TFR.  These relations have an rms
	scatter of 0.54 mag for W1 and 0.56 mag for W2.}
	\label{fig_all}
\end{figure}

Our resulting universal slopes are -9.56 $\pm$ 0.12 (W1) and -9.74 $\pm$
0.12 (W2).  The universal slopes and the shifted cluster ensembles are
shown in Figure~\ref{fig_all} which is annotated with the zero point
offsets relative to Virgo for each cluster.  The agreement in these offsets
between the W1 and W2 data are quite good.  The universal slope is also
shown as the solid lines in Figures~\ref{fig_virgo} and \ref{fig_others}.

\subsection{Zero Point and Absolute Distances\label{SEC:ZEROPOINT}}

\begin{figure}
	\includegraphics[width=\linewidth]{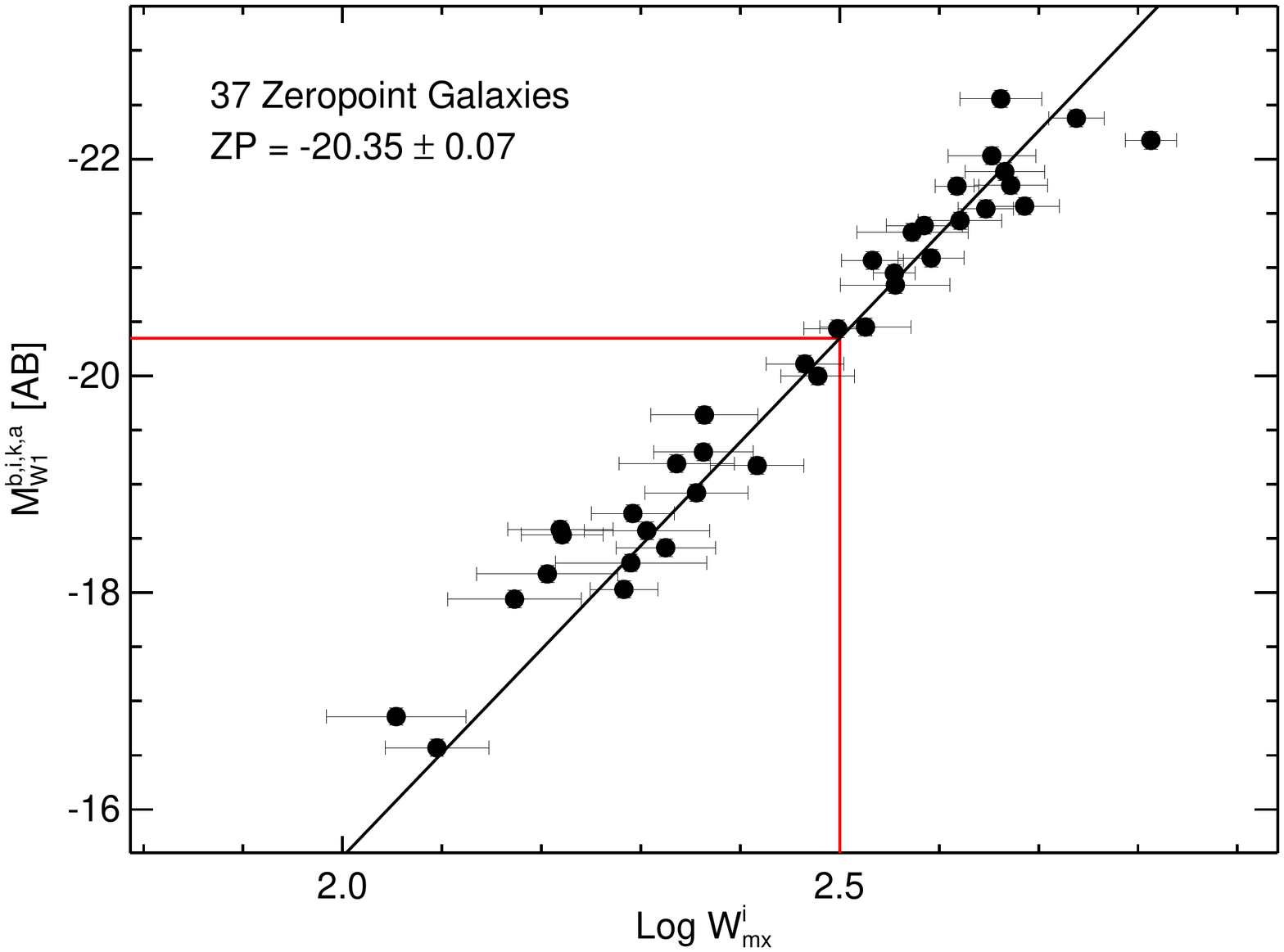}
	\includegraphics[width=\linewidth]{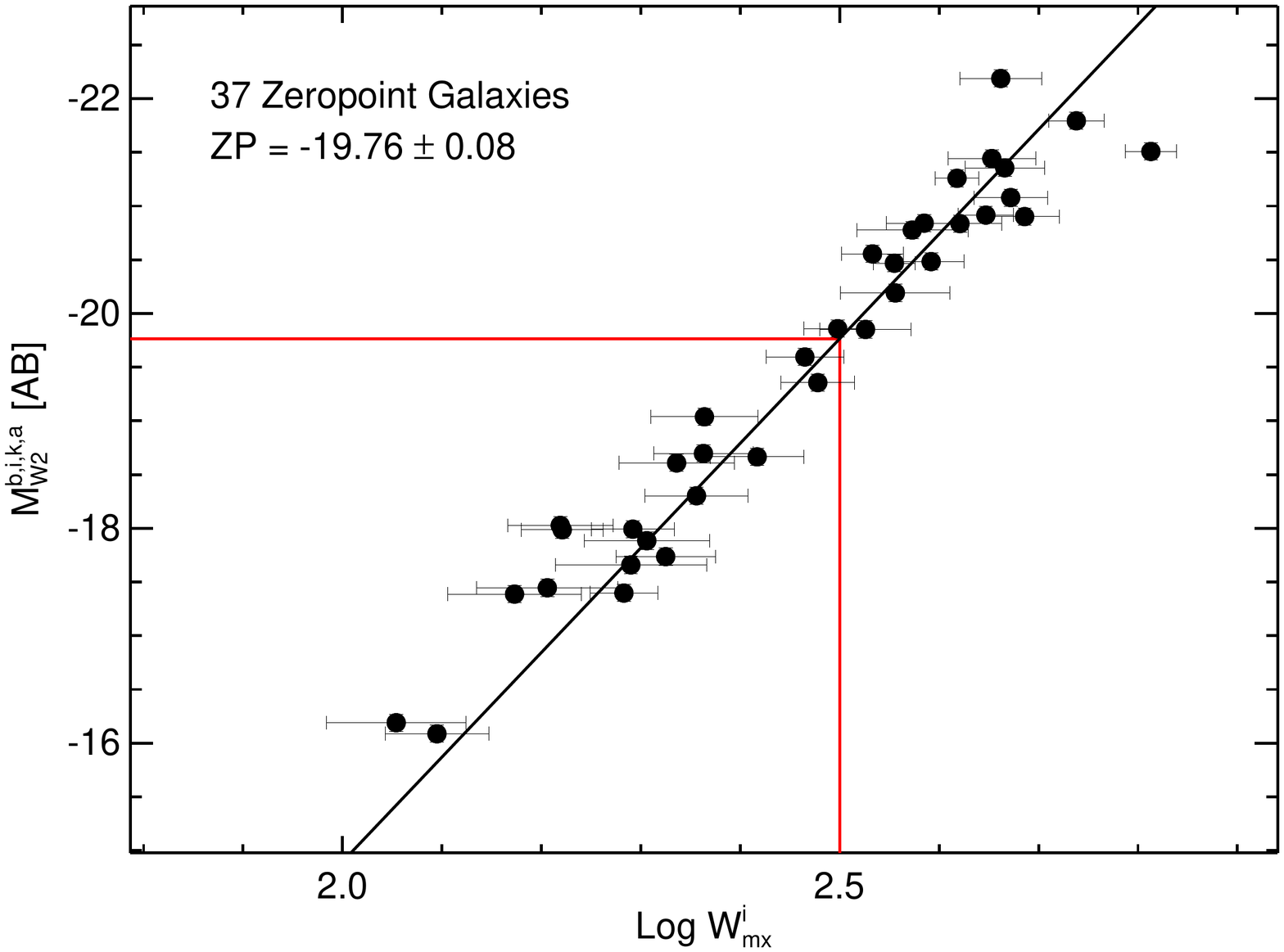}
	\caption{Linear TFR for the 37 galaxies with distances established
	by observations of Cepheid variables or the TRGB for the W1 (top)
	and the W2 (bottom). The  solid black line is the least-squares fit
	with the slope established by the 13 cluster template. The zero
	point of the TFR is set at the value of this fit at log$W^i_{mx} =
	2.5$, as indicated by the solid (red) vertical and horzontal lines.
	The zero-point fits have an rms scatter of 0.45 mag for W1 and 0.49
	mag for W2.}
	\label{fig_zp}
\end{figure}

There are 37 nearby galaxies in our zero point sample (see
Table~\ref{tab_data}) that pass our selection criteria and for which there
are good, independent distances from either the Cepheid period-luminosity
method or the TRGB method.  The distance moduli used are from
\citet[Table~2]{Tully:12:78}.  Since WISE is an all-sky data set, we are
able to measure the total asymptotic W1 and W2 magnitudes for all of them
and calibrate our distances in an absolute sense.  We use the universal
slope and the independent absolute magnitudes as input to our least-squares
fit and allow only the zero point to vary.  The resultant fits are shown in
Figure~\ref{fig_zp}.

The measured zero points are -20.35 $\pm$ 0.07 for W1 and -19.76 $\pm$ 0.08
for W2.  As was pointed out in \citet{Sorce:13:94}, NGC2841 is the fastest
rotator and the biggest outlier.  There is still no good reason to exclude
this galaxy from the zero point sample, so it is included here.

\begin{figure}
	\includegraphics[width=\linewidth]{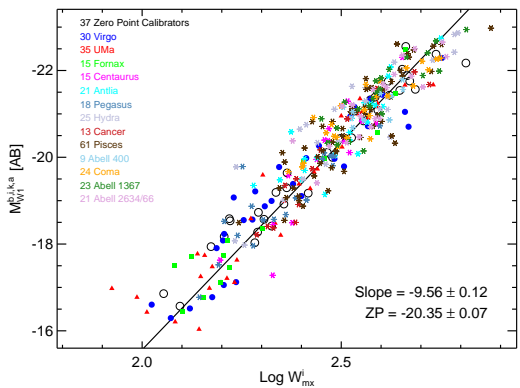}
	\includegraphics[width=\linewidth]{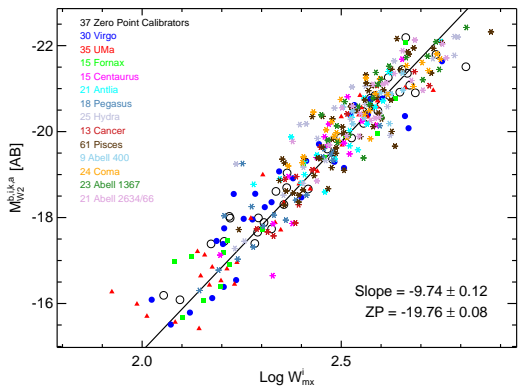}
	\caption{Linear TFR with slope fit to the galaxies in 13 clusters
	and the absolute magnitude scale set by 37 zero-point calibrators
	for the W1 (top) and the W2 (bottom).}
	\label{fig_zpall}
\end{figure}

These zero points allow us to put the {\it WISE} TFR on an absolute scale.
Since we have already calculated the cluster distances relative to Virgo,
we need only calculate the offset between the constrained zero points in
Figure~\ref{fig_all} and the absolute zero points in Figure~\ref{fig_zp}
and apply these offsets to our W1 and W2 ensembles and combine them with
the zero point calibrators, which we do in Figure~\ref{fig_zpall}.  The
zero point calibration allows us to express the TFR as
\begin{subequations}
	\label{eq_itfr}
	\begin{align}
		\begin{split}
		\mathcal{M}^{b,i,k,a}_{W1} = & - (20.35 \pm 0.07) \\
			& - (9.56 \pm 0.12)(\log W^i_{mx} - 2.5), 
			\label{eq_itfr_w1}
		\end{split} \\
		\begin{split}
		\mathcal{M}^{b,i,k,a}_{W2} = & - (19.76 \pm 0.08) \\
			& - (9.74 \pm 0.12)(\log W^i_{mx} - 2.5).  
			\label{eq_itfr_w2}
		\end{split}
	\end{align}
\end{subequations}
We adopt a convention here and throughout the paper, that TFR predicted
values are given in script, hence our TFR predicted absolute magnitudes are
given as $\mathcal{M}^{b,i,k,a}_{W1,2}$.  To derive the distance modulus
for a given galaxy based on pure W1 or W2 photometry, we subtract the
appropriate predicted TFR absolute magnitude from Equation~\ref{eq_itfr}
from the input corrected total magnitude:
\begin{equation}
	\mu_{W1,2} = W1,2^{b,i,k,a}_T - \mathcal{M}^{b,i,k,a}_{W1,2}.
	\label{eq_tfmu_wise}
\end{equation}

\begin{figure}
	\includegraphics[width=\linewidth]{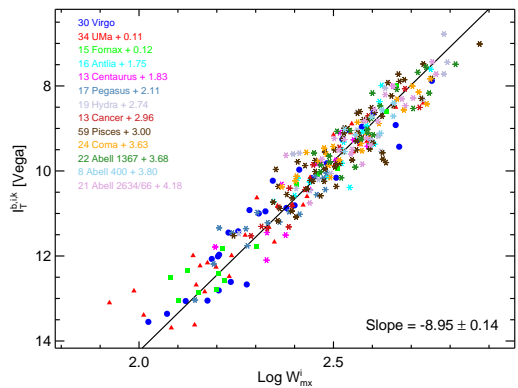}
	\includegraphics[width=\linewidth]{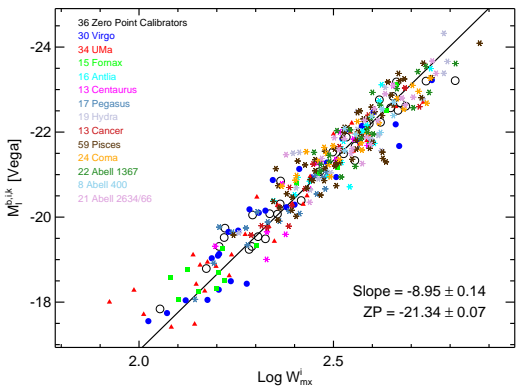}
	\caption{Linear TFR for I-band (Vega) using galaxies shifted to the
	apparent distance of Virgo (top) and on the absolute magnitude
	scale set by 36 zero-point calibrators (bottom).  This relation has
	an rms scatter of 0.46 mag and a zero-point rms scatter of 0.40
	mag.}
	\label{fig_icalibration}
\end{figure}

The rms scatter about the mean TFR will allow us to assess the usefulness
of this relation for distance measurement.  In order to do this, we use the
zero point and the Virgo offset for each cluster to shift each galaxy
magnitude in a given cluster onto the absolute magnitude scale.  We compare
this ensemble absolute magnitude, $M^{ens}_{W1,2}$, with the predicted
magnitudes from Equation~\ref{eq_itfr} to derive the residual for every
galaxy in the sample as follows:
\begin{equation}
	\Delta M_{W1,2} = M^{ens}_{W1,2} - \mathcal{M}^{b,i,k,a}_{W1,2}. 
	\label{eq_resid}
\end{equation}
We calculate the rms scatter of the resulting ensemble of residuals.  We
define the scatter in this case to be the standard deviation of the
residuals, or the square root of the second moment of the residuals.  The
distribution of the residuals is approximately Gaussian, therefore, one can
take these values as a $1\sigma$ error, i.e., 68\% of the galaxies fall
within this $1\sigma$ envelope.  The W1 calibration has a scatter of 0.54
magnitudes, while the W2 calibration has a scatter of 0.56 magnitudes
representing distance errors of 27\% and 28\%, respectively.  The scatter
in the zero point fits are slightly better at 0.45 mag (W1) and 0.49 mag
(W2).  We point out that the formal errors on the zeropoint values are much
smaller at 0.07 mag (W1) and 0.08 mag (W2).

We expect the scatter in the W1 and W2 passbands to exceed that in the I-band.
Since, at a given linewidth, red and blue galaxies separate in magnitude in different passbands, the TFR rms scatter must change with passband.
We expect the scatter in the TFR to reach a minimum where metallicity
and young population effects are minimized.  The empirical evidence suggests
the minimum is near the peak of the stellar light for disk galaxies around
$1~\mu$m.  The I-band is much closer to $1~\mu$m than are the W1 and W2
bands.

Nonetheless, these are the pure {\it WISE}
W1 and W2 TFRs requiring no other photometry to derive distance moduli to
any galaxy.  For a sample that may not have complete I-band coverage, one
may decide that the statistical benefit of a larger sample outweighs the
larger scatter of these pure {\it WISE} TFRs.  \citet{Sorce:13:94} discuss
the sources of this scatter in the MIR TFR and conclude that the most
significant arises due to a color term in the TFR.  We explore the
analogous color terms for the {\it WISE} W1 and W2 data in
\S\ref{SEC:COLOR_TERM}.  In addition, when comparing cluster distances
derived from the pure {\it WISE} and the I-band TFRs, there is evidence for
a systematic offset that may be the result of curvature in the pure MIR TFR
relation.  We discuss this in \S\ref{SEC:CURVE}, but first we derive a new
I-band TFR.

\section{I-band Calibration\label{SEC:ICALIBRATION}}

We use the identical procedure to calibrate the I-band TFR as we did for
the {\it WISE} calibration except we adjust the linewidth errors to account
for the larger I-band photometric errors.  This adjustment is carried out
as follows.  We use a preliminary TFR derived with the original linewidth
errors to project the I-band photometric errors onto the linewidth axis.
This generates a linewidth error due only to the photometric errors.  This
photometric linewidth error is then added in quadrature with the original
linewidth errors and the TFR is re-generated.  The result of this final
fitting is shown in Figure~\ref{fig_icalibration}.  The individual cluster
fits and results for each of the calibration clusters are shown in
Table~\ref{tab_clusters_i}.  The error adjustment flattens the TFR slightly
from a slope of -8.97 to -8.95.  The final TFR calibration from the I-band
data can be expressed as:
\begin{equation}
	\begin{split}
	\mathcal{M}^{b,i,k}_{I} = &- (21.34 \pm 0.07) \\
		&- (8.95 \pm 0.14)(\log W^i_{mx} - 2.5). 
	\end{split}
		\label{eq_itfr_i}
\end{equation}
The scatter is calculated in the same way as for the {\it WISE} calibration
and results in a value of 0.46 mag rms, smaller than for the {\it WISE} bands
as expected.  The formula for the distance modulus using the I-band TFR is:
\begin{equation}
	\mu_I = I^{b,i,k}_T - \mathcal{M}^{b,i,k}_I.
	\label{eq_tfmu_i}
\end{equation}

\begin{deluxetable}{lrrrr}
	\tablewidth{0in}
	\tabletypesize{\scriptsize}
	\setlength{\tabcolsep}{0.03in}
	\tablecaption{I-band Cluster Fit Properties\label{tab_clusters_i}}
	\tablehead{\colhead{Cluster\tablenotemark{1}} &
	\colhead{N\tablenotemark{2}} &
	\colhead{Slope\tablenotemark{3}} &
	\colhead{ZP\tablenotemark{4}} &
	\colhead{rms\tablenotemark{5}}
	}
\startdata
Virgo           &  30 &  -8.75 $\pm$  0.39 &  9.77 $\pm$ 0.09 & 0.51 \\
U Ma            &  34 &  -8.46 $\pm$  0.36 &  9.88 $\pm$ 0.10 & 0.57 \\
Fornax          &  15 &  -8.64 $\pm$  0.54 &  9.88 $\pm$ 0.11 & 0.42 \\
Antlia          &  16 & -11.26 $\pm$  1.38 & 11.52 $\pm$ 0.10 & 0.40 \\
Centaurus       &  13 & -11.09 $\pm$  1.22 & 11.60 $\pm$ 0.13 & 0.48 \\
Pegasus         &  17 &  -7.54 $\pm$  0.70 & 11.88 $\pm$ 0.10 & 0.43 \\
Hydra           &  19 &  -9.03 $\pm$  0.62 & 12.51 $\pm$ 0.13 & 0.55 \\
Pisces          &  59 &  -9.63 $\pm$  0.46 & 12.76 $\pm$ 0.06 & 0.46 \\
Cancer          &  13 &  -8.85 $\pm$  0.69 & 12.73 $\pm$ 0.09 & 0.32 \\
A400            &   8 &  -9.58 $\pm$  3.21 & 13.56 $\pm$ 0.10 & 0.28 \\
A1367           &  22 &  -9.70 $\pm$  0.93 & 13.45 $\pm$ 0.09 & 0.41 \\
Coma            &  24 &  -7.00 $\pm$  0.49 & 13.39 $\pm$ 0.08 & 0.39 \\
A2634/66        &  21 &  -9.26 $\pm$  0.83 & 13.94 $\pm$ 0.10 & 0.44 \\
\enddata
\tablenotetext{1}{\ Cluster name}
\tablenotetext{2}{\ Number of galaxies measured in cluster}
\tablenotetext{3}{\ Slope of the fit to individual clusters}
\tablenotetext{4}{\ Zero-point with universal slope (mag)}
\tablenotetext{5}{\ Scatter about universal slope (mag)}
\end{deluxetable}

\begin{figure}
	\includegraphics[width=\linewidth]{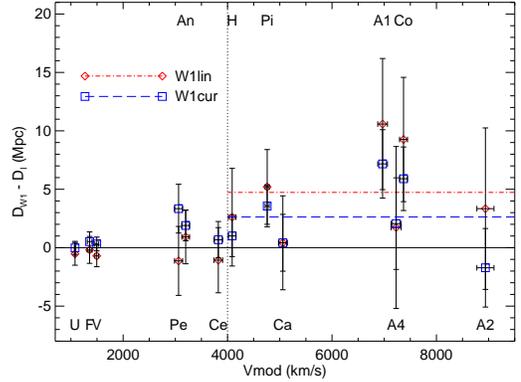}
	\includegraphics[width=\linewidth]{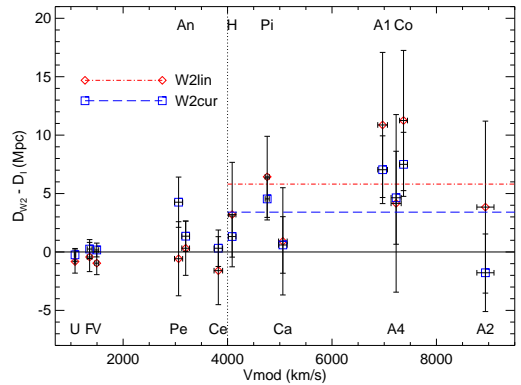}
	\caption{Distance offsets in Mpc relative to the I-band linear TFR
	of the calibration clusters for W1 (top) and W2 (bottom) using the
	linear TFR, shown by (red) diamonds and the curved TFR, shown by
	(blue) squares.  The dashed (blue) lines show the average offset
	for clusters beyond 50 Mpc ($>4000$ \kms) for the curved TFRs,
	while the dash-dot (red) lines show the average offset for the
	linear TFRs.}
	\label{fig_wise_curlin}
\end{figure}

The calibration cluster distances derived from the single-band uncorrected
TFRs in the {\it WISE} bands and the I-band are listed in columns five
through seven of Table~\ref{tab_tfr_distances}.  The cluster distance
offsets for the {\it WISE} linear TFR, relative to the I-band TFR, are
illustrated by the (red) open diamonds in Figure~\ref{fig_wise_curlin}.
The particular cluster is indicated with the code listed in column two of
Table~\ref{tab_tfr_distances}.  We note that, compared to our I-band
distances, the linear {\it WISE} TFR predicts distances that are lower for
nearby clusters and higher for more distant clusters.  If there is
curvature in the MIR TFR, it could manifest itself in just this fashion
\citep[see \S V.a.i in][]{Aaronson:86:536}.  Since we must use more distant
clusters to estimate $H_0$ from the TFR, such a systematic deviation from a
linear relation would bias the distances larger and produce a smaller
$H_0$.  Thus it behooves us to consider this possible curvature in more
detail.

\begin{deluxetable*}{llccccccccc}
	\tablewidth{0in}
	\tabletypesize{\scriptsize}
	\setlength{\tabcolsep}{0.03in}
	\tablecaption{TFR Cluster Distance Comparison\tablenotemark{1}\label{tab_tfr_distances}}
	\tablehead{
	 & & & & \multicolumn{7}{c}{This work} \\
	 \cline{5-11} \\
	\colhead{Cluster} &
	\colhead{Code\tablenotemark{2}} &
	\colhead{\citet{Tully:12:78}} &
	\colhead{\citet{Sorce:13:94}} &
	\colhead{I-band} &
	\colhead{W1} &
	\colhead{W2} &
	\colhead{W1cur} &
	\colhead{W2cur} &
	\colhead{W1cc} &
	\colhead{W2cc} \\
	}
\startdata
Virgo      & V & $15.9 \pm 0.8$ & $14.7 \pm 0.9$ & 
	$16.6 \pm 0.8$ &
	$15.9 \pm 0.9$ & $15.6 \pm 1.0$ &
	$16.9 \pm 0.6$ & $16.8 \pm 0.6$ & $16.2 \pm 0.8$ & $16.2 \pm 0.9$ \\
U Ma       & U & $17.4 \pm 0.9$ & $18.0 \pm 0.9$ & 
	$17.5 \pm 0.9$ &
	$16.9 \pm 1.0$ & $16.6 \pm 1.0$ &
	$17.4 \pm 0.6$ & $17.2 \pm 0.6$ & $17.2 \pm 0.9$ & $17.1 \pm 0.9$ \\
Fornax     & F & $17.3 \pm 1.0$ & $17.4 \pm 1.2$ & 
	$17.5 \pm 1.0$ &
	$17.4 \pm 1.2$ & $17.1 \pm 1.2$ &
	$18.1 \pm 0.8$ & $17.8 \pm 0.8$ & $17.5 \pm 1.0$ & $17.4 \pm 1.0$ \\
Antlia     & An & $37 \pm 2$ & $37 \pm 2$ & 
	$38 \pm 2$ &
	$39 \pm 2$ & $38 \pm 2$ &
	$40 \pm 1$ & $39 \pm 1$ & $39 \pm 2$ & $39 \pm 2$ \\
Centaurus  & Ce & $38 \pm 3$ & $39 \pm 4$ & 
	$39 \pm 3$ &
	$38 \pm 3$ & $37 \pm 3$ &
	$39 \pm 2$ & $39 \pm 2$ & $39 \pm 3$ & $38 \pm 3$ \\
Pegasus    & Pe & $43 \pm 3$ & $45 \pm 3$ & 
	$44 \pm 2$ &
	$43 \pm 3$ & $43 \pm 3$ &
	$47 \pm 2$ & $48 \pm 2$ & $44 \pm 2$ & $44 \pm 2$ \\
Hydra      & H & $59 \pm 4$ & $56 \pm 4$ & 
	$59 \pm 4$ &
	$62 \pm 4$ & $62 \pm 4$ &
	$60 \pm 2$ & $60 \pm 2$ & $62 \pm 4$ & $62 \pm 4$ \\
Pisces     & Pi & $64 \pm 2$ & $65 \pm 3$ & 
	$67 \pm 3$ &
	$72 \pm 3$ & $73 \pm 3$ &
	$71 \pm 2$ & $71 \pm 2$ & $68 \pm 3$ & $68 \pm 3$ \\
Cancer     & Ca & $65 \pm 3$ & $67 \pm 4$ & 
	$66 \pm 3$ &
	$66 \pm 4$ & $67 \pm 5$ &
	$66 \pm 2$ & $66 \pm 2$ & $64 \pm 3$ & $64 \pm 4$ \\
A400       & A4 & $94 \pm 5$ & $97 \pm 5$ & 
	$100 \pm 5$ &
	$102 \pm 7$ & $105 \pm 8$ &
	$103 \pm 4$ & $105 \pm 4$ & $100 \pm 5$ & $100 \pm 6$ \\
A1367      & A1 & $94 \pm 5$ & $96 \pm 6$ & 
	$94 \pm 5$ &
	$104 \pm 6$ & $105 \pm 6$ &
	$101 \pm 3$ & $101 \pm 3$ & $98 \pm 5$ & $98 \pm 5$ \\
Coma       & Co & $90 \pm 4$ & $95 \pm 6$ & 
	$90 \pm 4$ &
	$99 \pm 5$ & $101 \pm 6$ &
	$96 \pm 3$ & $97 \pm 3$ & $94 \pm 5$ & $94 \pm 5$ \\
A2634/66   & A2 & $121 \pm 7$ & $112 \pm 7$ & 
	$117 \pm 6$ &
	$121 \pm 7$ & $121 \pm 7$ &
	$116 \pm 3$ & $116 \pm 3$ & $117 \pm 6$ & $117 \pm 6$ \\
\enddata
\tablenotetext{1}{\ all distances in Mpc}
\tablenotetext{2}{\ see Figure\ref{fig_bias}}
\end{deluxetable*}

\section{Curvature in the MIR TFR\label{SEC:CURVE}}

Curvature in the near-IR TFR has been seen before using H-band luminosities, as
described in \citet[and references therein]{Aaronson:86:536}.  We adopt the
same strategy for dealing with the curvature, namely we take an empirical
approach rather than attempt to correct the magnitudes or linewidths.
Quadratic fits are also used in \citet[Appendix C]{Sakai:00:698} for the
$BVRIH_{-0.5}$ bands which show an increase in the curvature term with
wavelength.

We treat the curvature of the MIR TFR as an additional bias, or
perturbation term on top of the linear relation seen in optical TFRs.  By
adding a curvature term, we are fitting the relation with a quadratic and
as such the curvature of a quadratic requires that we fit with the
dependent variable in magnitudes.  We attempted to fit an inverted
quadratic, but the curvature does not follow the data well, thus we are
forced to fit the direct TFR.  We minimize the Malmquist bias by using the
ensemble of cluster galaxies shifted to the distance of Virgo with the
linear TFR as the input for the fit.  Fitting the direct relation with a
least-squares fitter means that our errors will be on the magnitude axis,
however we have already stated that the linewidth errors dominate,
especially compared to the {\it WISE} photometry.  We therefore use the
linear TFR to project the linewidth errors onto the magnitude axis and use
these projected magnitude errors in the fitting.

\begin{figure}
	\includegraphics[width=\linewidth]{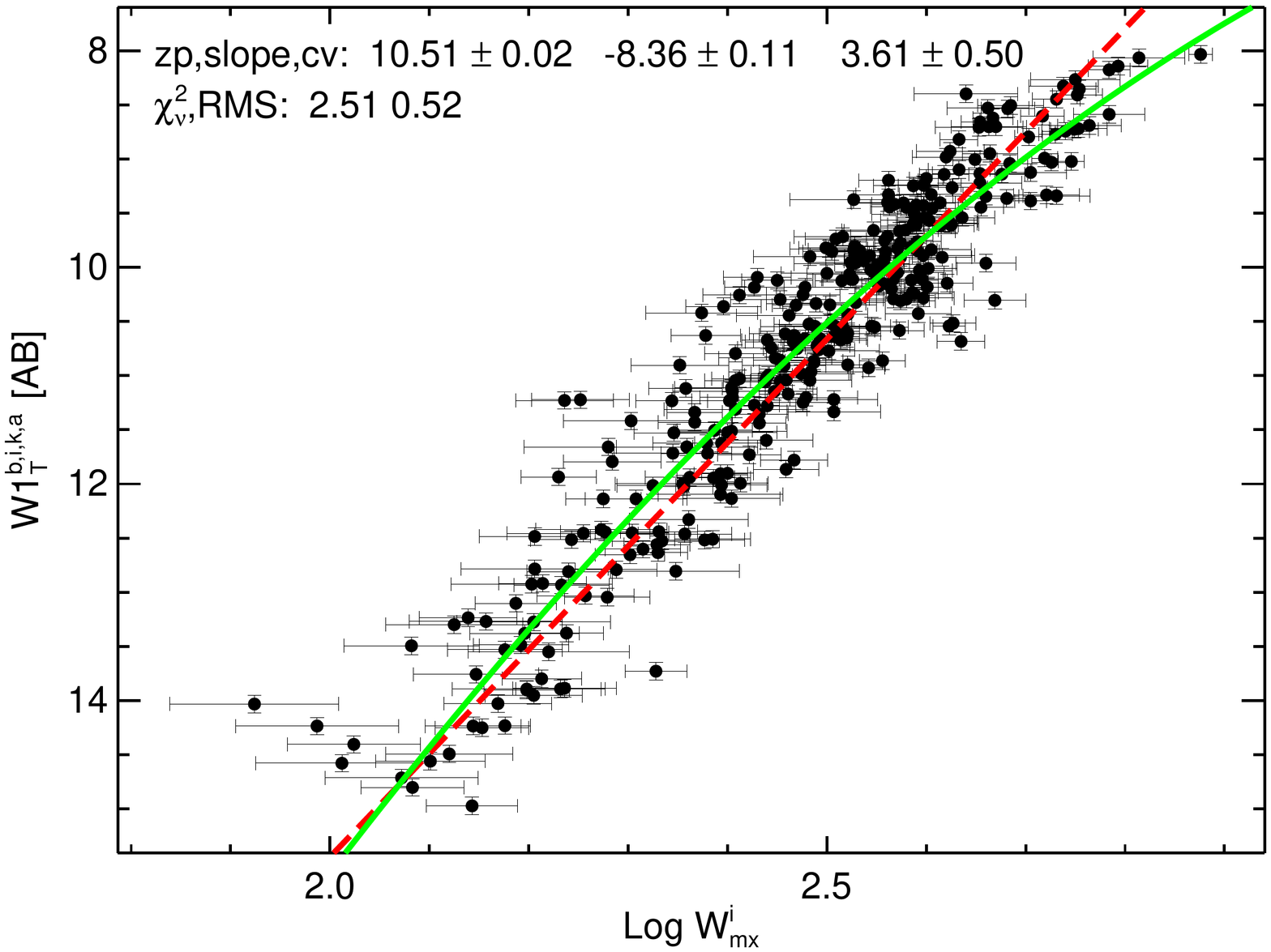}
	\includegraphics[width=\linewidth]{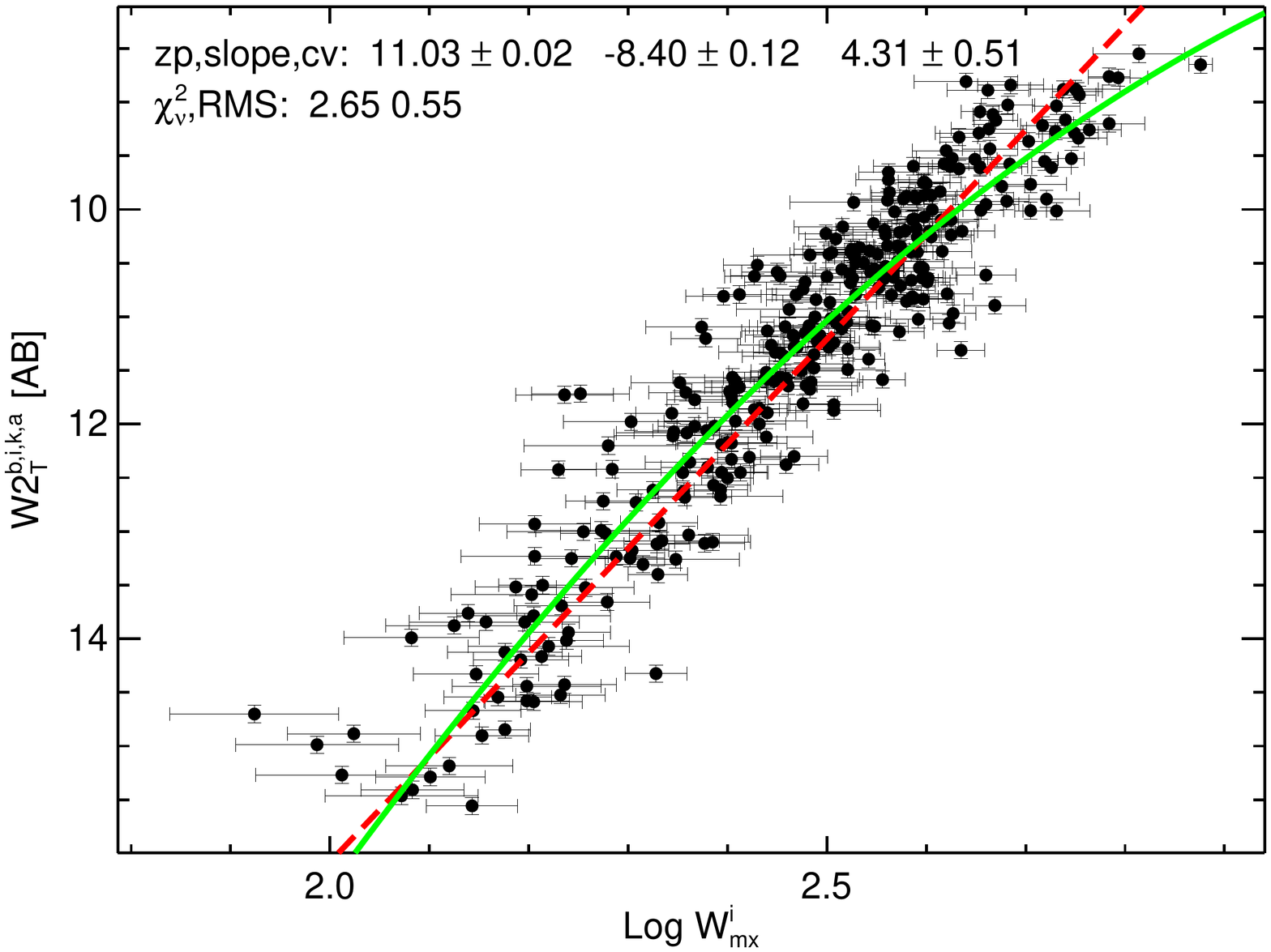}
	\caption{Fits to the ensemble of 13 clusters shifted to the
	distance of Virgo for the {\it WISE} W1 (top) and W2 (bottom).  The
	solid (green) line is the quadratic error-weighted fit to the
	direct TFR with errors entirely in linewidth, projected onto the
	magnitude axis using the linear TFR.  The dashed (red) line is the
	linear fit to the inverse TFR.  The two fits are very similar,
	especially at the faint end.  The annotations are for the curved
	direct fit and show an improvement in both rms scatter and
	$\chi^2_{\nu}$ over the linear fit.}
	\label{fig_ens_curve}
\end{figure}

We fit the same ensemble created from the linear fit to derive a universal
curve.  The results of these fits for both W1 and W2 are shown by the solid
(green) lines in Figure~\ref{fig_ens_curve}, while the linear fits are
shown by the dashed (red) lines \citep[compare these to Figure 5
in][]{Aaronson:86:536}.  We notice that the curved fits are close to the
linear fits, especially at the faint end of the relations.  The curved fits
reduce the rms scatter from 0.54 to 0.52 mag in W1 and from 0.56 to 0.55
mag in W2.  This brings the distance errors down to 26\% in W1 and down to
27\% in W2.  In addition, the curved fits improve $\chi^2_{\nu}$, which
goes from 3.1 to 2.5 in W1 and from 3.4 to 2.6 in W2.

\begin{figure}
	\includegraphics[width=\linewidth]{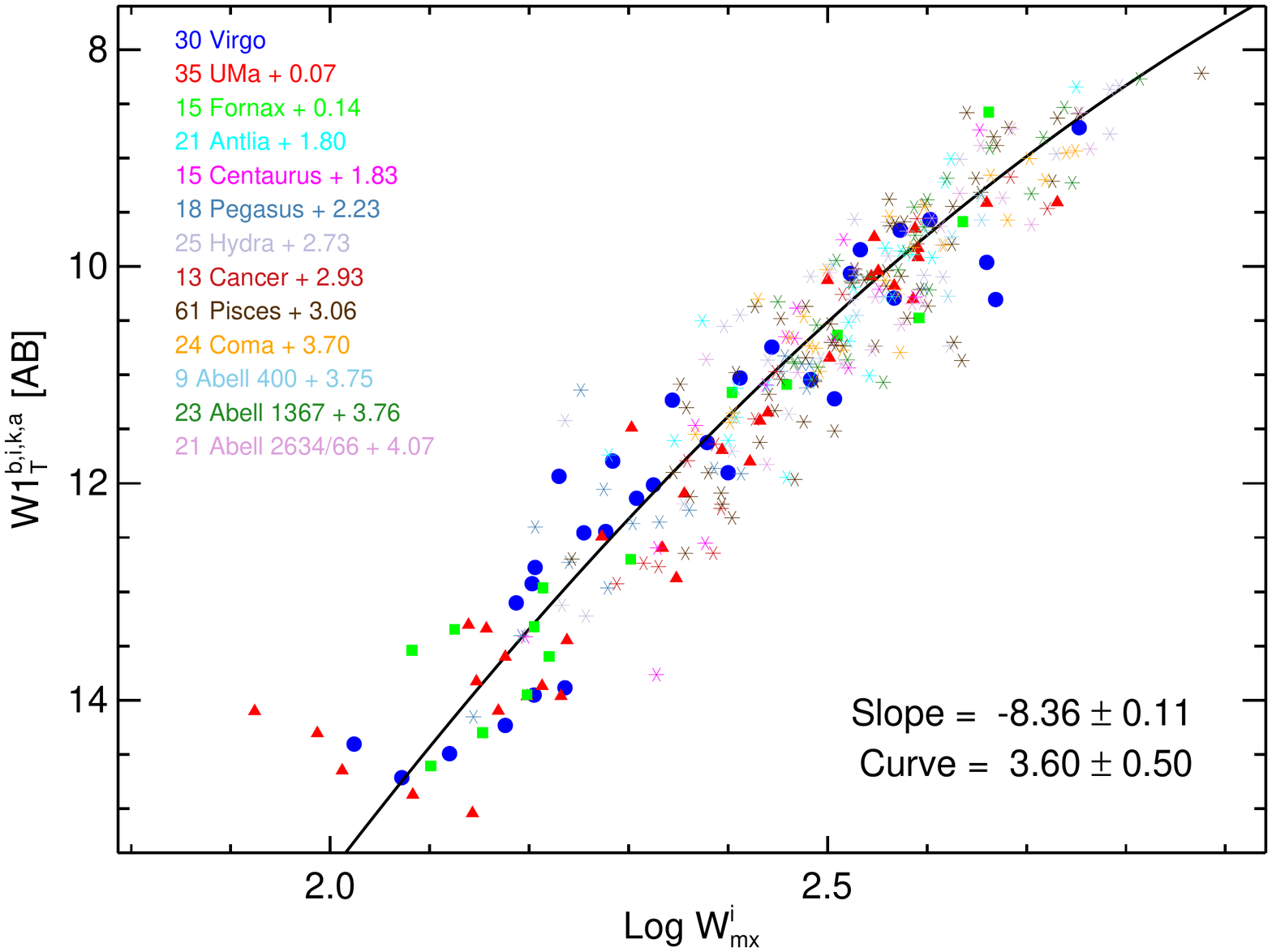}
	\includegraphics[width=\linewidth]{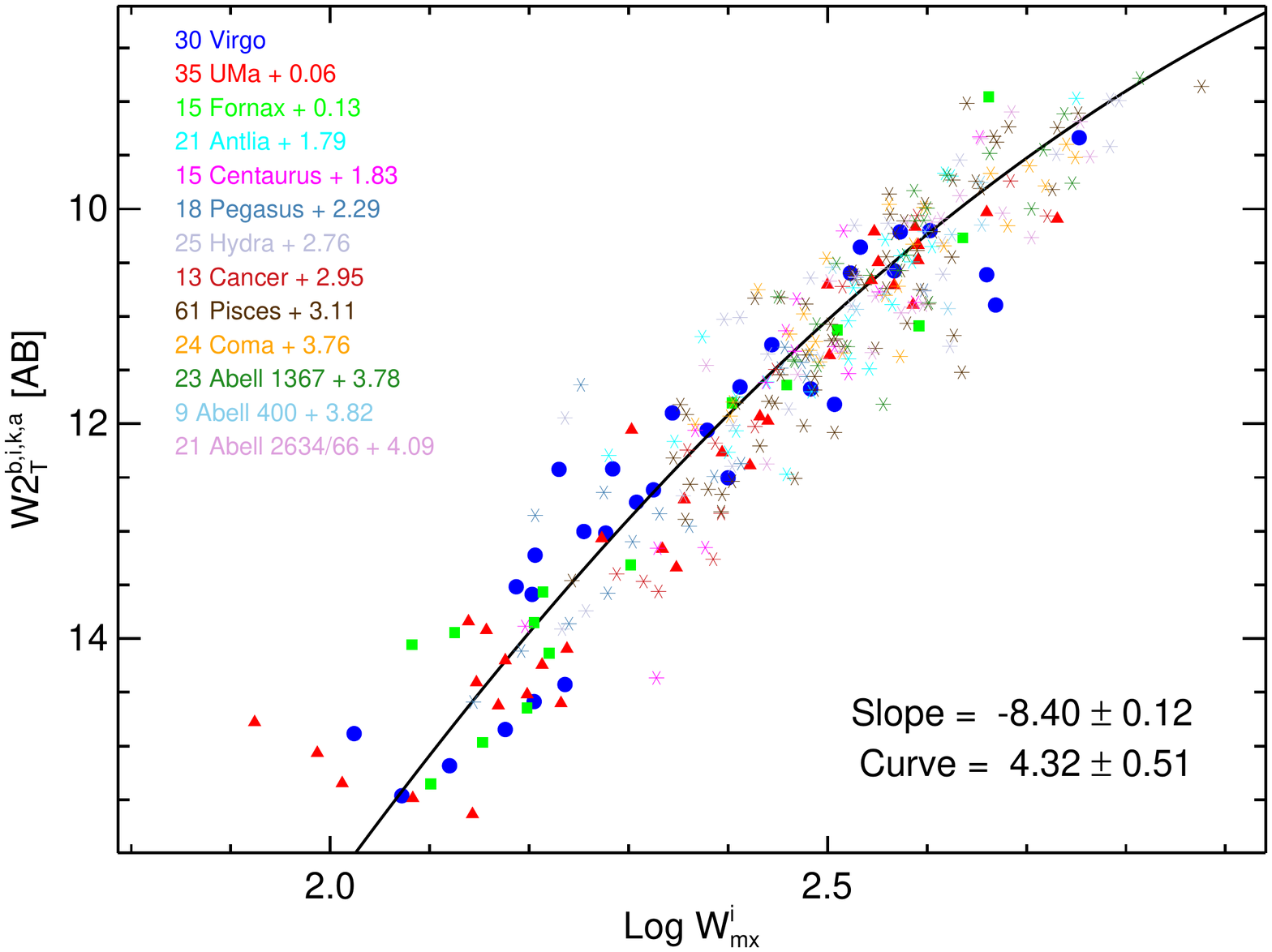}
	\caption{Curved TFR in the {\it WISE} W1 (top) and W2 (bottom)
	bands obtained from the galaxies in 13 clusters.  Offsets given
	with respect to the Virgo Cluster represent distance modulus
	differences between each cluster and Virgo. The solid line is the
	least-squares fit to all of the offset shifted galaxies with errors
	entirely in linewidths, projected onto the magnitude axis using the
	linear TFR.  the relations have an rms scatter of 0.52 mag for W1
	and 0.55 mag for W2.}
	\label{fig_all_curve}
\end{figure}

We present the annotated ensemble in Figure~\ref{fig_all_curve} for W1 in
the top panel and for W2 in the bottom panel.  Both of the fits have
similar slope terms of $-8.36 \pm 0.11$ for W1 and $-8.40 \pm 0.12$ for W2.
The curvature terms are $3.60 \pm 0.50$ for W1 and $4.32 \pm 0.51$ for W2.
We could compare these curvatures to the one found in
\citet{Aaronson:86:536}, for the H-band, however, they use a different
velocity measure for their TFR fitting.  The distance modulus offsets
from Virgo listed on
the annotation for the figure are in reasonable agreement with those for
the I-band shown in Figure~\ref{fig_icalibration}.  The cluster distances
shown in column five of Table~\ref{tab_tfr_distances} for the I-band are
also in agreement with the distances for the curved {\it WISE} TFRs shown
in columns eight and nine.  The distance offsets relative to the I-band TFR
distances are shown graphically in Figure~\ref{fig_wise_curlin} by the
(blue) open squares.  It is clear that the curved fits reduce the
systematic relative to the I-band.

\begin{figure}
	\includegraphics[width=\linewidth]{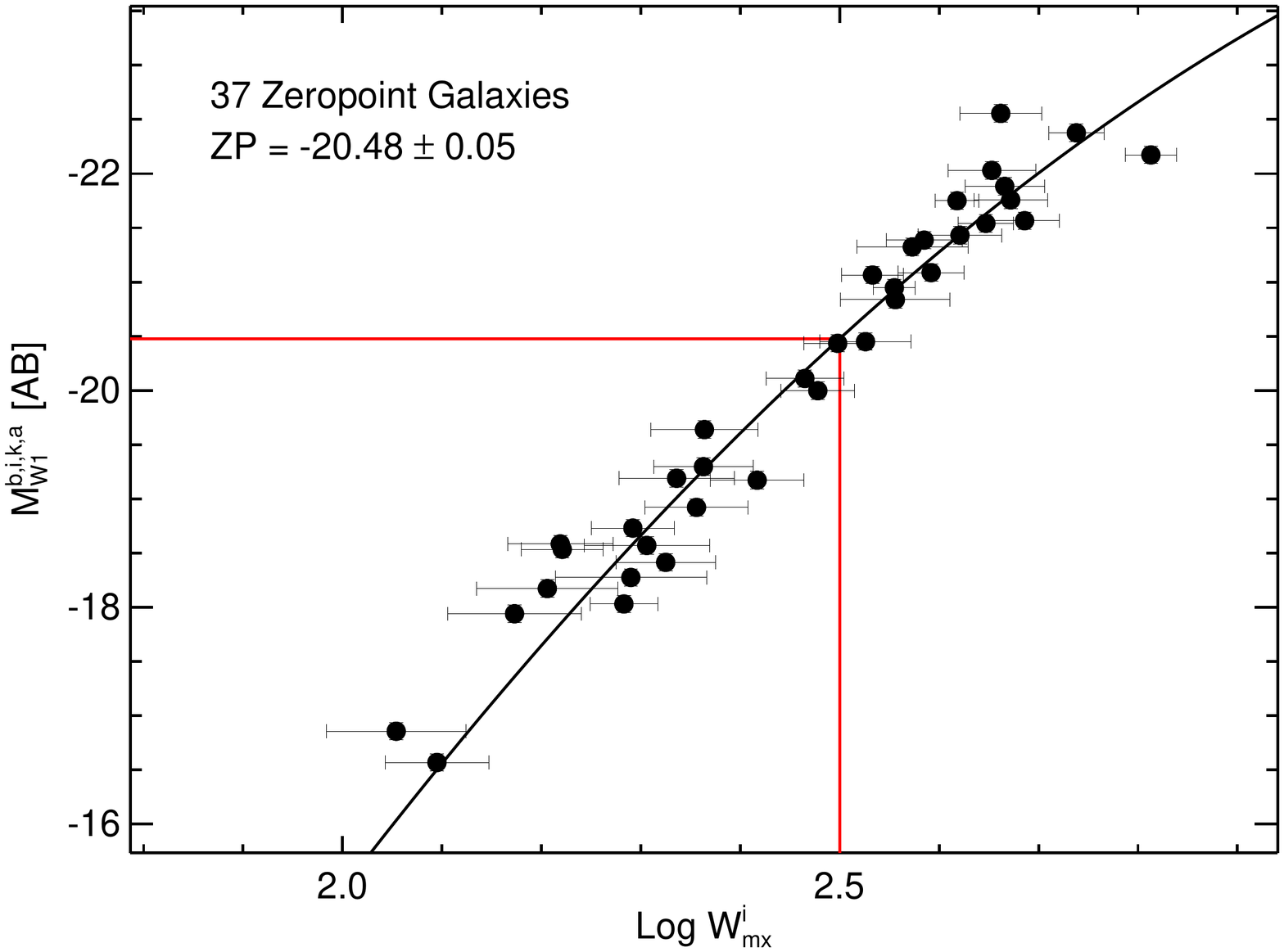}
	\includegraphics[width=\linewidth]{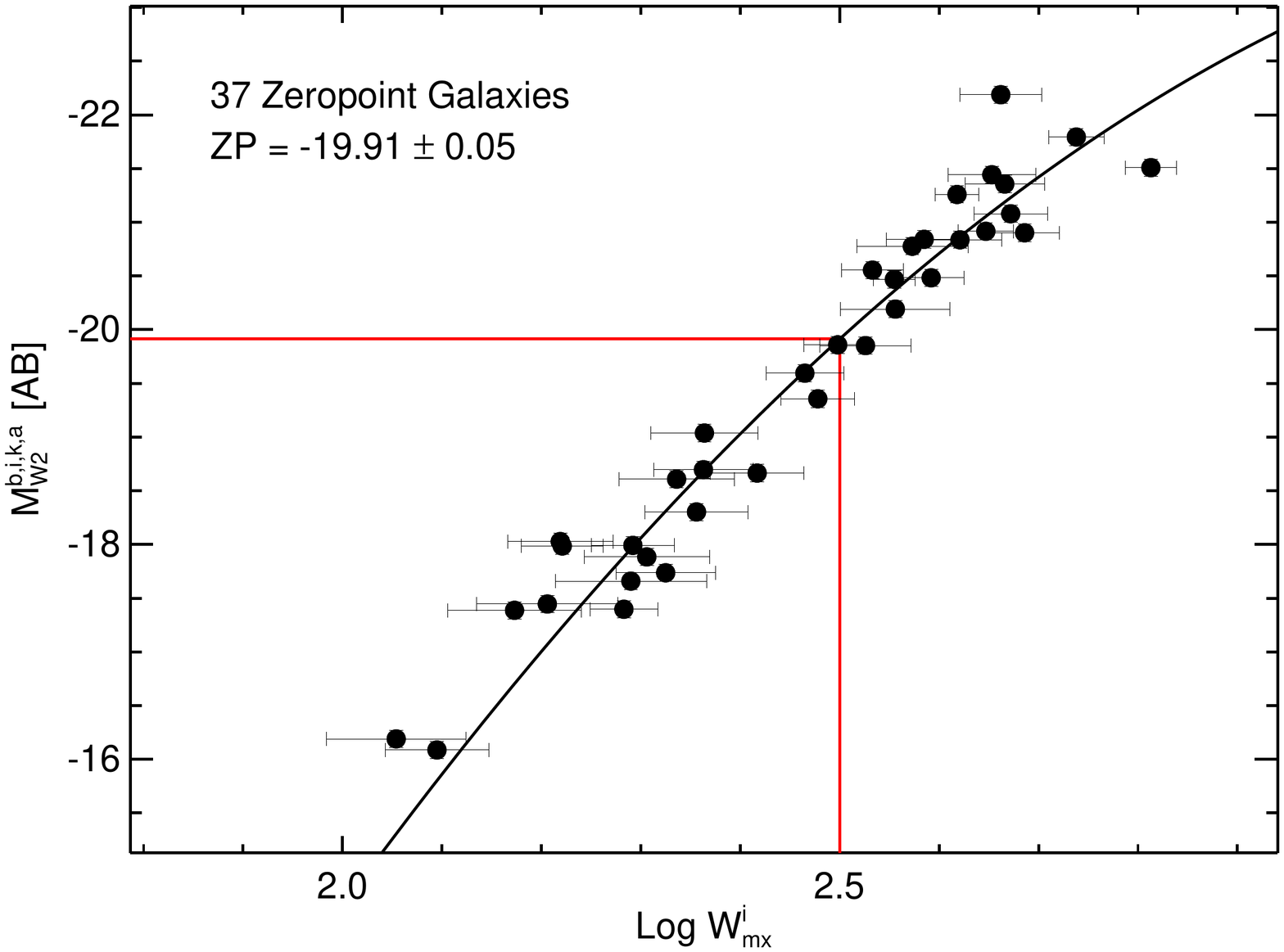}
	\caption{Curved TFR for the {\it WISE} W1 (top) and W2 (bottom)
	band using the 37 galaxies with distances established by
	observations of Cepheid variables or the TRGB. The solid line is
	the least-squares fit with the coefficients established by the 13
	cluster template. The zero point of the TFR is set at the value of
	this fit at log$W^i_{mx} = 2.5$ as indicated by the solid (red)
	vertical and horzontal lines.  The zero-point fits have a scatter
	of 0.39 mag in W1 and 0.43 mag in W2.}
	\label{fig_zp_curve}
\end{figure}

\begin{figure}
	\includegraphics[width=\linewidth]{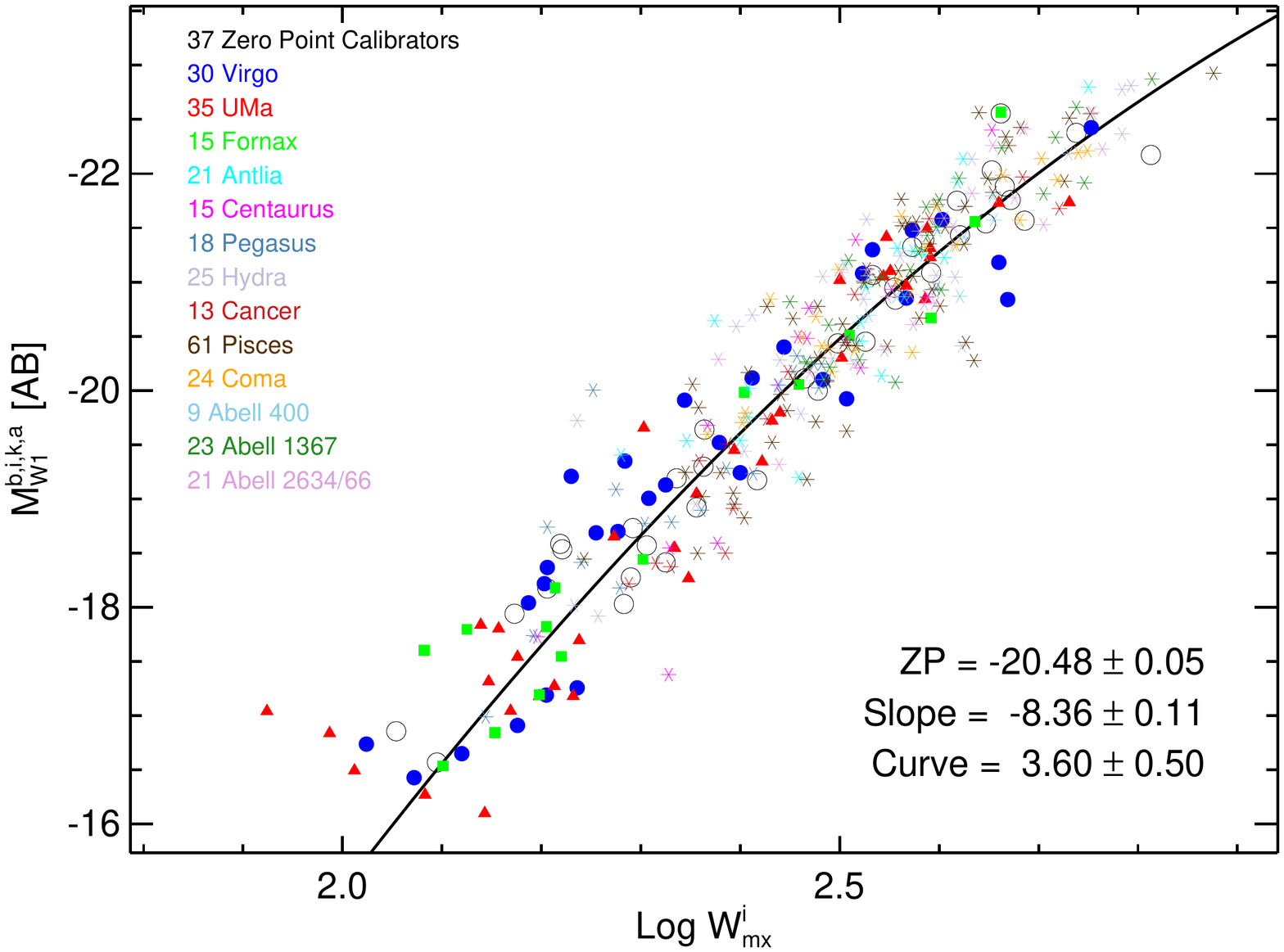}
	\includegraphics[width=\linewidth]{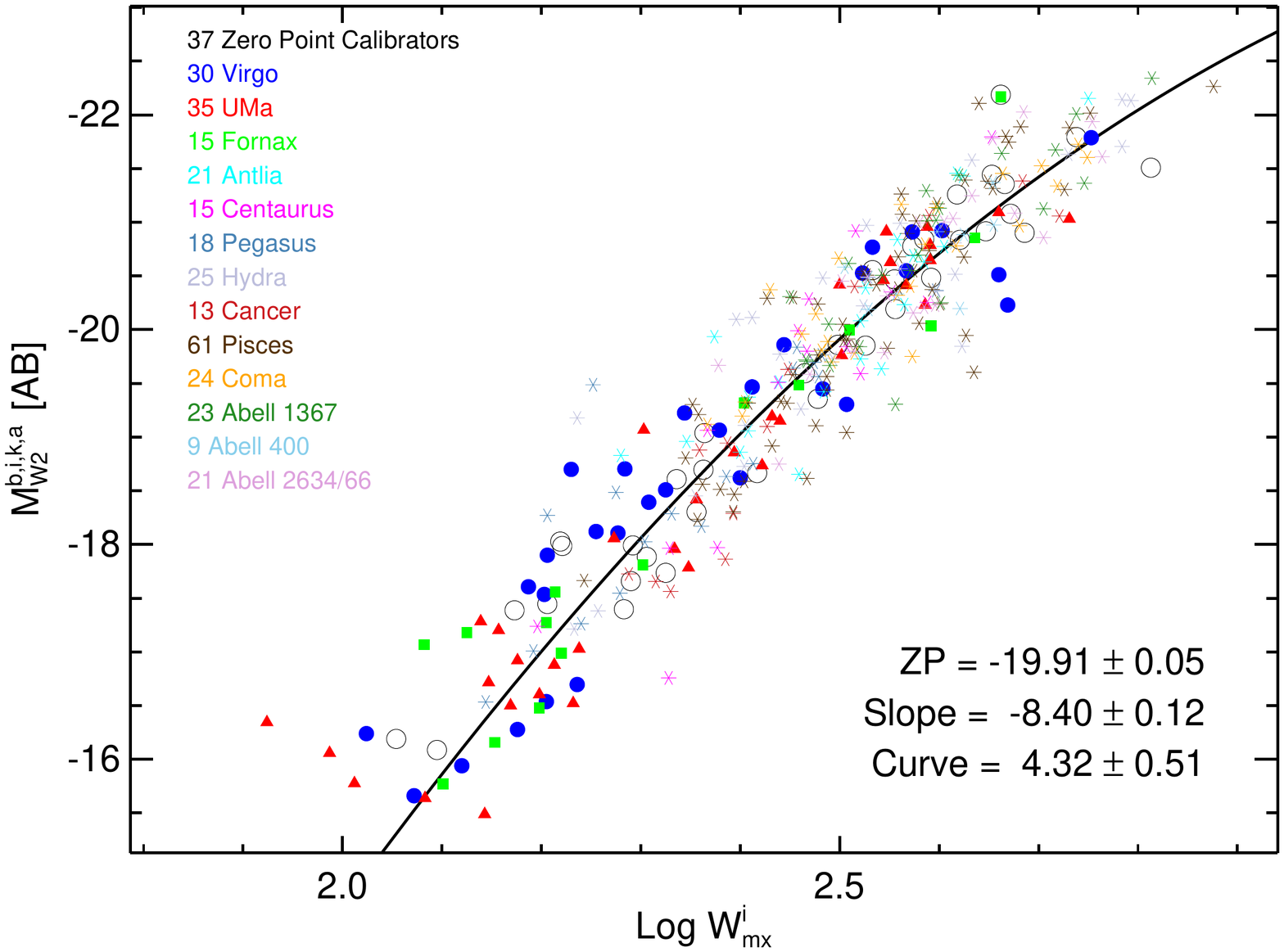}
	\caption{Curved TFR with the curve fit to the galaxies in 13
	clusters and the absolute magnitude scale set by 37 zero-point
	calibrators for the W1 (top) and the W2 (bottom).}
	\label{fig_zpall_curve}
\end{figure}

Now that we have a universal curve, we can use the same procedure that was
used for the linear TFR to find the zero-point of the curved relation.  We
present the results of the zero-point fitting in Figure~\ref{fig_zp_curve}.
The formal errors on the zero-point values are smaller relative to the
linear fits as is the rms scatter which is reduced from 0.45 to 0.39 mag in
W1 and from 0.49 to 0.43 mag in W2.  We point out that NGC2841 is no longer
such a large outlier as it was with the linear TFR.  The final curved TFR
is presented for both W1 and W2 in Figure~\ref{fig_zpall_curve}.  We
express the curved {\it WISE} TFR as
\begin{subequations}
	\label{eq_itfr_curve}
	\begin{align}
		\begin{split}
		\mathcal{M}^{b,i,k,a}_{W1} = &- (20.48 \pm 0.05) \\
			&- (8.36 \pm 0.11)(\log W^i_{mx} - 2.5) \\
			&+ (3.60 \pm 0.50)(\log W^i_{mx} - 2.5)^2, 
			\label{eq_itfr_w1_curve}
		\end{split} \\
		\begin{split}
		\mathcal{M}^{b,i,k,a}_{W2} = &- (19.91 \pm 0.05) \\
			&- (8.40 \pm 0.12)(\log W^i_{mx} - 2.5) \\
			&+ (4.32 \pm 0.51)(\log W^i_{mx} - 2.5)^2.  
			\label{eq_itfr_w2_curve}
		\end{split}
	\end{align}
\end{subequations}
The distance modulus can then be calculated as before with
Equation~\ref{eq_tfmu_wise}.

\section{Optical - MIR Color Term\label{SEC:COLOR_TERM}}

\begin{figure*}
	\includegraphics[width=0.5\linewidth]{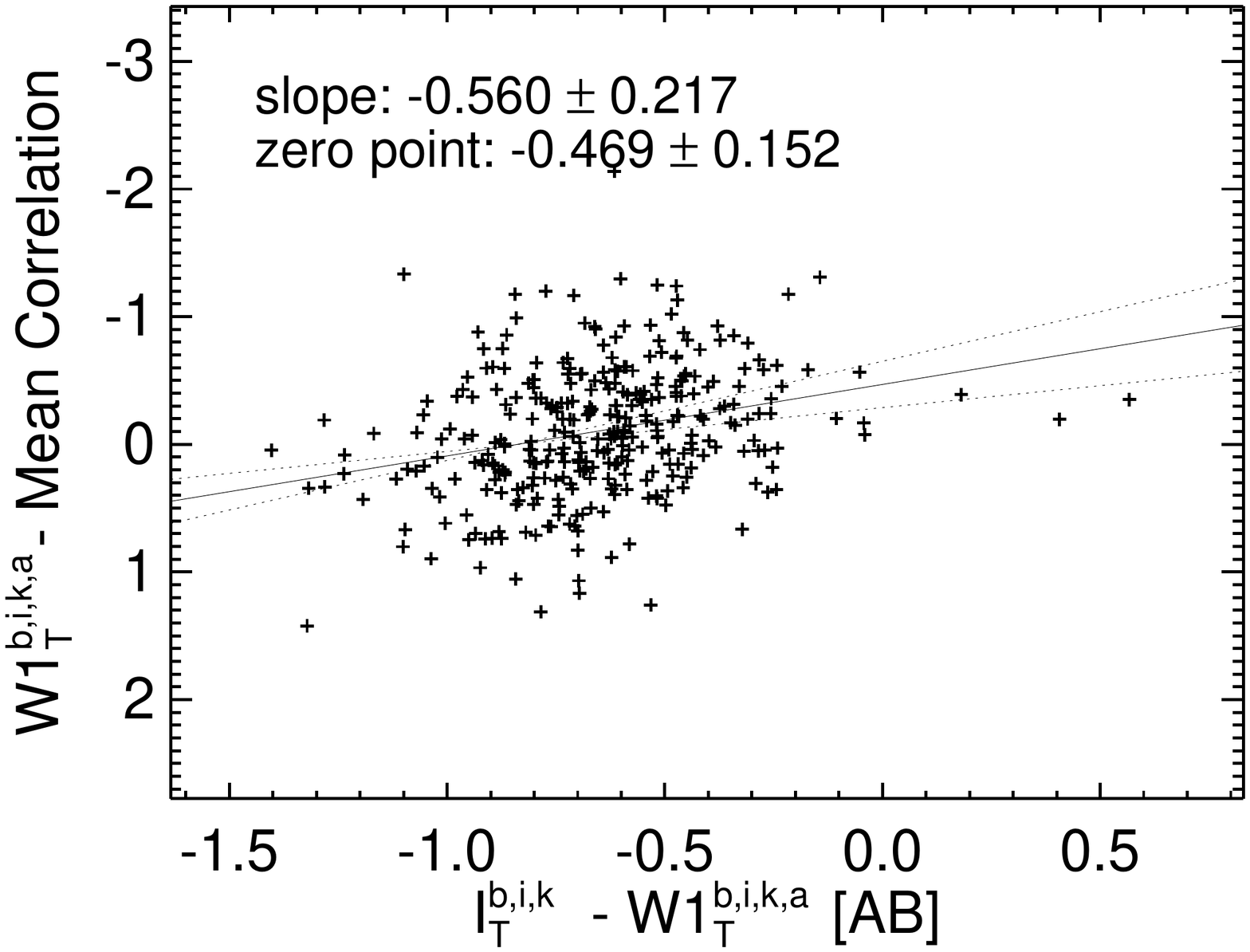}
	\includegraphics[width=0.5\linewidth]{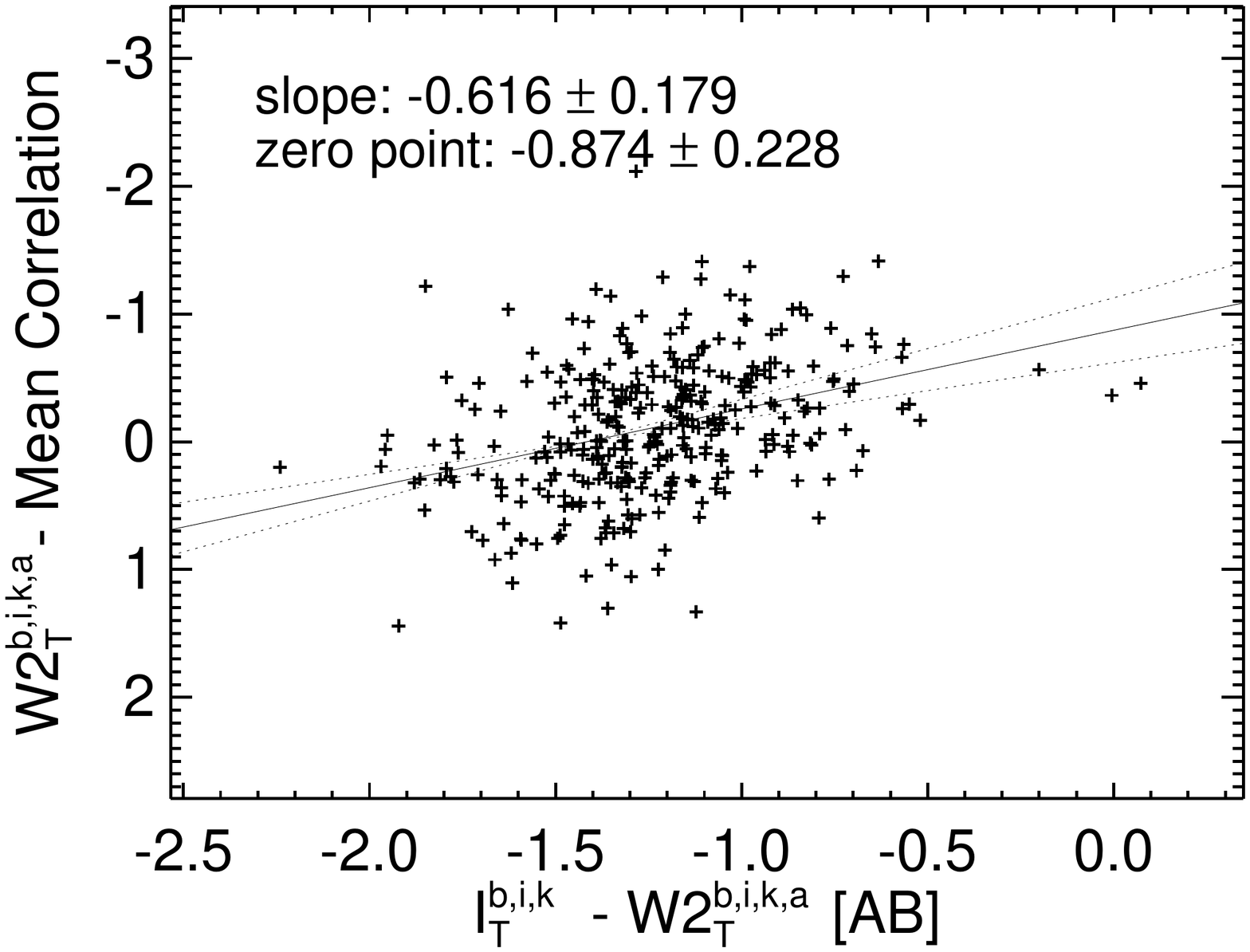}
	\caption{Deviations from the mean linear {\it WISE} TFRs as a
	function of $I - W1$ (left) and $I - W2$ (right) color.}
	\label{fig_color_term}
\end{figure*}

\begin{figure}
\begin{center}
	\includegraphics[width=\linewidth]{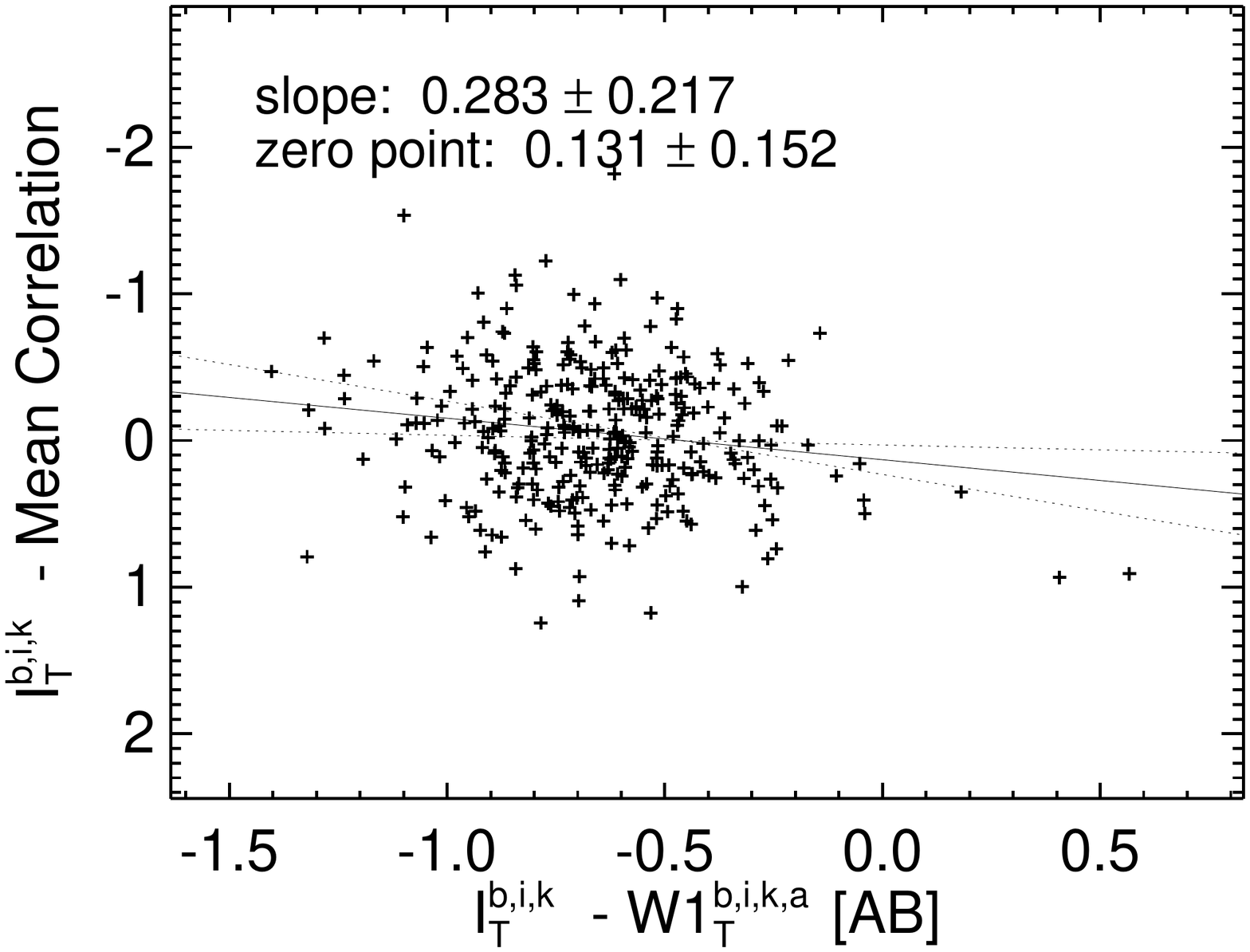}
	\end{center}
	\caption{Deviations from the mean I-band TFR as a function of $I -
	W1$ color (note that the slope is significant only at $1.2\sigma$
	and the zero point is consistent with zero).}
	\label{fig_color_term_i}
\end{figure}

We do not repeat the color term discussion from \citet{Sorce:13:94},
however, we remind the reader that there is good reason to suspect that a
color term might exist because the TFR relation steepens with wavelength.
Indeed, such a color term was detected in \citet{Sorce:13:94} (see their
\S3.3 and their Figures~6 through 8) and was used to reduce their scatter
from 0.49 to 0.44 magnitudes.  The {\it WISE} data also show correlations
between the optical to MIR color and the mean linear TFR residuals as shown
in Figure~\ref{fig_color_term}.  Figure~\ref{fig_color_term_i} shows an
attempt to find a similar trend in the I-band residuals, but the slope of
our fit is consistent with an insignificant ($1.2\sigma$) correlation.  We
note that in this section we are using the linear, not the curved, {\it
WISE} TFR.

\begin{figure}
	\includegraphics[width=\linewidth]{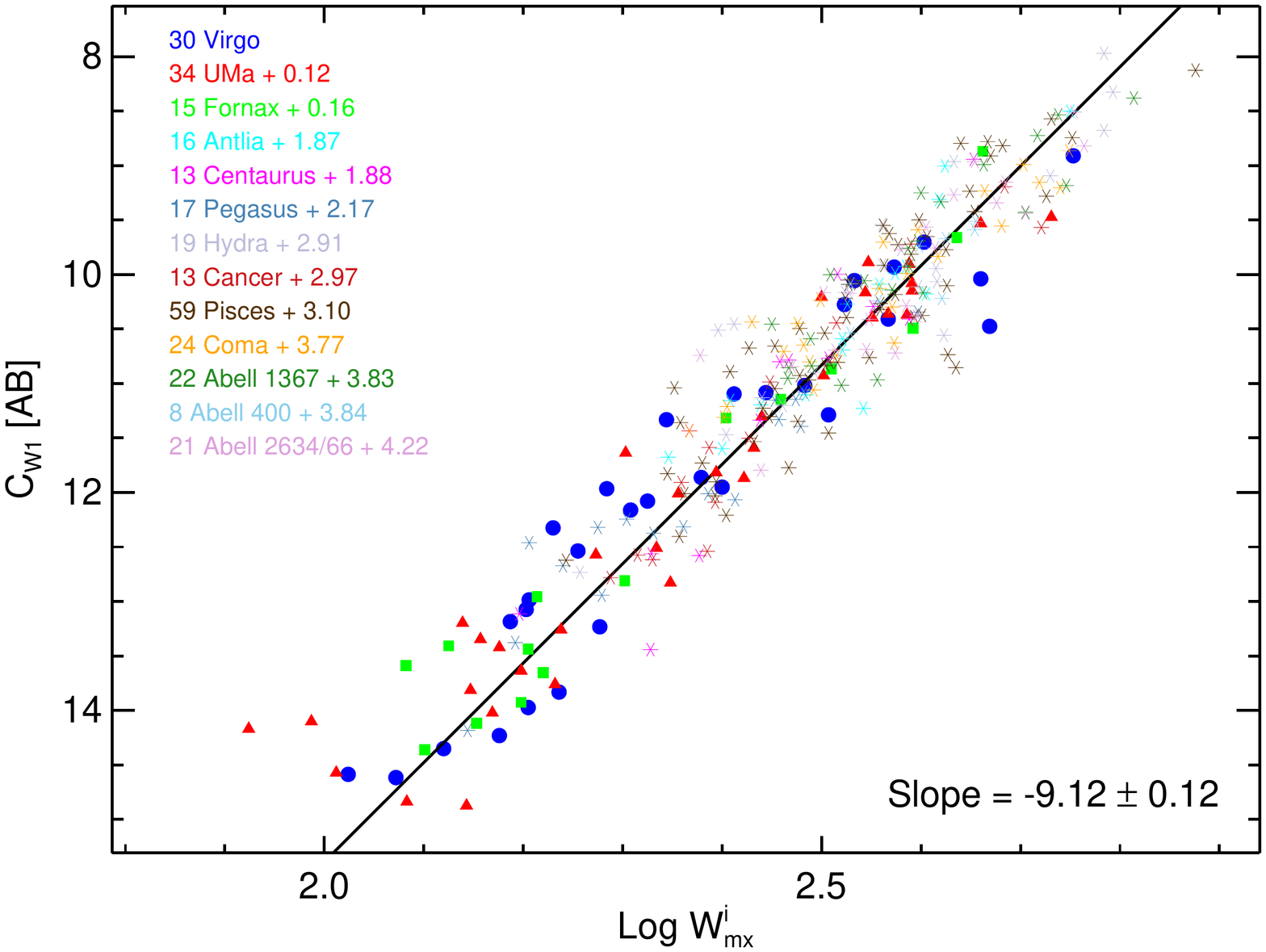}
	\includegraphics[width=\linewidth]{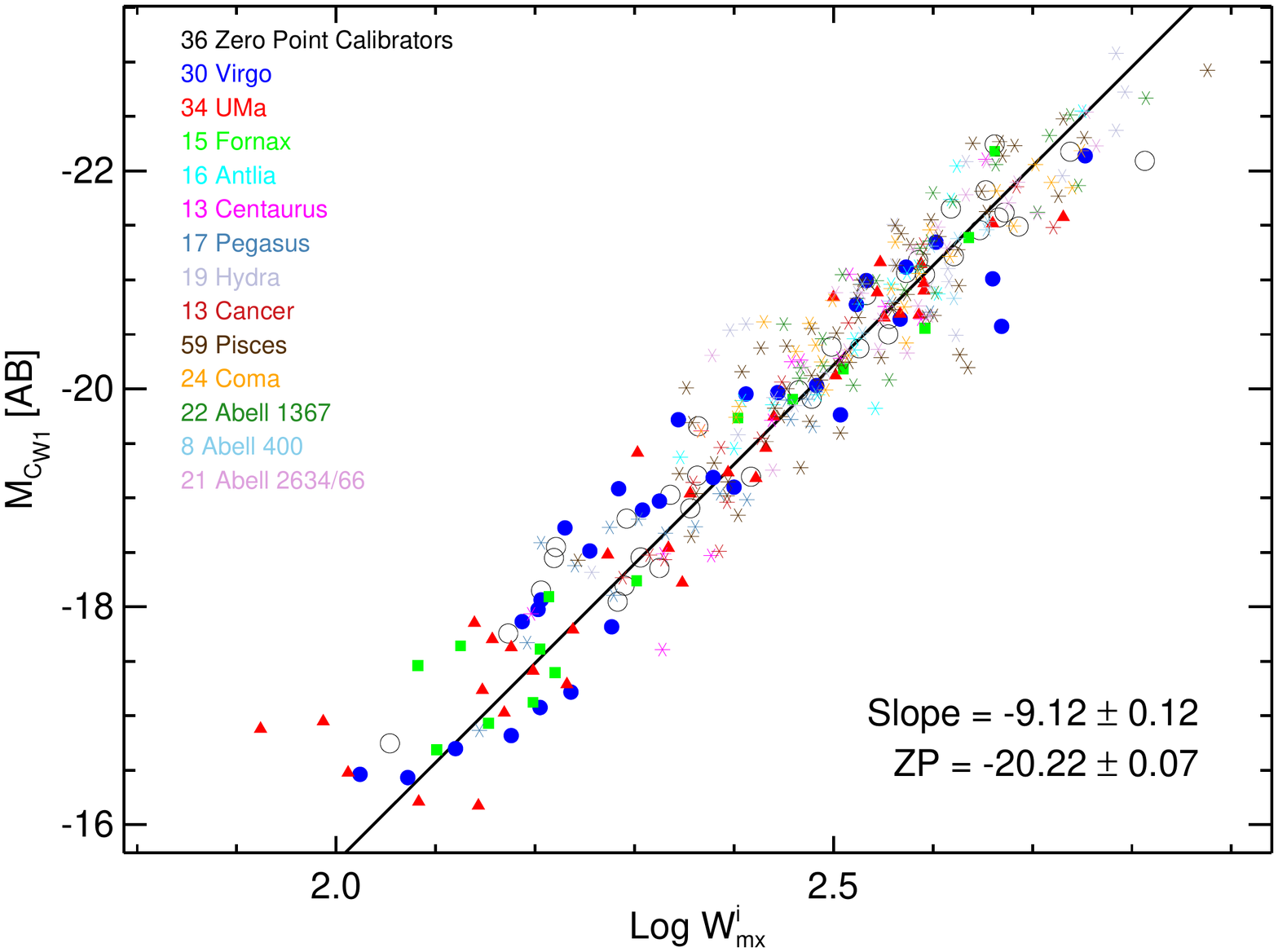}
	\caption{Linear TFR for W1 after adjustments for the color term
	with galaxies shifted to the apparent distance of Virgo (top) and
	on the absolute magnitude scale set by 37 zero-point calibrator
	galaxies (bottom).}
	\label{fig_w1_cc}
\end{figure}

\begin{figure}
	\includegraphics[width=\linewidth]{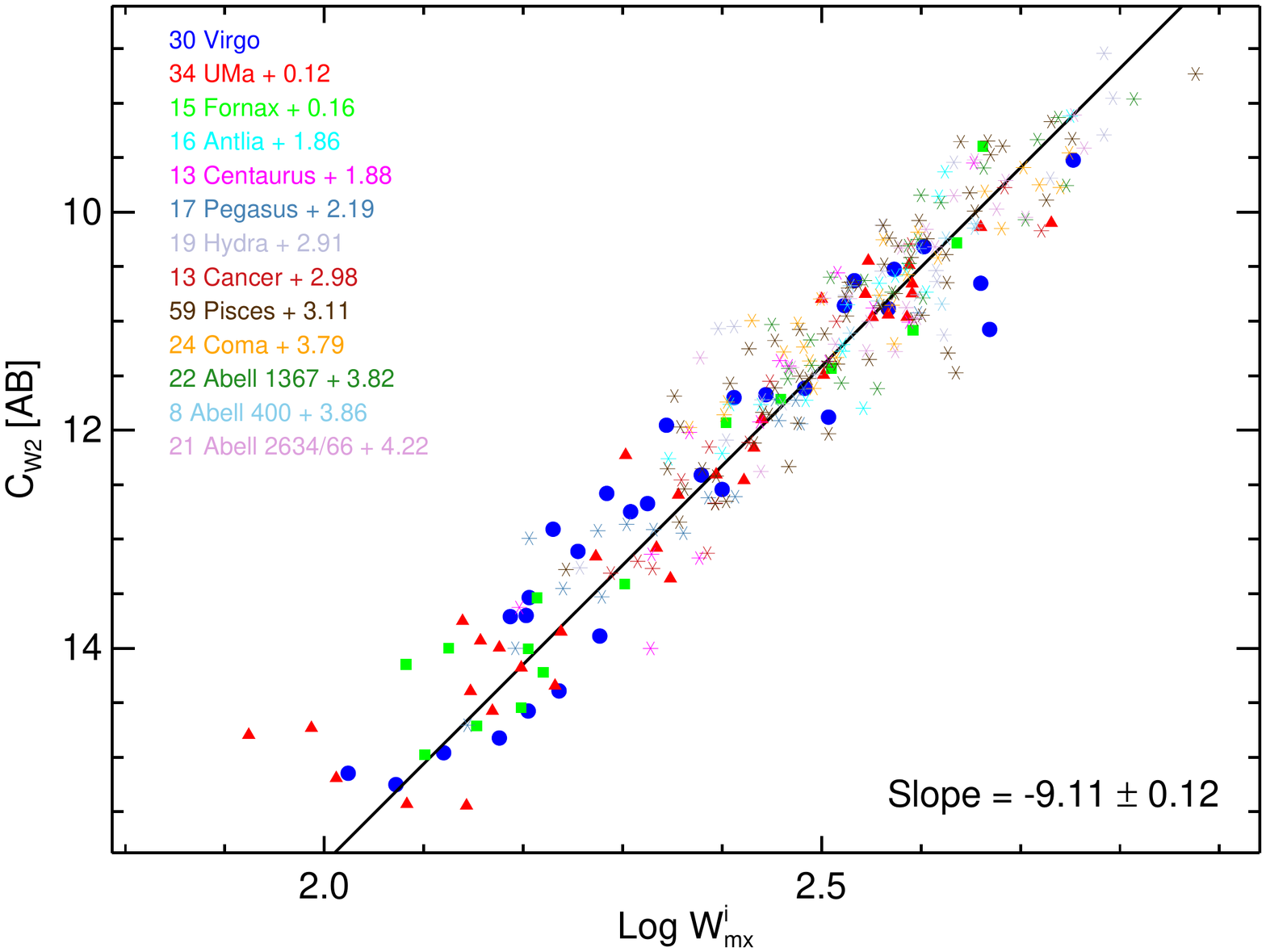}
	\includegraphics[width=\linewidth]{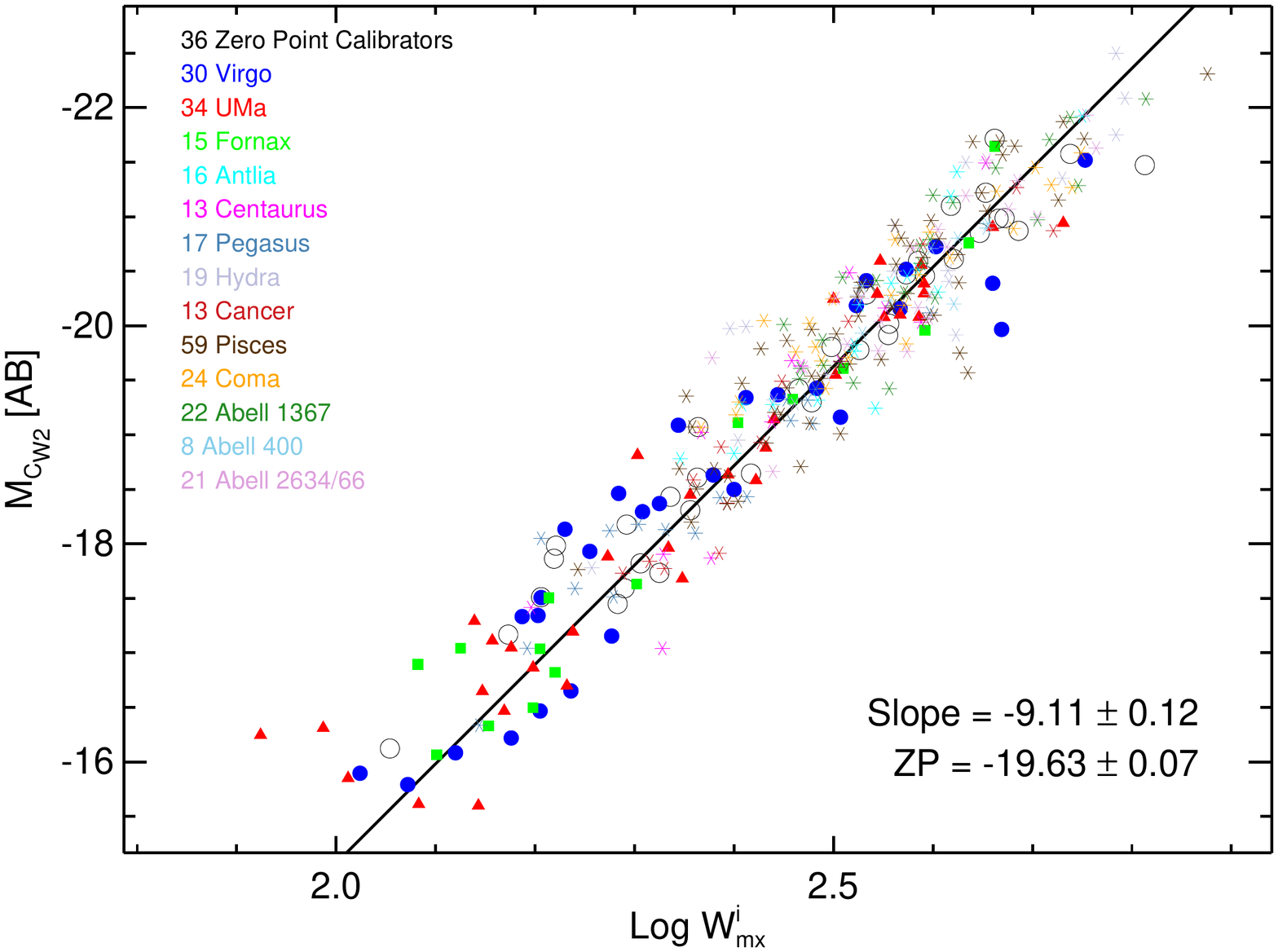}
	\caption{Linear TFR for W2 after adjustments for the color term
	with galaxies shifted to the apparent distance of Virgo (top) and
	on the absolute magnitude scale set by 37 zero-point calibrator
	galaxies (bottom).}
	\label{fig_w2_cc}
\end{figure}

We use the I-band minus {\it WISE} band color to correct the magnitudes 
and improve the scatter of the fits following \citet{Sorce:13:94}.  We
use the residuals calculated in Equation~\ref{eq_resid} and then we fit 
the correlation between the I-band to {\it WISE} band color and the 
residuals as shown in Figure~\ref{fig_color_term}.  Thus we derive the 
correction to the magnitude that will produce an absolute  magnitude from
the TFR with the least scatter:
\begin{subequations}
	\label{eq_cc}
	\begin{align}
	\Delta W1^{color} & = -0.470 - 0.561(I^{b,i,k}_{T} -
		W1^{b,i,k,a}_{T}), \label{eq_cc_w1} \\
	\Delta W2^{color} & = -0.874 - 0.617(I^{b,i,k}_{T} -
		W2^{b,i,k,a}_{T}).  \label{eq_cc_w2}
	\end{align}
\end{subequations}
These are then used to adjust the input magnitudes as follows:
\begin{equation}
	C_{W1,2} = W1,2^{b,i,k,a}_{T} - \Delta W1,2^{color}.
	\label{eq_corrected}
\end{equation}
We repeat the entire fitting process using $C_{W1}$ and $C_{W2}$ instead of
$W1^{b,i,k,a}_{T}$ and $W2^{b,i,k,a}_{T}$.  Using these pseudo-magnitudes
reduces the ensemble scatter from 0.54 for W1 and 0.56 for W2 to an
ensemble scatter of 0.46 magnitudes for both bands, which compares well
with the scatter of 0.44 magnitudes after color term correction found in
\citet{Sorce:13:94}.  The individual cluster zero-points and scatters for
the color-corrected pseudo-magnitudes are shown in columns seven and eight
of Tables~\ref{tab_clusters_w1} and \ref{tab_clusters_w2}.  The value of
0.46 mag for the color-corrected ensemble scatter corresponds to a distance
error of 23\% in both W1 and W2. In addition, the universal slopes for W1
and W2 are now nearly identical with a value of -9.12 for W1 and -9.11 for
W2, whereas prior to color correction they were -9.56 for W1 and -9.74 for
W2.  The scatter in the zero point sample was also reduced from 0.45 mag in
W1 and 0.49 mag in W2 to 0.41 mag for W1 and 0.42 mag W2.  Compare these
with a color term corrected scatter of 0.37 mag for the zero point sample
in \citet{Sorce:13:94}.  Figures~\ref{fig_w1_cc} and \ref{fig_w2_cc} show
the result of fitting these pseudo magnitudes.  Since the color correction
requires I-band photometry, the sample used for the color correction is
reduced from 310 to 291 galaxies.  The number of color-corrected galaxies
in each cluster is listed in column six of Tables~\ref{tab_clusters_w1} and
\ref{tab_clusters_w2}.

This pseudo-magnitude calibration can now be expressed as
\begin{subequations}
	\label{eq_itfr_cc}
	\begin{align}
	\begin{split}
		\mathcal{M}_{C_{W1}} = &- (20.22 \pm 0.07) \\
			&- (9.12 \pm 0.12)(\log W^i_{mx} - 2.5), 
			\label{eq_itfr_cc_w1} 
	\end{split} \\
	\begin{split}
		\mathcal{M}_{C_{W2}} = &- (19.63 \pm 0.07) \\
			&- (9.11 \pm 0.12)(\log W^i_{mx} - 2.5).  
			\label{eq_itfr_cc_w2}
	\end{split}
	\end{align}
\end{subequations}
In order to derive the distance modulus for a given galaxy, we subtract
Equation~\ref{eq_itfr_cc} from Equation~\ref{eq_corrected}:
\begin{equation}
	\mu_{C_{W1,2}} = C_{W1,2} - \mathcal{M}_{C_{W1,2}}.
\label{eq_tfmu}
\end{equation}

As a check for a color term in the I-band TFR, we plot the residuals with
respect to the mean TFR, $\Delta M_I = M^{ens}_I - \mathcal{M}^{b,i,k}_I$
(analogous to Equation~\ref{eq_resid}), as a function of the I-band minus
W1 color in Figure~\ref{fig_color_term_i}.  The formal error on the zero
point is larger than zero point itself.  The slope has a value which is
insignificant at $1.2\sigma$.  Thus, we conclude that an I-band color
correction would have little or no effect.

We point out that this color-correction has the effect of linearizing the
{\it WISE} TFR and thereby removing the curvature we found in the pure {\it
WISE} linear TFR.  This color-corrected TFR has the advantage of lower
scatter, while the curved TFR has the advantage that it does not rely on
any other source of photometry.  We refer to these color-corrected
magnitudes as W1cc and W2cc in plots and tables to distinguish them from
the pure {\it WISE} magnitudes W1 and W2.

\section{Comparison with Previous Calibrations\label{SEC:TFR_COMPARE}}

\begin{deluxetable*}{lrrlcccrlc}
	\tablewidth{0in}
	\tabletypesize{\scriptsize}
	\setlength{\tabcolsep}{0.03in}
	\tablecaption{TFR Parameter Comparison\label{tab_tfr_params}}
	\tablehead{
	& &
	\multicolumn{4}{c}{Universal Slope/Curve} & &
	\multicolumn{3}{c}{Zero Point} \\
	\cline{3-6} \cline{8-10} \\
	\colhead{Reference} &
	\colhead{Photometry} &
	\colhead{Ngal} &
	\colhead{Slope} &
	\colhead{Curve} &
	\colhead{rms} &
	&
	\colhead{Ngal} &
	\colhead{Mag} &
	\colhead{rms}
	}
\startdata
\citet{Tully:12:78} & I-band (Vega) & 267 & $-8.81 \pm 0.16$ & \ldots & 0.41 & & 36 & $-21.39 \pm 0.07$ & 0.36 \\
          This work & I-band (Vega) & 291 & $-8.95 \pm 0.14$ & \ldots & 0.46 & & 36 & $-21.34 \pm 0.07$ & 0.40 \\
          \hline
\citet{Sorce:13:94} & IRAC [3.6] (AB) & 213 & $-9.74 \pm 0.22$ & \ldots & 0.49 & & 26 & $-20.34 \pm 0.10$ & 0.44 \\
          This work & W1 (AB) & 310 & $-9.56 \pm 0.12$ & \ldots & 0.54 & & 37 & $-20.35 \pm 0.07$ & 0.45 \\
          This work & curved W1 (AB) & 310 & $-8.36 \pm 0.11$ & $3.60 \pm 0.50$ & 0.52 & & 37 & $-20.48 \pm 0.05$ & 0.39 \\
          \hline
\citet{Sorce:13:94} & $M_{C3.6\mu{m}}$ (AB) & 213 & $-9.13 \pm 0.22$ & \ldots & 0.44 & & 26 & $-20.34 \pm 0.08$ & 0.37 \\
          This work & $M_{CW1}$ (AB)& 291 & $-9.12 \pm 0.12$ & \ldots & 0.46 & & 36 & $-20.22 \pm 0.07$ & 0.41 \\
\citet{Lagattuta:13:88} & $M_{corr}$ (AB) & 568 & $-10.05$ & \ldots & 0.69 & & \ldots & $-19.54$ & \ldots \\
          \hline
          This work & W2 (AB) & 310 & $-9.74 \pm 0.12$ & \ldots & 0.56 & & 37 & $-19.76 \pm 0.08$ & 0.49 \\
          This work & $M_{CW2}$ (AB)& 291 & $-9.11 \pm 0.12$ & \ldots & 0.46 & & 36 & $-19.63 \pm 0.07$ & 0.42 \\
          This work & curved W2 (AB) & 310 & $-8.40 \pm 0.12$ & $4.32 \pm 0.51$ & 0.55 & & 37 & $-19.91 \pm 0.05$ & 0.43 \\
\enddata
\end{deluxetable*}

We compare our results with previous TFR calibrations in
Table~\ref{tab_tfr_params}.  In the I-band, the new calibration agrees with
that from \citet{Tully:12:78} to well within the formal errors on the
parameters.  Our scatter is a little higher perhaps due to adding fainter
galaxies.  Comparing our W1 calibration to the IRAC [3.6] result in
\citet{Sorce:13:94} shows consistency, both in the pure linear calibration
parameters and in the color corrected parameters.  When we restrict our
sample to the same galaxies used in \citet{Sorce:13:94}, we obtain the
exact same scatter for the pure linear W1 TFR (0.49 mag).  The only
deviation of note is the color-corrected zero point which is fainter for
the W1.  The zero-point samples are not identical and it is possible the
color corrections couple with the I-band in a different way due to
differences in filter responses between the IRAC [3.6] and W1 bandpasses.

The curved {\it WISE} TFR offers an improvement over the linear {\it WISE}
TFR, although the rms scatter is still not as good as the color-corrected
linear {\it WISE} TFR.  We point out that the formal errors on the
zero-points for both the curved W1 and W2 TFRs are the lowest of all the
fits and the scatter on the zero-point curved W1 calibration is marginally
lower than for the color-corrected linear W1 TFR.

Lastly, we compare our results to the calibration in
\citet{Lagattuta:13:88}.  This calibration was derived from {\it WISE}
catalog photometry and not derived by the authors from their own photometry
of the W1 images, as we have done here.  The extended photometry is based
on 2MASS apertures with a correction applied to account for the shallowness
of the 2MASS survey as compared to the {\it WISE} survey.  No errors on the
individual luminosity-linewidth correlation parameters are given, so we can
only compare the scatter which is greater by 50\% than the calibration
presented here.  The zero-point from that paper has been converted from
Vega to AB magnitudes in Table~\ref{tab_tfr_params}.

\section{The Hubble Constant, $H_0$\label{SEC:HUBBLE}}

We can now use the TFR relation to derive distances and, with cosmological
model-corrected recession velocities, estimate the local Hubble constant
($H_0$).  Before we do this, we must consider any residual bias in our
distance estimates due to our sample.

\subsection{Distance Bias\label{SEC:BIAS}}

The residual bias pointed out by \citet{Willick:94:1} and discussed in
detail in \citet{Sorce:13:94} is mitigated to some extent here since the
current sample was selected using the 2MASS redshift survey complete to
K=11.75 \citep{Huchra:12:26}, effectively bringing the sample selection
wavelength much closer to the {\it WISE} bands than the original sample,
which was selected in the B-band.  However, we still must account for the
fact that with a faint-end limit, more faint galaxies will be scattered
into the sample than bright galaxies out of the sample.

\begin{figure}
	\includegraphics[width=\linewidth]{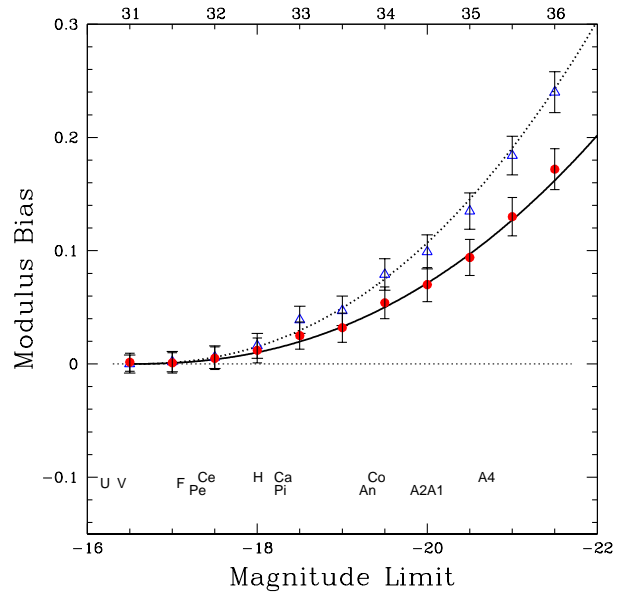}
	\caption{Bias $\langle M \rangle_{measured}$ for the pure {\it
	WISE} TFR (blue open triangles) and color-corrected {\it WISE} TFR
	(solid red circles) as a function of absolute magnitude limit which
	increases with distance.  The solid curve is the empirical bias fit
	to the color-corrected {\it WISE} points which has the form $b =
	0.004(\mu - 31)^{2.3}$. The dotted curve is the empirical bias fit
	to the pure {\it WISE} data which has the form $b = 0.006(\mu -
	31)^{2.3}$.  Letters at the bottom are codes for the 13 calibrating
	clusters (see column two of Table~\ref{tab_tfr_distances}).  Their
	horizontal positions indicate sample limits and the vertical
	intercepts with the solid curve give the corresponding biases.}
	\label{fig_bias}
\end{figure}

The bias analysis carried out in \citet{Sorce:13:94} and
\citet{Tully:12:78} is repeated here, but using a \citet{Schechter:76:297}
function with $\alpha = -1.0$ instead of $-0.9$.  This value of $\alpha$
was arrived at by fitting the {\it WISE} W1 luminosity function of the
combined nearest three calibration clusters: Virgo-Fornax-UMa \citep[see
their \S3.1,]{Tully:12:78}.  The bright end characteristic magnitude for
the {\it WISE} is the same as was used for the IRAC [3.6] magnitudes:
$M^{\star}_{W1} = -22$.  Augmenting our sample selection with the 2MASS
redshift survey allows us to assume a flat cutoff in the magnitudes as a
function of linewidth.  The driving factor in calculating the bias is the
observed scatter in the TFR.  Since the scatter in the color-corrected {\it
WISE} TFR (for both W1 and W2) is the same as the I-band TFR (0.46 mag), we
can use a simulation with this scatter to characterize the bias for all
three TFRs.  For the pure {\it WISE} linear and curved TFRs we use a
scatter of 0.54 mag.  A simulated TFR having the appropriate scatter is
generated from these parameters and randomly sampled at a range of cutoff
magnitude, $M^{lim}$, which slides to brighter limits linearly as distance
increases.  The bias $\langle M \rangle_{measured}$ is determined at
intervals of $M^{lim}$ corresponding to increasing distance.  At each
cutoff limit, a random set of galaxies brighter than $M^{lim}$ is drawn
from the simulated TFR and used to calculate a new TFR.  The average
deviation from the input (true) TFR is the bias $\langle M
\rangle_{measured}$.  This bias is plotted in Figure~\ref{fig_bias} for
both the pure (open blue triangles) and color-corrected (solid red circles)
{\it WISE} TFR.  The solid and dotted curves are normalized to zero at a
distance modulus of $\mu = 31$ (Virgo) where we are assumed to be complete.
These curves are described by the formulae
\begin{subequations}
	\label{eq_bias}
	\begin{align}
		b_{pure} = 0.006(\mu - 31)^{2.3}
		\label{eq_bias_wise} \\
		b_{cc} = 0.004(\mu - 31)^{2.3}
		\label{eq_bias_cc_wise}
	\end{align}
\end{subequations}
where $\mu$ is the distance modulus to the object (galaxy or cluster)
derived using one of Equation~\ref{eq_tfmu_wise}, \ref{eq_tfmu_i} or
\ref{eq_tfmu}.  This bias function is slightly steeper than that seen in
\citet{Sorce:13:94} having an exponent of 2.3 instead of 2 due to an
increase in the assumed scatter from 0.40 to 0.46 magnitudes.  The letter
codes in Figure~\ref{fig_bias} show the cutoff magnitudes for the
calibration clusters (see column two of Table~\ref{tab_tfr_distances}) by
their horizontal placement and the resulting bias by the vertical
intersection with the solid line.  For a galaxy in the field, the corrected
distance modulus is thus
\begin{subequations}
	\label{eq_dmod}
	\begin{align}
		\begin{split}
		\mu^c_{pure} =&\ (W1,2_T^{b,i,k,a} - \mathcal{M}_{W1,2}^{b,i,k,a})\ + \\
			&\ 0.006[(W1,2_T^{b,i,k,a} - \mathcal{M}_{W1,2}^{b,i,k,a}) - 31]^{2.3}, 
		\end{split} \\
		\begin{split}
		\mu^c_{cc} =&\ (C_{W1,2} - \mathcal{M}_{C_{W1,2}})\ + \\
			&\ 0.004[(C_{W1,2} - \mathcal{M}_{C_{W1,2}}) - 31]^{2.3}.
		\end{split}
	\end{align}
\end{subequations}

We list the biases for the pure {\it WISE} and the color-corrected {\it
WISE} magnitudes in column three of Table~\ref{tab_h0_data}.  Now we turn
to using our calibrating clusters to estimate $H_0$.
\begin{deluxetable*}{lrrrrrl}
	\tablewidth{0in}
	\tabletypesize{\scriptsize}
	\tablecaption{Cluster Distances and $H_0$ Data\label{tab_h0_data}}
	\tablehead{
	\colhead{Cluster\tablenotemark{1}} &
	\colhead{$V_{mod}$\tablenotemark{2}} &
	\colhead{Bias\tablenotemark{3}} &
	\colhead{DM\tablenotemark{4}} &
	\colhead{$D_{Mpc}$\tablenotemark{5}} &
	\colhead{$V_{mod}/D_{Mpc}$\tablenotemark{6}} &
	\colhead{TFR Band\tablenotemark{7}}
	}
\startdata
\multirow{5}{*}{Virgo} & \multirow{5}{*}{1495. $\pm$  37.} & 0.000 & 31.14 $\pm$ 0.08 &  16.93 $\pm$  0.59 &  88.32 $\pm$  5.47 &  W1cur \\
 & & 0.000 & 31.12 $\pm$ 0.08 &  16.78 $\pm$  0.59 &  89.09 $\pm$  5.52 &  W2cur \\
 & & 0.000 & 31.05 $\pm$ 0.12 &  16.21 $\pm$  0.85 &  92.22 $\pm$  7.51 &  W1cc \\
 & & 0.000 & 31.04 $\pm$ 0.12 &  16.16 $\pm$  0.86 &  92.52 $\pm$  7.59 &  W2cc \\
 & & 0.000 & 31.10 $\pm$ 0.11 &  16.61 $\pm$  0.85 &  90.00 $\pm$  7.23 &  I-band \\
\hline
\multirow{5}{*}{U Ma} & \multirow{5}{*}{1079. $\pm$  14.} & 0.000 & 31.21 $\pm$ 0.07 &  17.45 $\pm$  0.56 &  61.83 $\pm$  2.88 &  W1cur \\
 & & 0.000 & 31.18 $\pm$ 0.07 &  17.22 $\pm$  0.55 &  62.66 $\pm$  2.91 &  W2cur \\
 & & 0.000 & 31.17 $\pm$ 0.12 &  17.16 $\pm$  0.92 &  62.90 $\pm$  4.44 &  W1cc \\
 & & 0.000 & 31.16 $\pm$ 0.12 &  17.09 $\pm$  0.92 &  63.13 $\pm$  4.45 &  W2cc \\
 & & 0.000 & 31.21 $\pm$ 0.12 &  17.47 $\pm$  0.93 &  61.78 $\pm$  4.32 &  I-band \\
\hline
\multirow{5}{*}{Fornax} & \multirow{5}{*}{1358. $\pm$  45.} & 0.000 & 31.28 $\pm$ 0.10 &  18.06 $\pm$  0.83 &  75.21 $\pm$  6.21 &  W1cur \\
 & & 0.000 & 31.25 $\pm$ 0.10 &  17.79 $\pm$  0.81 &  76.33 $\pm$  6.31 &  W2cur \\
 & & 0.000 & 31.21 $\pm$ 0.13 &  17.48 $\pm$  1.04 &  77.68 $\pm$  7.63 &  W1cc \\
 & & 0.000 & 31.21 $\pm$ 0.13 &  17.43 $\pm$  1.05 &  77.93 $\pm$  7.75 &  W2cc \\
 & & 0.000 & 31.22 $\pm$ 0.13 &  17.54 $\pm$  1.00 &  77.43 $\pm$  7.42 &  I-band \\
\hline
\multirow{5}{*}{Antlia} & \multirow{5}{*}{3198. $\pm$  74.} & 0.060 & 33.00 $\pm$ 0.07 &  39.81 $\pm$  1.31 &  80.33 $\pm$  4.67 &  W1cur \\
 & & 0.060 & 32.97 $\pm$ 0.07 &  39.26 $\pm$  1.30 &  81.45 $\pm$  4.73 &  W2cur \\
 & & 0.040 & 32.96 $\pm$ 0.11 &  38.99 $\pm$  1.97 &  82.01 $\pm$  6.37 &  W1cc \\
 & & 0.040 & 32.94 $\pm$ 0.11 &  38.78 $\pm$  1.96 &  82.47 $\pm$  6.40 &  W2cc \\
 & & 0.040 & 32.89 $\pm$ 0.12 &  37.91 $\pm$  2.04 &  84.35 $\pm$  6.87 &  I-band \\
\hline
\multirow{5}{*}{Centaurus} & \multirow{5}{*}{3823. $\pm$  82.} & 0.000 & 32.97 $\pm$ 0.09 &  39.30 $\pm$  1.56 &  97.28 $\pm$  6.19 &  W1cur \\
 & & 0.000 & 32.95 $\pm$ 0.09 &  38.94 $\pm$  1.56 &  98.18 $\pm$  6.29 &  W2cur \\
 & & 0.000 & 32.93 $\pm$ 0.16 &  38.55 $\pm$  2.71 &  99.18 $\pm$  9.79 &  W1cc \\
 & & 0.000 & 32.92 $\pm$ 0.16 &  38.42 $\pm$  2.72 &  99.50 $\pm$  9.89 &  W2cc \\
 & & 0.000 & 32.93 $\pm$ 0.15 &  38.62 $\pm$  2.54 &  98.99 $\pm$  9.26 &  I-band \\
\hline
\multirow{5}{*}{Pegasus} & \multirow{5}{*}{3062. $\pm$  78.} & 0.000 & 33.37 $\pm$ 0.10 &  47.27 $\pm$  2.09 &  64.77 $\pm$  4.72 &  W1cur \\
 & & 0.000 & 33.42 $\pm$ 0.10 &  48.19 $\pm$  2.15 &  63.53 $\pm$  4.66 &  W2cur \\
 & & 0.000 & 33.22 $\pm$ 0.12 &  44.04 $\pm$  2.32 &  69.54 $\pm$  5.74 &  W1cc \\
 & & 0.000 & 33.23 $\pm$ 0.12 &  44.34 $\pm$  2.29 &  69.06 $\pm$  5.61 &  W2cc \\
 & & 0.000 & 33.21 $\pm$ 0.12 &  43.93 $\pm$  2.45 &  69.70 $\pm$  5.99 &  I-band \\
\hline
\multirow{5}{*}{Hydra} & \multirow{5}{*}{4088. $\pm$  72.} & 0.015 & 33.89 $\pm$ 0.07 &  59.95 $\pm$  1.77 &  68.19 $\pm$  3.32 &  W1cur \\
 & & 0.015 & 33.90 $\pm$ 0.06 &  60.26 $\pm$  1.76 &  67.84 $\pm$  3.28 &  W2cur \\
 & & 0.010 & 33.97 $\pm$ 0.14 &  62.09 $\pm$  3.91 &  65.84 $\pm$  5.67 &  W1cc \\
 & & 0.010 & 33.96 $\pm$ 0.14 &  62.06 $\pm$  3.95 &  65.87 $\pm$  5.72 &  W2cc \\
 & & 0.010 & 33.85 $\pm$ 0.14 &  58.94 $\pm$  3.75 &  69.36 $\pm$  6.02 &  I-band \\
\hline
\multirow{5}{*}{Pisces} & \multirow{5}{*}{4759. $\pm$  39.} & 0.030 & 34.23 $\pm$ 0.06 &  70.21 $\pm$  1.77 &  67.78 $\pm$  2.32 &  W1cur \\
 & & 0.030 & 34.26 $\pm$ 0.06 &  71.19 $\pm$  1.80 &  66.85 $\pm$  2.29 &  W2cur \\
 & & 0.020 & 34.17 $\pm$ 0.09 &  68.30 $\pm$  2.83 &  69.68 $\pm$  3.61 &  W1cc \\
 & & 0.020 & 34.17 $\pm$ 0.09 &  68.36 $\pm$  2.84 &  69.62 $\pm$  3.61 &  W2cc \\
 & & 0.020 & 34.12 $\pm$ 0.09 &  66.65 $\pm$  2.70 &  71.40 $\pm$  3.63 &  I-band \\
\hline
\multirow{5}{*}{Cancer} & \multirow{5}{*}{5059. $\pm$  82.} & 0.030 & 34.10 $\pm$ 0.08 &  66.10 $\pm$  2.43 &  76.54 $\pm$  4.21 &  W1cur \\
 & & 0.030 & 34.11 $\pm$ 0.08 &  66.28 $\pm$  2.42 &  76.32 $\pm$  4.17 &  W2cur \\
 & & 0.020 & 34.04 $\pm$ 0.12 &  64.24 $\pm$  3.41 &  78.75 $\pm$  5.76 &  W1cc \\
 & & 0.020 & 34.04 $\pm$ 0.12 &  64.24 $\pm$  3.54 &  78.75 $\pm$  5.95 &  W2cc \\
 & & 0.020 & 34.09 $\pm$ 0.11 &  65.68 $\pm$  3.28 &  77.03 $\pm$  5.37 &  I-band \\
\hline
\multirow{5}{*}{A400} & \multirow{5}{*}{7228. $\pm$  97.} & 0.165 & 35.05 $\pm$ 0.08 & 102.52 $\pm$  3.92 &  70.50 $\pm$  3.79 &  W1cur \\
 & & 0.165 & 35.11 $\pm$ 0.08 & 105.10 $\pm$  3.98 &  68.77 $\pm$  3.67 &  W2cur \\
 & & 0.110 & 35.00 $\pm$ 0.12 & 100.00 $\pm$  5.38 &  72.28 $\pm$  5.14 &  W1cc \\
 & & 0.110 & 35.01 $\pm$ 0.12 & 100.37 $\pm$  5.53 &  72.01 $\pm$  5.22 &  W2cc \\
 & & 0.110 & 35.01 $\pm$ 0.12 & 100.46 $\pm$  5.45 &  71.95 $\pm$  5.15 &  I-band \\
\hline
\multirow{5}{*}{A1367} & \multirow{5}{*}{6969. $\pm$  93.} & 0.120 & 35.02 $\pm$ 0.06 & 101.06 $\pm$  2.93 &  68.96 $\pm$  3.01 &  W1cur \\
 & & 0.120 & 35.02 $\pm$ 0.06 & 100.93 $\pm$  2.90 &  69.05 $\pm$  2.99 &  W2cur \\
 & & 0.080 & 34.96 $\pm$ 0.11 &  98.13 $\pm$  4.93 &  71.02 $\pm$  4.76 &  W1cc \\
 & & 0.080 & 34.95 $\pm$ 0.11 &  97.54 $\pm$  5.00 &  71.44 $\pm$  4.86 &  W2cc \\
 & & 0.080 & 34.86 $\pm$ 0.11 &  93.89 $\pm$  4.63 &  74.23 $\pm$  4.89 &  I-band \\
\hline
\multirow{5}{*}{Coma} & \multirow{5}{*}{7370. $\pm$  76.} & 0.060 & 34.91 $\pm$ 0.06 &  95.81 $\pm$  2.72 &  76.92 $\pm$  3.07 &  W1cur \\
 & & 0.060 & 34.94 $\pm$ 0.06 &  97.41 $\pm$  2.74 &  75.66 $\pm$  2.99 &  W2cur \\
 & & 0.040 & 34.86 $\pm$ 0.11 &  93.93 $\pm$  4.53 &  78.46 $\pm$  4.82 &  W1cc \\
 & & 0.040 & 34.87 $\pm$ 0.11 &  94.06 $\pm$  4.62 &  78.36 $\pm$  4.90 &  W2cc \\
 & & 0.040 & 34.77 $\pm$ 0.10 &  89.91 $\pm$  4.22 &  81.97 $\pm$  4.92 &  I-band \\
\hline
\multirow{5}{*}{A2634/66} & \multirow{5}{*}{8938. $\pm$ 164.} & 0.105 & 35.32 $\pm$ 0.06 & 115.72 $\pm$  3.36 &  77.24 $\pm$  3.77 &  W1cur \\
 & & 0.105 & 35.32 $\pm$ 0.06 & 115.66 $\pm$  3.32 &  77.28 $\pm$  3.74 &  W2cur \\
 & & 0.070 & 35.34 $\pm$ 0.12 & 116.90 $\pm$  6.12 &  76.46 $\pm$  5.71 &  W1cc \\
 & & 0.070 & 35.33 $\pm$ 0.12 & 116.63 $\pm$  6.10 &  76.64 $\pm$  5.71 &  W2cc \\
 & & 0.070 & 35.35 $\pm$ 0.12 & 117.44 $\pm$  6.16 &  76.11 $\pm$  5.69 &  I-band \\
\enddata
\tablenotetext{1}{\ Cluster name}
\tablenotetext{2}{\ Mean cluster cosmology-corrected velocity in CMB frame (\kms)}
\tablenotetext{3}{\ Bias, $b$ (mag)}
\tablenotetext{4}{\ Bias-corrected distance modulus (mag)}
\tablenotetext{5}{\ Cluster distance (Mpc)}
\tablenotetext{6}{\ Hubble parameter \kms\ Mpc$^{-1}$}
\tablenotetext{7}{\ Source photometry}
\end{deluxetable*}

\subsection{$H_0$ From Clusters\label{SEC:H0_CLUSTERS}}

\begin{figure*}
	\includegraphics[width=0.5\linewidth]{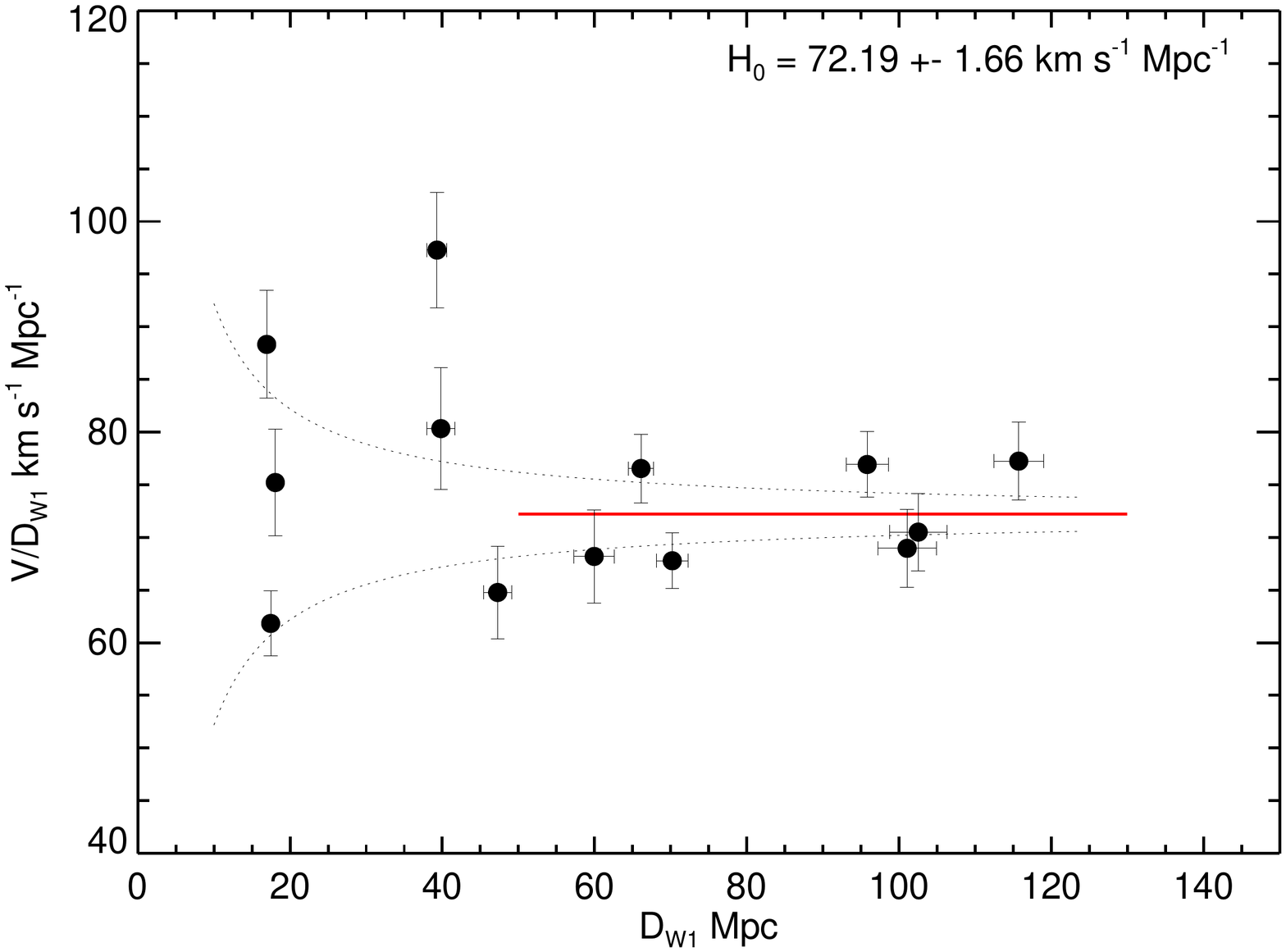}
	\includegraphics[width=0.5\linewidth]{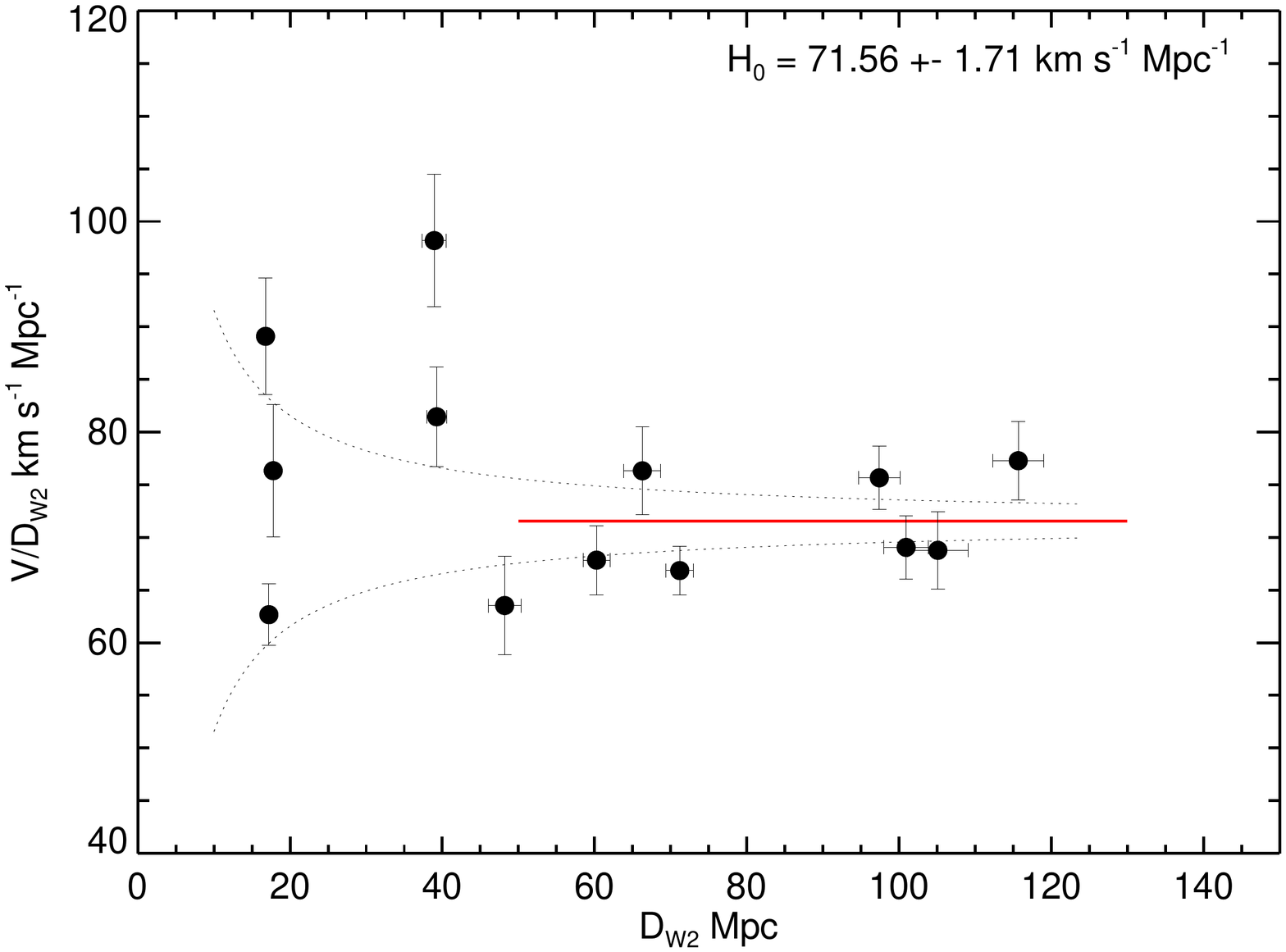}
	\includegraphics[width=0.5\linewidth]{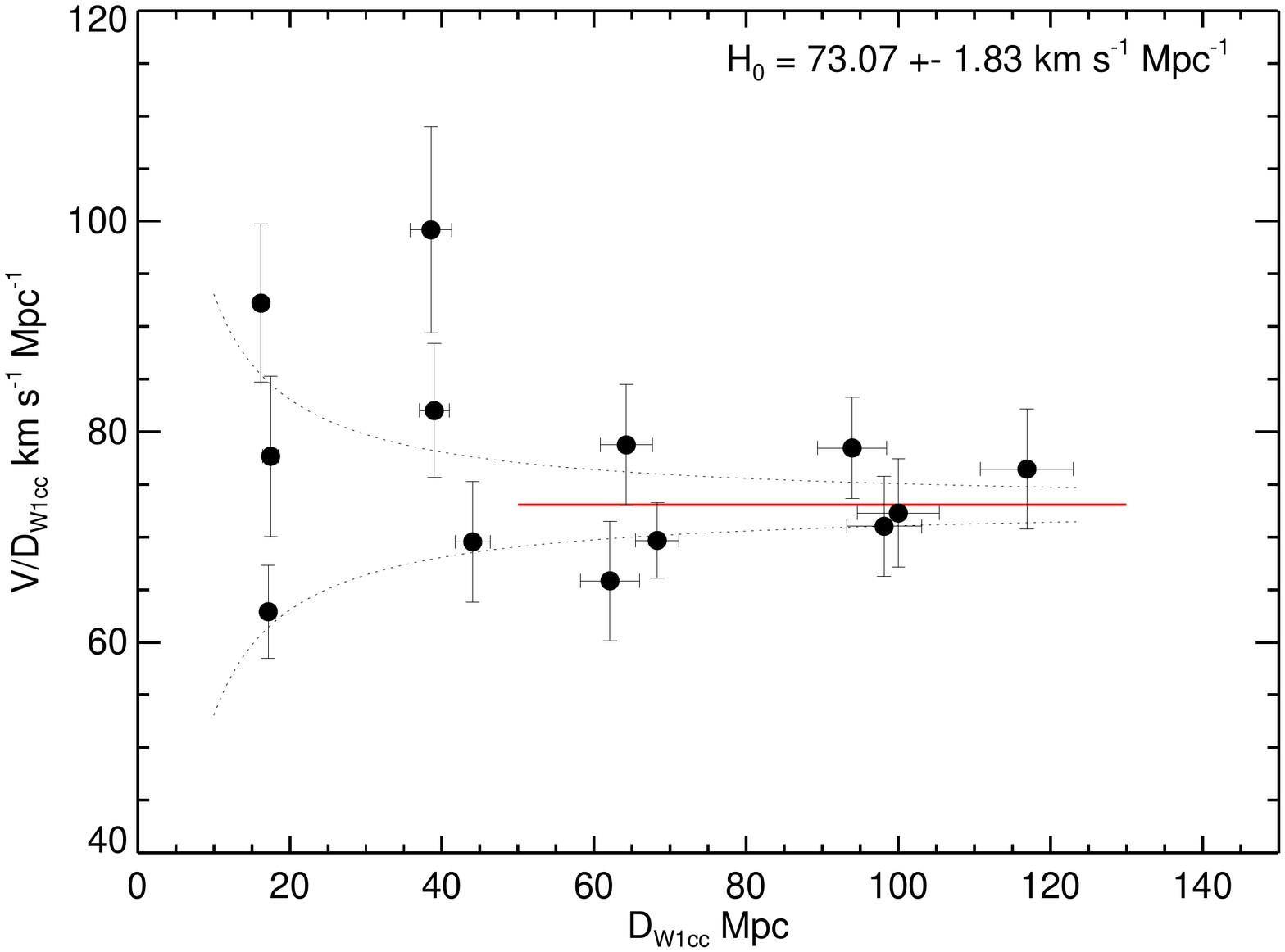}
	\includegraphics[width=0.5\linewidth]{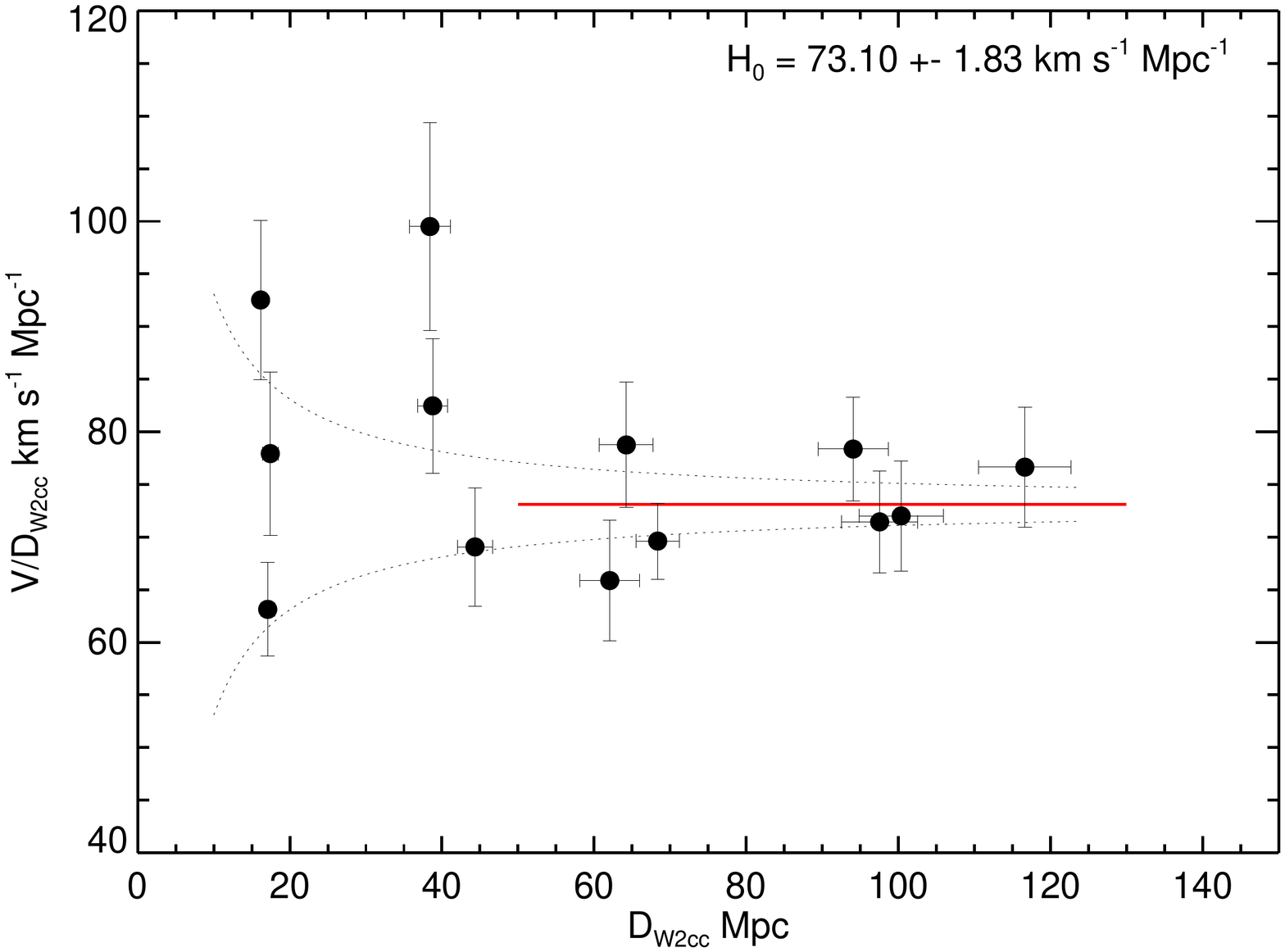}
	\caption{Hubble parameter as a function of distance for the {\it
	WISE} W1 curved TFR (top left), W2 curved TFR (top right), W1
	color-corrected TFR (W1cc, bottom left), and W2 color-corrected TFR
	(W2cc, bottom right).}
	\label{fig_h0_wise}
\end{figure*}

\begin{figure}
	\includegraphics[width=\linewidth]{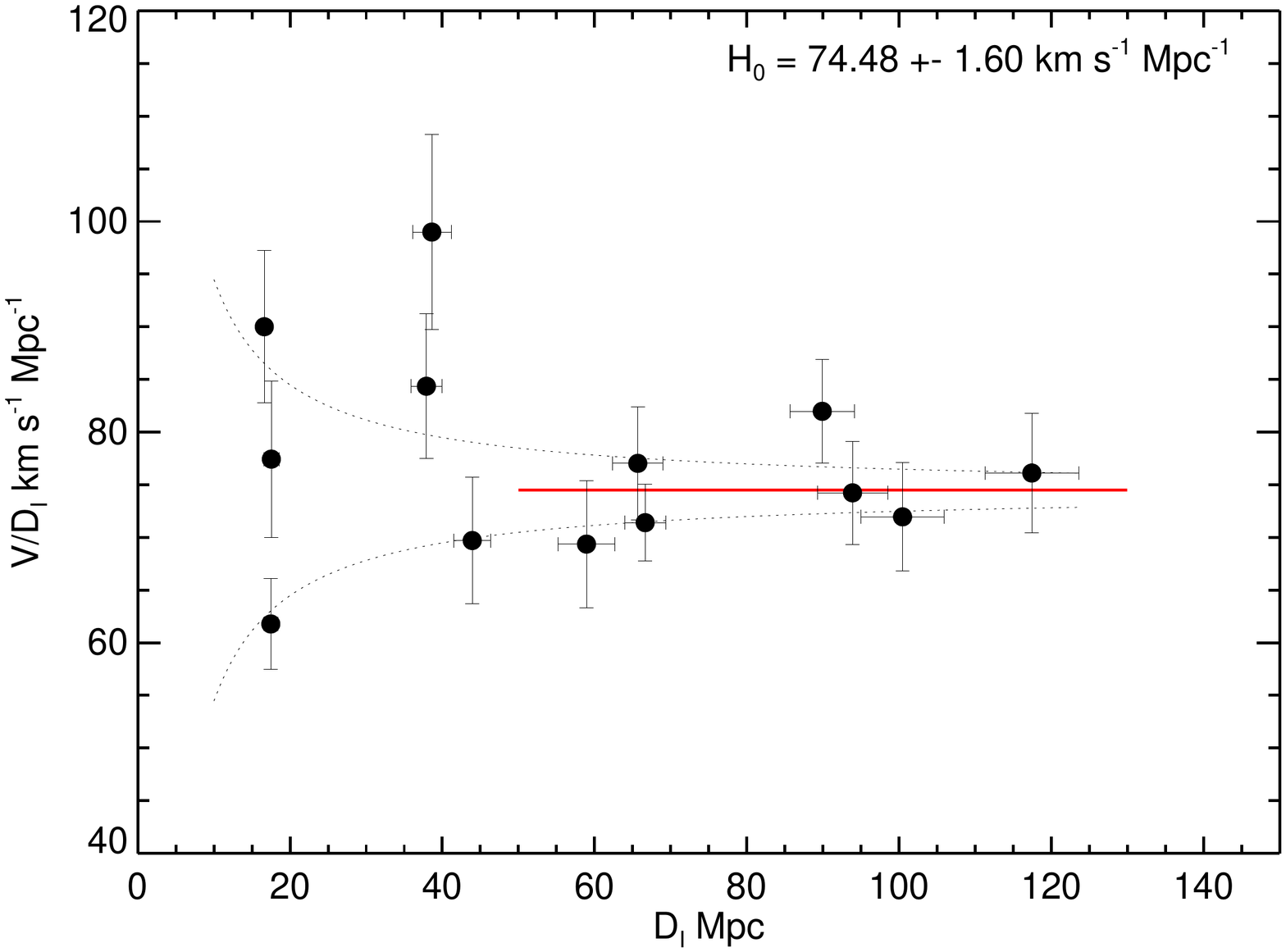}
	\caption{Hubble parameter as a function of distance for the I-band.}
	\label{fig_h0_i}
\end{figure}

As we have already pointed out, we expect a systematic problem with using the
linear, uncorrected {\it WISE} TFR (see \S\S\ref{SEC:ICALIBRATION} and
\ref{SEC:CURVE}) to calculate $H_0$.  We present these values to illustrate
this systematic, but we concentrate on the curved {\it WISE} TFR or the
color-corrected {\it WISE} TFR for calculating distances used to derive
$H_0$.  With distance moduli and hence distances for each cluster derived
from the ensemble of galaxies used to calibrate the cluster (see
Table~\ref{tab_h0_data}), we can use the ensemble velocity to calculate a
Hubble constant, $H_0$, for each cluster.  We use the bi-weight method
described in \citet{Beers:90:32} to derive a robust ensemble velocity for
each cluster.  These velocities are then shifted to the cosmic microwave
background (CMB) frame and adjusted based on a cosmological model which
assumes $\Omega_{m} = 0.27$ and $\Omega_{\Lambda} = 0.73$.  The
cosmological adjustments are admittedly small, but not insignificant.
These velocities and associated errors are listed in
Table~\ref{tab_h0_data} in column two labeled $V_{mod}$
\citep[see][equation 14]{Tully:13:86} to indicate the adjustment for the
cosmological model specified previously.  We calculate $H_0 =
V_{mod}/D_{Mpc}$ for each cluster as shown in column seven of the
aforementioned table and plotted for {\it WISE} and the I-band in
Figures~\ref{fig_h0_wise} and \ref{fig_h0_i}.

Examining Figures~\ref{fig_h0_wise} and \ref{fig_h0_i}, we see that the
nearer clusters have a large scatter while those beyond 50 Mpc ($V_{mod} >
4000$ \kms) have a smaller scatter.  This is simply the result of the
peculiar motions induced by local structures in our supercluster complex
(the transformation of velocities to the cosmic microwave background frame
gives all nearby galaxies large peculiar velocities).  We plot an error
envelope of 200 \kms\ as a dotted line in each figure to show the effect of
peculiar velocities on $H_0$ as a function of distance.  In order to derive
an estimate of the universal Hubble constant, we consider only clusters
beyond 50 Mpc and we average the log of their resulting $H_0$ values since
the errors are predominantly in the distance and symmetric about the
distance modulus.  We find an error-weighted, logarithmic averaged Hubble
constant of $H_0 = 73.1 \pm 1.8$ \kms\ Mpc$^{-1}$ for both W1cc and W2cc.
For the curved pure {\it WISE} TFR, we get $H_0 = 72.2 \pm 1.7$ \kms\ 
Mpc$^{-1}$ for W1 and $H_0 = 71.6 \pm 1.7$ \kms\ Mpc$^{-1}$ for W2.  For
the I-band we get a larger value of $H_0 = 74.5 \pm 1.6$ \kms\ Mpc$^{-1}$.
This amounts to a range of $\sim\pm$1.5\% from a logarithmic average of
72.9 \kms\ Mpc$^{-1}$ derived from all five of these cluster TFR $H_0$
values.

Using the linear, pure {\it WISE} TFR, we derive values of $H_0 = 70.6 \pm
1.6$ \kms\ Mpc$^{-1}$ for W1 and $H_0 = 69.8 \pm 1.6$ \kms\ Mpc$^{-1}$ for W2.
These values are low compared with either the curved or the color-corrected
TFR values, as expected from a systematic that biases distant clusters toward
larger distances.

All the errorbars listed above are formal statistical errorbars and do not
account for systematics.  The calibration clusters may still be strongly
influenced by local large-scale structures and thus may not provide the
most robust estimate of $H_0$.  In addition, there are only seven clusters
beyond 50 Mpc and small number statistics may play a role.  Systematic
errors are discussed in detail in the next subsection when we extend our
reach well beyond local structures and use a larger sample of supernova
host galaxies to estimate $H_0$.

\subsection{$H_0$ From Supernovae\label{SEC:H0_SUPERNOVAE}}

The precision of distances derived from Type Ia supernovae (SNe Ia) offers
a better avenue for determining $H_0$ free from the small number statistics
that influence the determination of $H_0$ from seven nearby galaxy
clusters.  In order to exploit the reach of SNe Ia, which is well beyond
the local velocity perturbations we see in our cluster $H_0$ estimations,
we must tie the SN Ia distance scale to the distance scale established by
the TFR.  Even though there are few SN Ia that have been detected in nearby
galaxies, there are 56 SNe Ia that have been detected in host galaxies
within the Cosmic Flows sample (see \S\ref{SEC:HILINEWIDTHS}).  These
galaxies also have I-band photometry allowing the color-corrected MIR TFR
to be used in addition to the curved pure {\it WISE} TFR.  This permits an
accurate determination of the offset between the SN Ia and TFR distance
moduli.  

We use the UNION2 sample of SNe Ia \citep{Amanullah:10:712} for this
comparison.  This sample encompasses distances out to beyond $z \sim 1$,
and includes all the SNe Ia hosts from the Cosmic Flows galaxy sample.  We
can improve our statistical error in the situation where there are multiple
SNe Ia within a cluster.  Of the 13 clusters used to calibrate the TFR,
eight have had one or more SN Ia erupt within one or more member galaxies
\citep[see Table~2 of][]{Sorce:12:L12}.  We use the same bi-weight method
from \citet{Beers:90:32} to derive robust averages for the group velocities
and distance moduli based on the SNe Ia and based on the TFR.  We also use
SN Ia hosts not in clusters.  These individual hosts will have lower weight
by virtue of their higher statistical error, however the ensemble will help
to constrain the offset.

\begin{figure*}
	\includegraphics[width=0.5\linewidth]{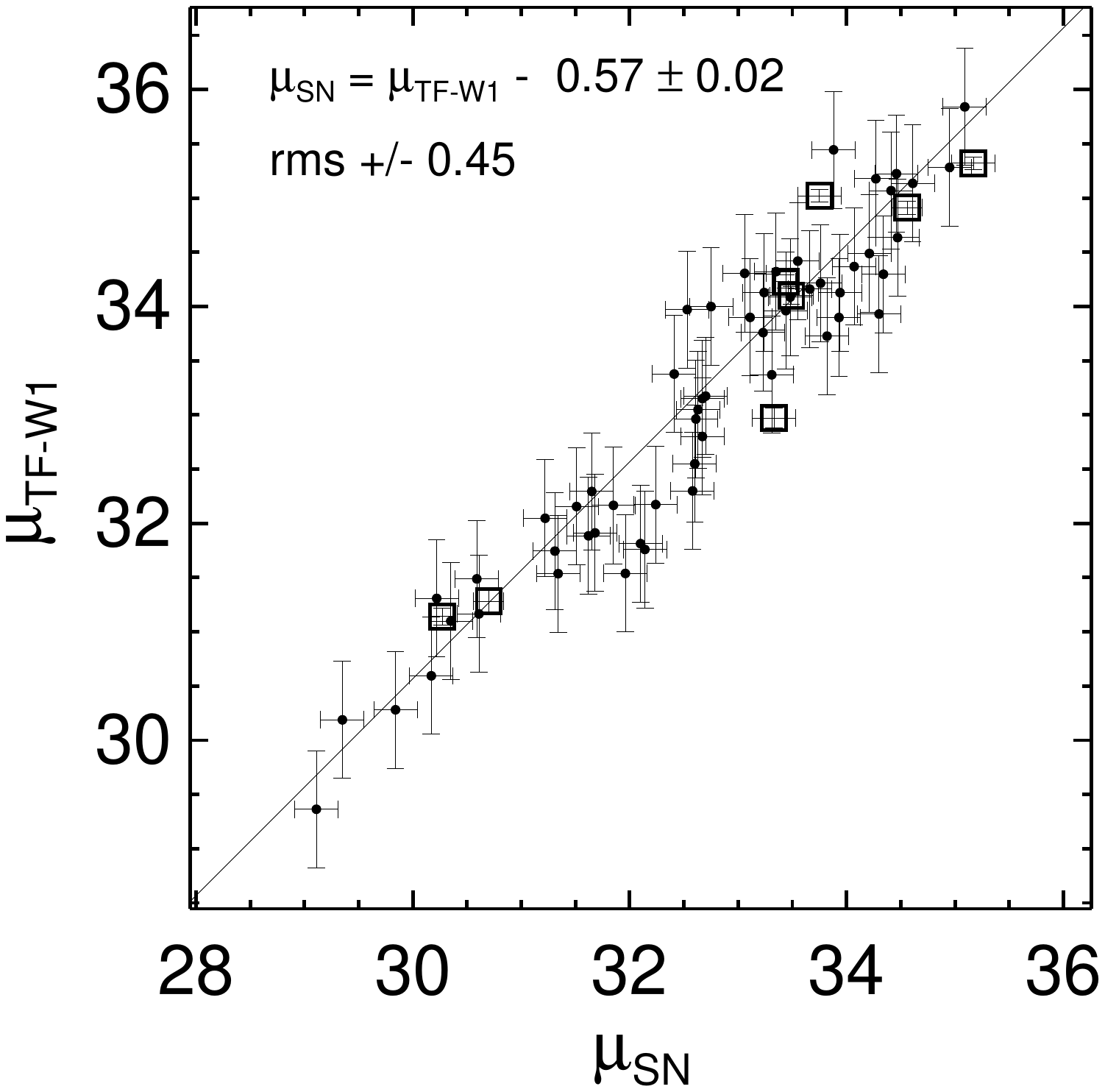}
	\includegraphics[width=0.5\linewidth]{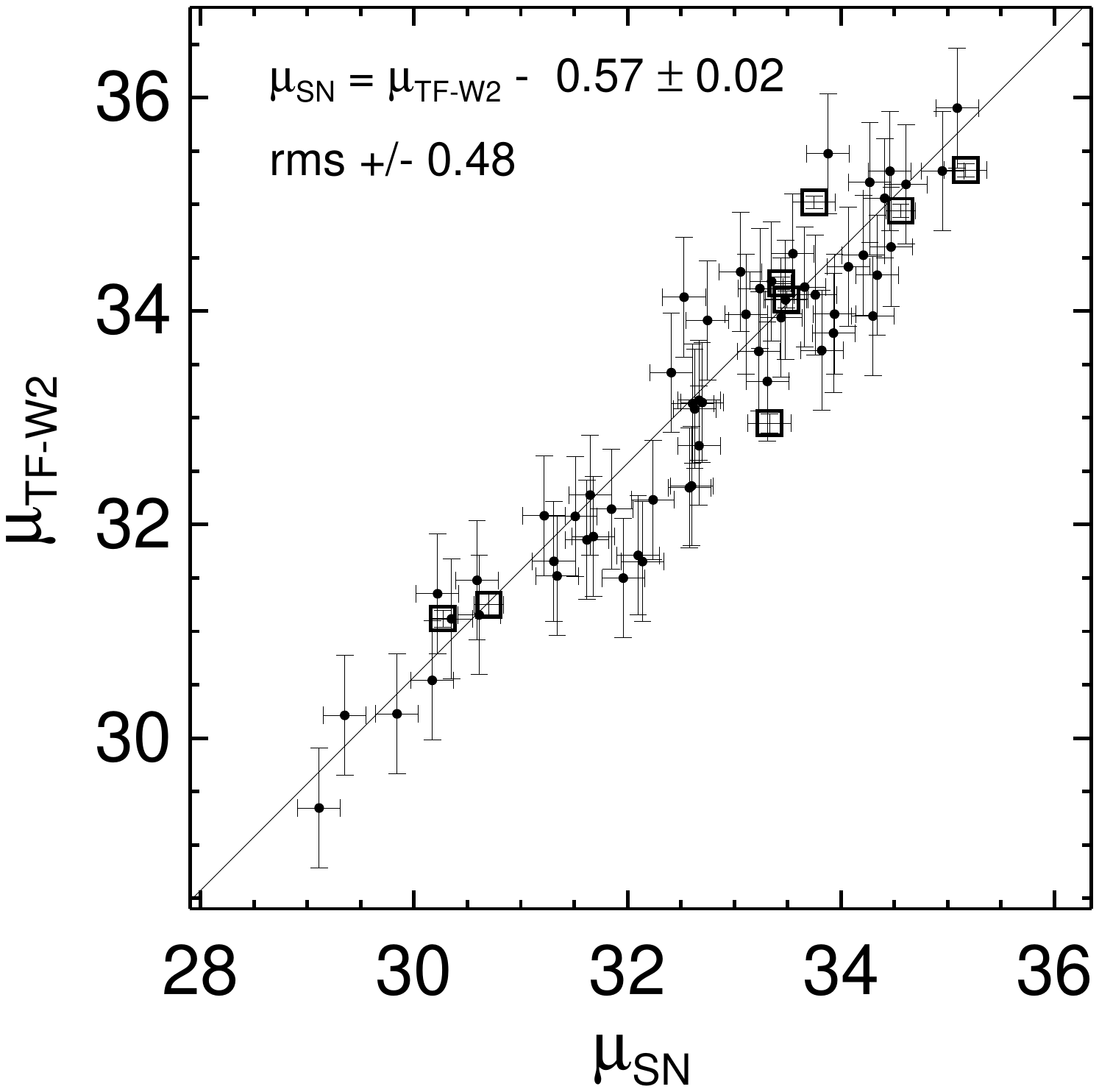}
	\includegraphics[width=0.5\linewidth]{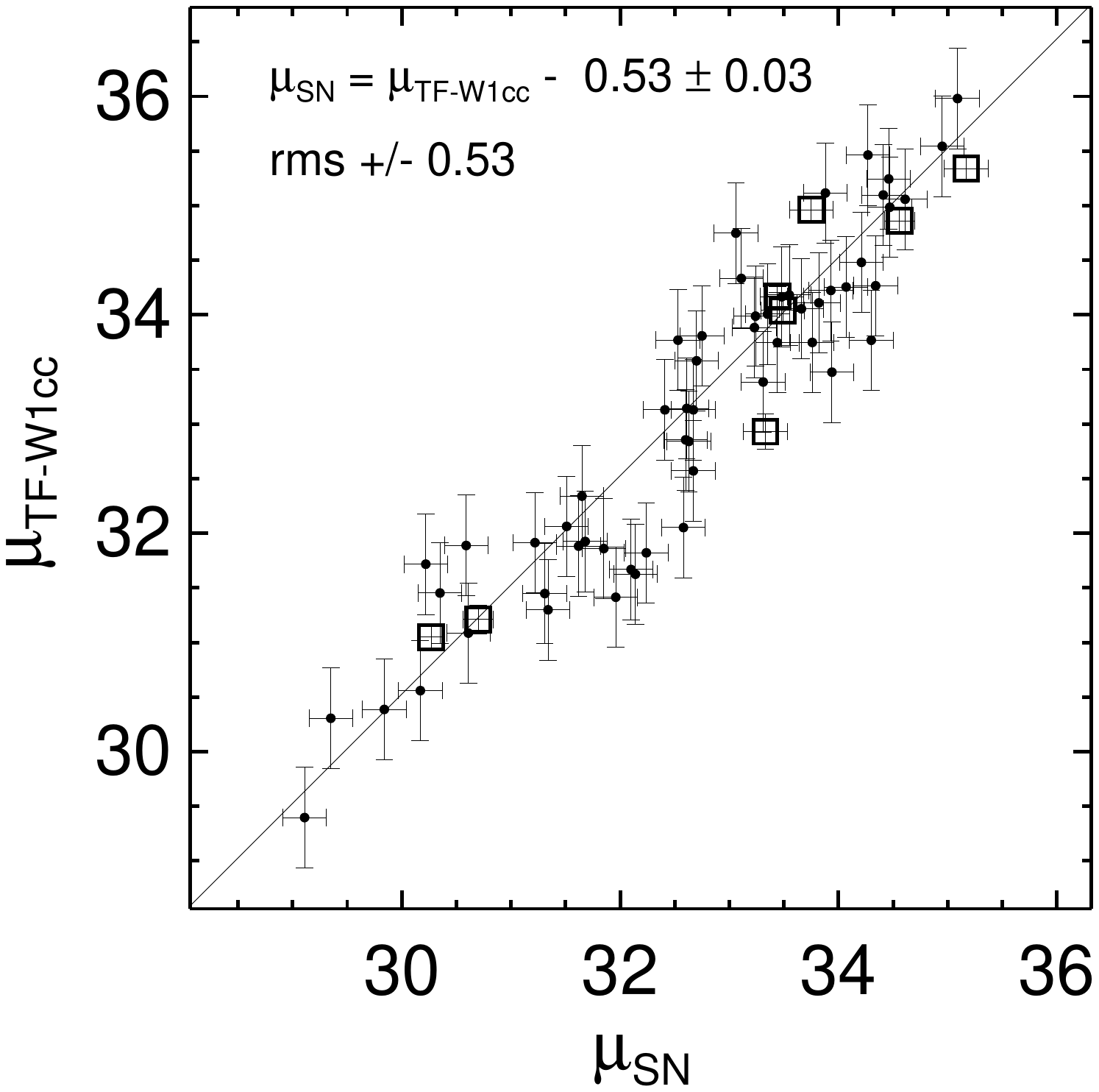}
	\includegraphics[width=0.5\linewidth]{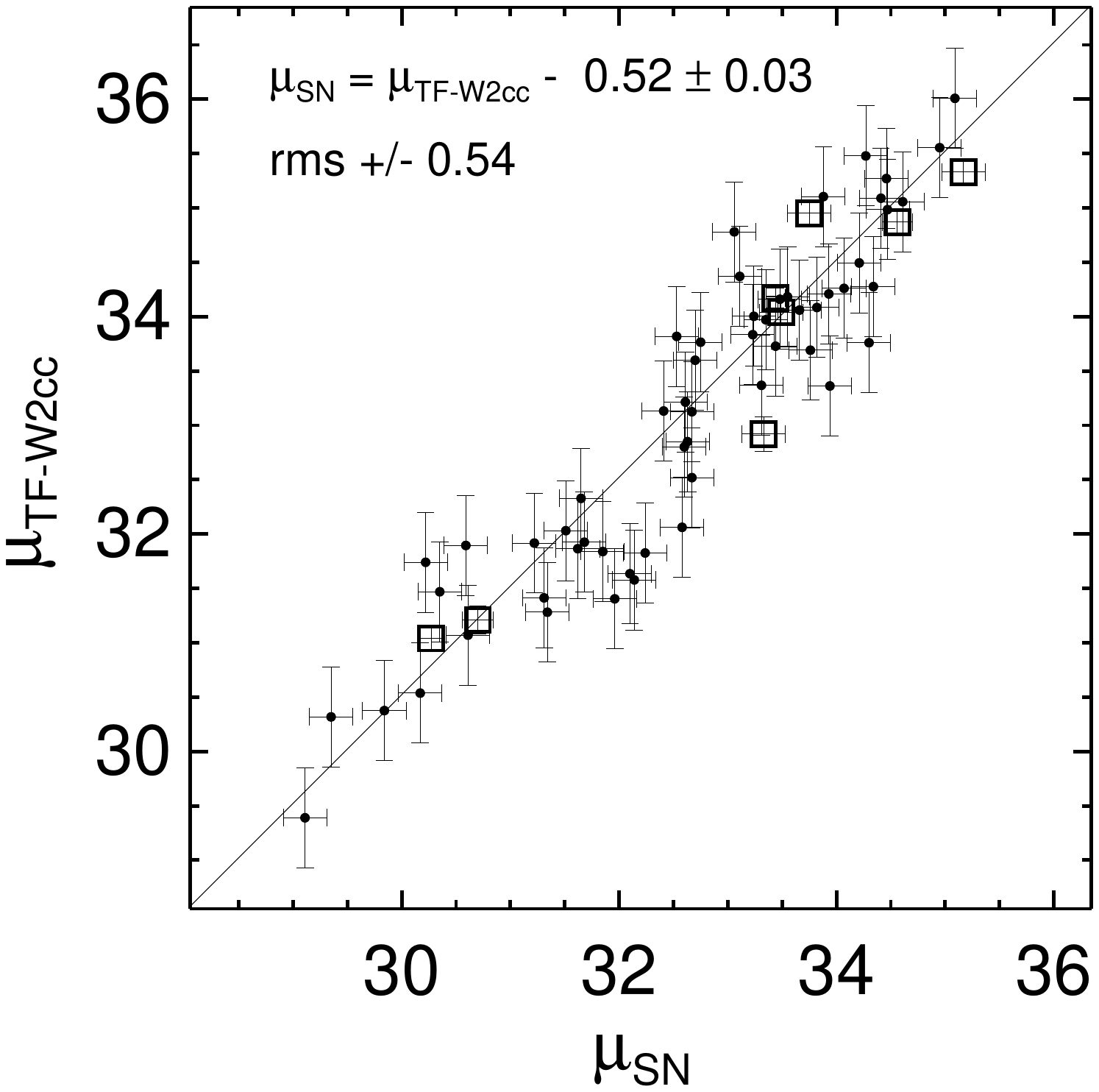}
	\caption{The distance modulus offsets derived from the SN Ia sample
	and from the TF relation for the {\it WISE} W1 curved TFR (top
	left), and W2 curved TFR (top right), W1 color-corrected TFR
	(bottom left) and W2 color-corrected TFR (bottom right).  Open
	squares indicate the ensemble robust averages for the eight
	clusters that have had one or more SN Ia erupt within member
	galaxies.  Solid circles indicate individual hosts within which a
	SN Ia from the UNION2 sample has erupted.  Note the small scatters
	and low formal errors using the curved pure {\it WISE} TFRs.}
	\label{fig_mucomp_wise}
\end{figure*}

\begin{figure}
	\includegraphics[width=\linewidth]{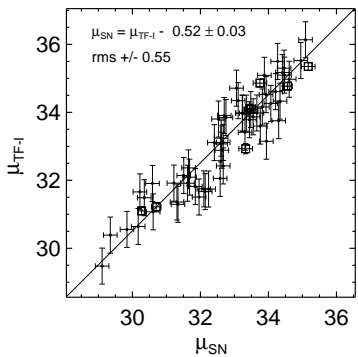}
	\caption{The distance modulus offset derived from the SN Ia sample
	and from the TF relation for the I-band. Symbols have the same
	meaning as in Figure~\ref{fig_mucomp_wise}.}
	\label{fig_mucomp_i}
\end{figure}

Figure~\ref{fig_mucomp_wise} presents the comparison of the TFR distance
moduli derived from the {\it WISE} passbands and the SN Ia distance moduli
for the eight clusters and 56 individual galaxies, while
Figure~\ref{fig_mucomp_i} shows the same comparison for the I-band.  The
distance modulus offsets are derived from error-weighted fits with the
slope fixed at a value of 1.  The clusters have the largest influence on
these offsets due to their low statistical errors, yet the resultant fits
appear to bisect the distributions for the individual galaxies as well.
The resulting distance modulus offsets are identical for the curved TFR for
W1 and W2: $0.57 \pm 0.02$ mag.  The rms values are calculated only from
the individual galaxy residuals and are 0.45 mag for W1 and 0.48 mag for
W2.  For the color-corrected TFR, we find an offset of $0.53 \pm 0.03$ mag
for W1cc and $0.52 \pm 0.03$ mag for W2cc and rms values of 0.53 mag for
W1cc and 0.54 mag for W2cc.  For the I-band the offset is $0.51 \pm 0.03$
mag with a scatter of 0.55 mag which is very close to the values shown in
the top panel of Figure~2 from \citet{Courtois:12:174}, which is also
derived only from the SN Ia - TFR offset.

As was pointed out by \citet{Courtois:12:174}, the scatters in the linear
TFR offset data are $\sim 10$\% larger than expected from the combination
of the individual scatters in the linear TFRs (0.46 mag for the
color-corrected W1,2 and the I-band) and the SN Ia (0.20 mag) distance
scales.  We don't present the offset data for the pure {\it WISE} linear
TFR, but these scatters are even larger at 0.56 mag for W1 and 0.58 mag for
W2.  For the curved pure {\it WISE} TFRs, the SN Ia - TFR distance modulus
scatter is actually less than expected when adding the TFR scatter (0.52
mag for curved W1 and 0.55 mag for curved W2) in quadrature with the SN Ia
distance modulus scatter. In fact, the scatter in the curved W1 is 20\%
less than the scatter in the I-band.  The scatter in all of these offsets
have been calculated with the exact same sample and using the exact same
method.  With a sample of 56 galaxies, it is hard to explain this away with
small number statistics.  It is possible that this results from a better
alignment between the clusters and the individual galaxies using the curved
TFRs, although this is not obvious from Figure~\ref{fig_mucomp_wise}.  As
a test of the TFR, this lower scatter is strong evidence in favor of using
the curved pure {\it WISE} TFR for deriving distances.

The formal errors on the curved TFR W1 and W2 distance modulus offsets
correspond to an error in $H_0$ of 0.7 \kms\ Mpc$^{-1}$, while the formal
errors on the color-corrected and I-band offsets correspond to an error in
$H_0$ of 1.1 \kms\ Mpc$^{-1}$.  Using the curved TFR W1 and W2 distance
modulus offsets represents a 30\% reduction in the $H_0$ error budget.

\begin{figure}
	\includegraphics[width=\linewidth]{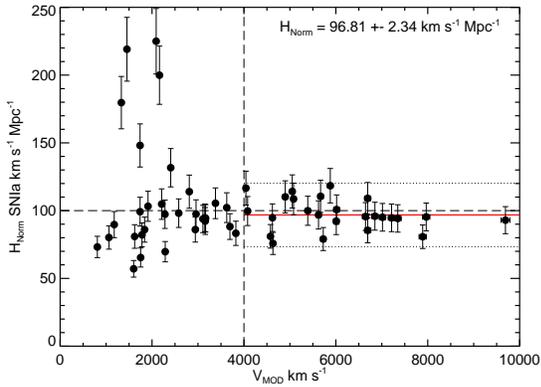}
	\caption{Hubble parameter normalization as a function of recession
	velocity using only the UNION2 SN Ia distance moduli
	\citep{Amanullah:10:712} for galaxies in common with the Cosmic
	Flows sample.  Here we are only verifying the normalization of this
	sample at 100 \kms\ Mpc$^{-1}$ for our sub-sample of SN Ia host
	galaxies.  This plot shows that, for our sub-sample, the
	normalization is less than the nominal value and thus we need to
	apply an offset of $-3.2 \pm 2.3$ \kms\ Mpc$^{-1}$ when we
	re-normalize the SN Ia hosts with the TFR distances.}
	\label{fig_h0sne}
\end{figure}

Figure~\ref{fig_h0sne} shows the calculation of the normalization of the
Hubble constant, $H_{Norm}$, using only the SNe Ia from the UNION2 sample
that overlap with the Cosmic Flows sample.  The overlap zero point is not,
in fact, $H_{Norm} = 100$ \kms\ Mpc$^{-1}$, but slightly less: $H_{Norm} =
96.8 \pm 2.3$ \kms\ Mpc$^{-1}$, an offset of 3.2 \kms\ Mpc$^{-1}$.  A
similar offset was found by \citet{Courtois:12:174} when setting the
distance zero-point using the I-band TFR.  Once we apply this normalization
offset, we derive values of $H_0 = 73.7 \pm 2.4$ \kms\ Mpc$^{-1}$ for W1
and W2, and $H_0 = 75.2 \pm 2.5$ \kms\ Mpc$^{-1}$ for W1cc and $H_0 = 75.5
\pm 2.5$ \kms\ Mpc$^{-1}$ for W2cc, and $H_0 = 75.9 \pm 2.5$ \kms\
Mpc$^{-1}$ for the I-band.  These errors are the combination in quadrature
of the SN Ia -- TFR offset $H_0$ error (stated in the previous paragraph)
and the UNION2 normalization offset uncertainty shown in
Figure~\ref{fig_h0sne}.  Since the two curved pure {\it WISE} values and
the I-band value are independent from one another, we are allowed to
perform a log average of these three values.  This gives our best estimate
of the Hubble constant of $H_0 = 74.4 \pm 1.4$ \kms\ Mpc$^{-1}$.  Here the
error is the statistical error in the mean value.

For completeness we present the values derived with the linear pure {\it WISE}
TFR.  Using the same SN Ia hosts, we derive $H_0 = 73.0 \pm 2.7$ \kms\ 
Mpc$^{-1}$ for W1 and $H_0 = 72.7 \pm 2.7$ \kms\ Mpc$^{-1}$ for W2.  These
values are low as expected from the systematic bias caused by using a
linear fit on a curved TFR.

Systematic errors need to be accounted for in the $H_0$ calculations.  By
using IR photometry we reduce the uncertainties due to dust significantly.
Since we have used three bandpasses for these calculations, we can use the
differences to estimate the systematic errors between bands.  Based on the
range of $H_0$ we derive for all three bandpasses, we estimate the
inter-band systematic to be $\pm1.1$ \kms\ Mpc$^{-1}$.  The other source of
systematic error is the error on the distance to the LMC, which forms the
basis for our TFR distance scale and has a systematic of $\pm0.033$ mag
\citep{Freedman:12:24}, which also corresponds to an error on $H_0$ of
$\pm1.1$ \kms\ Mpc$^{-1}$.  Another error is the formal TFR zero-point
error (see Table~\ref{tab_tfr_params}, column eight) which is 0.05 mag,
which corresponds to an error on $H_0$ of $\pm1.8$ \kms\ Mpc$^{-1}$.
Adding these in quadrature gives a systematic error on $H_0$ of $\pm2.4$
\kms\ Mpc$^{-1}$.  Thus our best value is $H_0 = 74.4 \pm1.4$(stat)
$\pm\ 2.4$(sys) \kms\ Mpc$^{-1}$.  We can add the statistical and
systematic errors in quadrature to give a total error of $\pm2.8$ \kms\ 
Mpc$^{-1}$, which amounts to a percentage error of $\sim4$\%.

\subsection{Comparison with Previous $H_0$ Results\label{SEC:H0_COMPARE}}

\begin{deluxetable}{llcc}
	\tablewidth{0in}
	\tabletypesize{\scriptsize}
	\tablecaption{Hubble Constant Comparison\label{tab_h0_compare}}
	\tablehead{
	\colhead{Reference} &
	\colhead{TFR band\tablenotemark{1}} &
	\colhead{Clusters\tablenotemark{2,3}} &
	\colhead{SNe Ia\tablenotemark{3,4}}
	}
\startdata
\citet{Tully:12:78}     & I-band   & $75.1 \pm 1.0$ & \ldots \\
\citet{Courtois:12:174} & I-band   & \ldots         & $75.9 \pm 3.8$ \\
This work               & I-band   & $74.5 \pm 1.6$ & $75.9 \pm 2.5$ \\
\citet{Sorce:13:94}     & [3.6]cc & $74 \pm 4$     & \ldots \\
\citet{Sorce:12:L12}    & [3.6]cc & \ldots         & $75.2 \pm 3.0$ \\
This work               & W1lin    & $71 \pm 2$ & $73.0 \pm 2.7$ \\
This work               & W2lin    & $70 \pm 2$ & $72.7 \pm 2.7$ \\
This work               & W1cc     & $73 \pm 2$ & $75.1 \pm 2.5$ \\
This work               & W2cc     & $73 \pm 2$ & $75.1 \pm 2.5$ \\
This work               & W1cur  & $72 \pm 2$ & $73.7 \pm 2.4$ \\
This work               & W2cur  & $72 \pm 2$ & $73.7 \pm 2.4$ \\
\hline
This work               & $<$W12cur,I$>$ & $73 \pm 1$ & $74.4 \pm 2.8$\tablenotemark{5} \\
\enddata
\tablenotetext{1}{\ `lin' indicates linear TFR, `cc' indicates optical - MIR color-corrected photometry, `cur' indicates curved TFR}
\tablenotetext{2}{\ seven clusters with $V_{mod} > 4000$ \kms}
\tablenotetext{3}{\ \kms\ Mpc$^{-1}$}
\tablenotetext{4}{\ offsets applied to UNION2 SN Ia sample \citep{Amanullah:10:712}}
\tablenotetext{5}{\ includes statistical and systematic errors}
\end{deluxetable}

Our cluster $H_0$ values compare well with previous determinations for
clusters calibrated with the TFR as shown in column three of
Table~\ref{tab_h0_compare}.  Here we are listing statistical errors only.
For the I-band, \citet{Tully:12:78} find $H_0
= 75.1 \pm 1.0$ \kms\ Mpc$^{-1}$, which agrees to within 1\% of our value
of $74.5 \pm 1.6$ \kms\ Mpc$^{-1}$.  \citet{Sorce:13:94} used the
color-corrected IRAC [3.6] band ([3.6]cc) to derive $H_0 = 73.8 \pm 1.1$
\kms\ Mpc$^{-1}$, which is well within the statistical errorbars (and also
within 1\%) of our color-corrected W1 (W1cc) value of $73.1 \pm 1.8$ \kms\
Mpc$^{-1}$.  Our pure {\it WISE} curved TFR cluster $H_0$ values are low
although still in statistical agreement with the other values.

Comparing the $H_0$ values derived by bringing the UNION2 sample onto the
TFR distance scale also shows good consistency, as can be seen in column
four of Table~\ref{tab_h0_compare}.  The I band value from
\citet{Courtois:12:174} of $H_0 = 75.9 \pm 3.8$ \kms\ Mpc$^{-1}$ is
identical to our value, although their errorbar includes systematic errors
and so appears larger than ours.
The color-corrected IRAC [3.6] value presented in
\citet{Sorce:12:L12} of $H_0 = 75.2 \pm 3.0$ \kms\ Mpc$^{-1}$ is less than
two-tenths of a percent different from our value of $75.1 \pm 2.5$ \kms\
Mpc$^{-1}$ derived from color-corrected W1.

What is new is using the uncorrected curved TFR to derive values of $H_0$.
While these values are low for the seven clusters used to calibrate the
TFR, when used to re-normalize the UNION2 SN Ia sample, the values agree
well with current best estimates of $H_0$ (see below).  The other
advantage of using the uncorrected curved TFR for {\it WISE} is that it is
truly independent of the I-band, unlike the color-corrected TFR values,
and thus we can average all three bands (in the logarithm) to derive a
more robust value of $H_0$.  This is presented in the last column of the
last row of Table~\ref{tab_h0_compare}, and the error includes both
statistical and systematic errors.

It is interesting to note that \citet{Sakai:00:698} give a value of
$H_0 = 71 \pm 4$ (stat) $\pm 7$ (sys) \kms\ Mpc$^{-1}$ using a
weighted average of their four band ($BVIH_{-0.5}$) TFRs fit with 
linear relations.  At the end of their \S5.2.1, they give a value of
$H_0 = 73 \pm 2$ (stat) \kms\ Mpc$^{-1}$ for a curved I-band TFR.
This is higher than the $H_0$ derived from linear TFRs and closer to
current estimates, including our own.

A current independent estimate for $H_0$ that is useful for comparison is
that presented in \citet{Freedman:12:24}: $H_0 = 74.3 \pm 1.5$(stat) $\pm\
2.1$(sys) \kms Mpc$^{-1}$ which has a percentage systematic error of 2.8\%.
All of the SN Ia derived $H_0$ values we present here agree with this value
within the errors.  Another value to compare with is that derived from the
{\it PLANCK} mission and presented in \citet{Collaboration:13:5076}: $H_0 =
67.3 \pm 1.2$ \kms\ Mpc$^{-1}$.  Our lowest value of $H_0$ is the one
derived from seven clusters in the uncorrected curved W2 band ($71.6 \pm
1.7$ \kms\ Mpc$^{-1}$).  This value is 3.6$\sigma$ high using their
errorbar and 2.5$\sigma$ high using our errorbar. Our best result of $H_0 =
74.4 \pm 2.8$(stat and sys) \kms\ Mpc$^{-1}$ is 5.9$\sigma$ high using
their errorbar and 2.5$\sigma$ high using our error estimate.  Our data do
not favor such a low value of $H_0$.  We can also compare with another CMD
$H_0$ value from \citet{Hinshaw:13:19} who quote $H_0 = 69.32 \pm 0.80$
(stat) \kms\ Mpc$^{-1}$ in their Table~4.  This value is closer to our value
but a tension still exists.  Relativistic corrections for foreground lensing
in the CMB analyses may resolve this tension \citep{Clarkson:14:7860}.

\section{Conclusions\label{SEC:CONCLUSIONS}}

We have derived a calibration of the absolute magnitude-linewidth relation
for the {\it WISE} W1 and W2 filters.  The raw, linear calibration, using
only {\it WISE} photometry that is aperture corrected, k-corrected, and has
been corrected for internal and external extinction gives:
\begin{subequations}
	\begin{align}
	\begin{split}
		M^{b,i,k,a}_{W1} = &- (20.35 \pm 0.07) \\
			&- (9.56 \pm 0.12)(\log W^i_{mx} - 2.5), 
	\end{split} \\
	\begin{split}
		M^{b,i,k,a}_{W2} = &- (19.76 \pm 0.08) \\
			&- (9.74 \pm 0.12)(\log W^i_{mx} - 2.5).
	\end{split}
	\end{align}
\end{subequations}
These calibrations show a scatter of 0.54 magnitudes in W1 and 0.56
magnitudes in W2.

The I-band sample grew by 24 galaxies (9\%) compared to the previous
calibration and so we updated it to:
\begin{equation}
	\begin{split}
	M^{b,i,k,e}_{I} = &- (21.34 \pm 0.07) \\
		&- (8.95 \pm 0.14)(\log W^i_{mx} - 2.5).
	\end{split}
\end{equation}
This calibration has a scatter of 0.46 magnitudes.

We find evidence for curvature in the MIR TFR based on a comparison between
calibration cluster distances generated using linear TFRs in the I-band and
in the {\it WISE} W1 and W2 bands.  We use the ensemble of cluster galaxies
shifted to have an apparent distance of Virgo to fit this curved TFR and
find the following curved TFRs for W1 and W2:
\begin{subequations}
	\begin{align}
	\begin{split}
		\mathcal{M}^{b,i,k,a}_{W1} = &- (20.48 \pm 0.05) \\
			&- (8.36 \pm 0.11)(\log W^i_{mx} - 2.5) \\
			&+ (3.60 \pm 0.50)(\log W^i_{mx} - 2.5)^2, 
	\end{split} \\
	\begin{split}
		\mathcal{M}^{b,i,k,a}_{W2} = &- (19.91 \pm 0.05) \\
			&- (8.40 \pm 0.12)(\log W^i_{mx} - 2.5) \\
			&+ (4.32 \pm 0.51)(\log W^i_{mx} - 2.5)^2.
	\end{split}
	\end{align}
\end{subequations}
These calibrations have a scatter of 0.52 mag for W1 and 0.55 mag for W2,
an improvement over the pure linear TFRs.  The formal errors on the
zero-point calibration are the smallest of all the calibrations derived
here.

Following previous work on calibrating the TFR in the MIR
\citep{Sorce:13:94}, we apply an optical - MIR color correction to our raw
W1 and W2 magnitudes in order to reduce the scatter.  The corrections have
the form:
\begin{subequations}
	\begin{align}
	\Delta W1^{color} & = -0.470 - 0.561(I^{b,i,k}_{T} -
		W1^{b,i,k,a}_{T}), \\
	\Delta W2^{color} & = -0.874 - 0.617(I^{b,i,k}_{T} -
		W2^{b,i,k,a}_{T}).
	\end{align}
\end{subequations}
Where $I^{b,i,k}_T$ values are derived from I-band imaging.  These are then
used to adjust the input magnitudes as follows:
\begin{equation}
	C_{W1,2} = W1,2^{b,i,k,a}_{T} - \Delta W1,2^{color}.
\end{equation}
We used these pseudo-magnitudes to generate color-corrected linear 
calibrations of the form:
\begin{subequations}
	\begin{align}
	\begin{split}
		M_{C_{W1}} = &- (20.22 \pm 0.07) \\
			&- (9.12 \pm 0.12)(\log W^i_{mx} - 2.5),
	\end{split} \\
	\begin{split}
		M_{C_{W2}} = &- (19.63 \pm 0.07) \\
			&- (9.11 \pm 0.12)(\log W^i_{mx} - 2.5).
	\end{split}
	\end{align}
\end{subequations}
These both show a scatter of 0.46 magnitudes, identical to the I-band
scatter.  These equations represent the most accurate calibration of the
luminosity-linewidth relation available for {\it WISE} data at this time.

We investigate a residual bias in the TFRs resulting from a flat magnitude
cutoff that varies with distance and produces more of a bias as the cutoff
samples the sparser upper end of the luminosity function.  We determine two
bias functions.  One for the pure {\it WISE} TFRs both curved and linear,
and one for the I-band and the color-corrected {\it WISE} TFRs:
\begin{subequations}
	\begin{align}
		b_{pure} &= 0.006(\mu - 31)^{2.3} \\
		b_{cc} &= 0.004(\mu - 31)^{2.3},
	\end{align}
\end{subequations}
where $\mu$ represents the distance modulus of a field galaxy.

From the calibrations we generate bias-corrected distances to the
calibrating clusters and derive a Hubble constant from the clusters far
enough away to be in the Hubble flow (D $> 50$ Mpc).  We derive $H_0 = 72.2
\pm 1.7$ \kms\ Mpc$^{-1}$ for the curved pure W1 TFR and $H_0 = 71.6 \pm
1.7$ \kms\ Mpc$^{-1}$ for the curved pure W2 TFR.  The color-corrected W1
and W2 TFRs give the same value of $H_0 = 73.1 \pm 1.8$ \kms\ Mpc$^{-1}$,
and we get $H_0 = 74.5 \pm 1.6$ \kms\ Mpc$^{-1}$ using the I-band TFR.

To leverage the redshift reach of SNe Ia, we measure the zero-point offset of
the UNION2 SN Ia sample by comparing the distances in 56 SN Ia hosts galaxies
in common with the Cosmic Flows 2 sample.  The measured offsets give $H_0 =
73.7 \pm 2.4$ using the curved W1 and W2 TFRs and $H_0 = 75.9 \pm 2.5$ using
the I-band linear TFR.  Taking the log average of these values gives a Hubble
constant of $H_0 = 74.4 \pm 1.4$ \kms\ Mpc$^{-1}$.  The total systematic error
on our measure of $H_0$ includes the systematic error in the calibration, the
zero-point error, the SN Ia distance error, and a band-to-band systematic
measured using the I-band and W1 and W2, and amounts to $\pm\ 2.4$ \kms\
Mpc$^{-1}$.  Thus our best value is $H_0 = 74.4 \pm1.4$(stat) $\pm\ 2.4$(sys)
\kms\ Mpc$^{-1}$.  Our estimates of $H_0$ do not favor the low values of $H_0$
presented in \citet{Collaboration:13:5076} and \citet{Hinshaw:13:19}, although
relativistic corrections may resolve this tension as suggested in
\citet{Clarkson:14:7860}.



\acknowledgments

We acknowledge useful conversations with the following people: Wendy
Freedman, Barry Madore, Eric Persson, and Andrew Monson.

JDN and MS acknowledge support from the NASA Astrophysical Data Analysis
Program under grant NNX12AE19G for the {\it WISE} Nearby Galaxies Atlas.

HC and JS acknowledge support from the Lyon Institute of Origins under
grant ANR-10-LABX-66 and from CNRS under PICS-06233.  RBT acknowledges
support from the US National Science Foundation award AST09-08846.  We
acknowledge the use of the HyperLeda database (http://leda.univ-lyon1.fr).

This publication makes use of data products from the {\it Wide-field
Infrared Survey Explorer (WISE)}, which is a joint project of the
University of California, Los Angeles, and the Jet Propulsion
Laboratory/California Institute of Technology, funded by the National
Aeronautics and Space Administration.

This research has made use of the NASA/IPAC Extragalactic Database (NED)
which is operated by the Jet Propulsion Laboratory, California Institute of
Technology, under contract with the National Aeronautics and Space
Administration.




\bibliographystyle{apj}
\bibliography{papers}



\end{document}